\documentclass[preprint,3p,times,12pt]{elsarticle}
\pdfoutput=1 

\usepackage[utf8]{inputenc}
\usepackage[english]{babel}
\usepackage{makeidx}
\usepackage{amsfonts}
\usepackage{hyperref, amsmath, amssymb, epsfig, graphicx}
\usepackage{enumerate}
\usepackage{mathrsfs}
\usepackage{tensor}
\usepackage{xcolor,tikz,pgfplots}
\usepackage[autostyle]{csquotes}

\usepackage{subcaption}
\usepackage{caption}
\usepackage{sidecap}

\usepackage{youngtab}

\usepackage{braket}
\usepackage[normalem]{ulem} 

\newcommand\subcap[1]{\phantomcaption%
       \caption*{\textbf{\figurename~\thefigure\thesubfigure:} #1}}

\usetikzlibrary{matrix,calc,positioning,decorations.markings,decorations.pathmorphing,decorations.pathreplacing}
\usetikzlibrary{arrows,cd,shapes}
\usepackage{bbding}
\tikzset{
    >=stealth',
    punkt/.style={
           rectangle,
           rounded corners,
           draw=black, very thick,
           text width=7.4em,
           minimum height=2em,
           text centered},
    punkt2/.style={
           rectangle,
           rounded corners,
           draw=black!20!red, very thick,
           text width=7em,
           minimum height=2em,
           text centered},
    punktL/.style={
           rectangle,
           rounded corners,
           draw=black!20!red, very thick,
           text width=8.8em,
           minimum height=2em,
           text centered},
    pil/.style={
           ->,
           thick,
           shorten <=2pt,
           shorten >=2pt,},
    pil2/.style={
           <->,
           thick,
           shorten <=2pt,
           shorten >=2pt,}}

\usepackage{float}
\usepackage{booktabs}

\usepackage{bm}

\journal{Nuclear Physics B}

\begin{document}

\begin{frontmatter}


\title{Suspended Fixed Points}

\author[label1]{Andrea Antinucci\fnref{1}}
\author[label2]{Massimo Bianchi\fnref{2}}
\author[label3,label4]{Salvo Mancani\fnref{3}}
\author[label4]{Fabio Riccioni\fnref{4}}

\address[label1]{SISSA, via Bonomea 265, 34136 Trieste, Italy}  
\address[label2]{Dipartimento  di  Fisica,  Universit\`a  di  Roma  ``Tor  Vergata'', \\
	Sezione INFN Roma ``Tor Vergata'', \\
	Via della Ricerca Scientifica 1, 00133, Roma, Italy}
\address[label3]{Dipartimento di Fisica, Universit\`a di Roma “La Sapienza”, Piazzale Aldo Moro 2, 00185 Roma, Italy}
\address[label4]{INFN Sezione Roma1, Dipartimento di Fisica, Universit\`a di Roma “La Sapienza”, Piazzale Aldo Moro 2, 00185 Roma, Italy}

\fntext[1]{aantinuc@sissa.it}
\fntext[2]{massimo.bianchi@roma2.infn.it}
\fntext[3]{salvo.mancani@uniroma1.it}
\fntext[4]{fabio.riccioni@roma1.infn.it}

\begin{abstract}
We study the orientifold of the ${\mathcal{N}} = 1$ superconformal field theories describing D3-branes probing the Suspended Pinch Point singularity, as well as the orientifolds of non-chiral theories obtained by a specific orbifold $\mathbb{Z}_n$ of SPP. We find that these models realize a mechanism analogous to the one recently found for the orientifold of the complex Calabi-Yau cone over the Pseudo del Pezzo surface PdP$_{3c}$: they all flow to a new IR fixed point such that the value of the $a$-charge is less than half the one of the oriented theory. We also find that the value of $a$ coincides with the charge of specific orientifolds of the toric singularities $L^{(\bar{n},\bar{n},\bar{n})}$ with $\bar{n}=3n/2$  for $n$ even or $L^{(\bar{n},\bar{n}+1,\bar{n})}$ with $\bar{n}=(3n{-}1)/2$  for $n$ odd, suggesting the existence of an IR duality.
\end{abstract}


\end{frontmatter}

\section{Introduction}

Gauge theories describing the world-volume of D3-branes probing the singularity of a toric Calabi-Yau (CY) cone~\cite{Beasley_2000} are quiver gauge theories, with unitary gauge groups and matter in bifundamental representations. These theories are expected to have a superconformal fixed point in the infrared, and  the AdS/CFT correspondence~\cite{Maldacena_1999, Gubser_1998, Witten:1998qj} relates this regime to IIB supergravity on an AdS background whose internal part is the base of the cone~\cite{Morrison:1998cs, Klebanov:1998hh}. The geometry of the singularity defines, up to Seiberg dualities, the gauge theory, determining the amount of supersymmetry, the number of gauge groups, the matter content and the superpotential. 
At the conformal fixed point, the superconformal $R$-symmetry allows to determine the anomalous dimensions of all gauge invariant operators. In general, if additional $U(1)$ flavour symmetries are present, the superconformal $R$-charges are not determined by symmetry arguments alone, but they are uniquely fixed by the requirement that they maximize the central charge~\cite{Intriligator:2003jj, Bertolini:2004xf}
\begin{equation}\label{eq:amax}
a = \frac{3}{32} \left( 3 \mathrm{Tr}R^3 - \mathrm{Tr}{R} \right) \;  \ .
\end{equation}
In holographic models $a=c\sim N^2$, where $N$ is the number of colours. At fixed $N$, $a=c$ is inversely proportional to the volume of the base of the CY cone~\cite{Gubser_1998amax}.  

In string theory one can consider the additional possibility of including orientifold planes ($\Omega$-planes)~\cite{Sagnotti:1987tw, Pradisi:1988xd, Bianchi:1990yu, Bianchi:1990tb, Polchinski:1995mt, Angelantonj:2002ct}, which induce a ${\mathbb{Z}_2}$ involution on the space and make the strings unoriented. On the gauge theory side, this results in  more general gauge theories, allowing orthogonal and symplectic groups, as well as matter content in symmetric and antisymmetric representations. The presence of $\Omega$-planes modifies the RG flow, and two different scenarios have been investigated in the literature.  In the first scenario there is a fixed point, and the $R$-charges of the operators that are not projected out by $\Omega$ are the same as the charges of the corresponding oriented theory (the {\it parent} theory) in the large $N$ limit. This results in a central charge $a^\Omega$ that is half the $a$ charge of the parent theory  in this limit. In the second scenario the unoriented
theory does not have a fixed point, and one can have a duality cascade~\cite{Argurio:2017upa} or conformal symmetry can be restored by the inclusion of flavour branes~\cite{Bianchi:2013gka}.

In~\cite{Antinucci:2020yki} a third possibility, which was dubbed {\it third scenario}, was shown to occur. In a specific model, namely the gauge theory corresponding to D3-branes probing 
the third Pseudo del Pezzo singularity PdP$_{3c}$~\cite{Feng_2003, Feng_2004, Hanany:2012hi}, one can construct an $\Omega$ projection in such a way that the resulting theory flows for any $N$ to an IR fixed point whose superconformal $R$-charges are different from those of the parent even at large $N$. The resulting central charge $a^\Omega_{PdP_{3c}}$ is less than half $a_{PdP_{3c}}$, the one of the oriented parent theory, in the large $N$ limit. This occurs because the number of flavour $U(1)$ charges that take part in determining the superconformal $R$-charge in the parent theory is larger than the analogous number in the orientifold. Specifically, in the parent theory the non-$R$ symmetry which mixes with the $R$-charge is $U(1)^3$, while in the orientifold one flavour $U(1)$ is broken and the remaining $U(1)^2$ mixes with the $R$-symmetry. 

Even more surprisingly, the analysis of~\cite{Antinucci:2020yki} shows that  the values of $a^\Omega_{PdP_{3c}}$ and of the $R$-charges coincide for any $N$ with the ones  of another unoriented theory, the one associated to  PdP$_{3b}$,  which is another   Pseudo del Pezzo singularity.
In turn, the PdP$_{3b}$ orientifold realizes the first scenario above, {\it i.e.} $a^\Omega_{PdP_{3b}} = {1\over 2} a_{PdP_{3b}}$, in the large $N$ limit, and the  non-$R$ symmetry which mixes with the $R$-charge is $U(1)^2$ both in the parent theory and in the orientifold.


The orientifold projection is usually believed to modify the $R$-charges only at subleading orders. In the specific model studied in~\cite{Antinucci:2020yki} these subleading corrections break the superconformal symmetry of the parent theory and the fact that $a$-maximization gives a new fixed point suggests that the theory flows to a new conformal fixed point in the infrared. In this sense, the third scenario stands as a novel possibility not considered before, and it is natural to investigate whether such scenario can occur in other orientifold models.

Note that, even if $a$ maximization gives an $a$-charge in agreement with this possibility, it is not necessarily guaranteed that this value corresponds to the endpoint of an allowed RG flow. As discussed in~\cite{Kutasov:2003iy}, what can invalidate the procedure is the emergence of some gauge-invariant chiral operator ${{\cal C}}$ such that the value of its $R$-charge, determined by $a$-maximation, is $R({{\cal C}}) \leq \frac{2}{3}$. This implies that before the theory reaches the fixed point, ${{\cal C}}$ becomes a free field ($\Delta({{\cal C}})= 1 = 3R({{\cal C}})/2$) and an accidental abelian symmetry is generated, therefore the whole maximization procedure has to be reconsidered. In a situation of this type, performing Seiberg dualities might lead to a better understanding of the physics. Unfortunately, in orientifold models, where there are gauge groups with matter in representations different from the (anti)fundamental and with specific superpotentials, the rules to construct Seiberg duals are not always known. Note that in the case of PdP$_{3c}$ the fact that the $a$ charge and $R$-charges are the same as in the PdP$_{3b}$ case guarantees that such issues do not occur. Indeed, the two theories have the same superconformal index and they only differ in the superpotential.

As far as the gravity side of the correspondence is concerned, the geometric interpretation of the infrared duality emerging from the third scenario, that would put the construction on a firmer ground, is presently lacking due to the complexity of the geometry.  It is therefore of great interest to find other examples in the third scenario, in the hope that a general geometric picture would emerge.
 
In this paper we show that an infinite class of unoriented toric theories, the orientifold projections of non-chiral theories resulting from orbifolds of the Suspended Pinch Point (SPP), realize the third scenario. We refer to them as $\mathrm{SPP}/\mathbb{Z}'_n$ for the parent and $(\mathrm{SPP}/\mathbb{Z}'_n)^{\Omega}$ for the orientifold. We first discuss the case of the orientifold of SPP, whose gauge symmetry is the product of a unitary group and a symplectic or an orthogonal group. By imposing the vanishing of the $\beta$-functions, we naively find a solution for arbitrary ranks of the two groups. We find that in general the resulting values of the $R$-charges lead to gauge invariant operators that become free fields and decouple from the dynamics, precisely as described above. 
 We discuss the correction to the central charge $a$ due to these operators becoming free. Fixing the rank of one of the gauge groups, we find that only for a finite number of choices of the other rank the resulting theory in the infrared has $\mathrm{Tr}R=0$, and we consider all the other cases unphysical. Unfortunately, in this case no Seiberg dualities are known with the superpotential at hand~\cite{Kutasov:1995np, Intriligator:1995ff, Leigh:1995qp, Intriligator:1995ax}, at least to our knowledge. We then study the orientifolds of the orbifold theories, and show how Seiberg dualities confirm the existence of the conformal point.

In analogy with~\cite{Antinucci:2020yki}, we then look for other orientifold theories that are in the first scenario and whose value of the superconformal $a$-charge coincides with the value that we find for $(\mathrm{SPP}/\mathbb{Z}_n)^{\Omega}$. The parent SPP theory can be obtained by mass deformation from the orbifold $\mathbb{C}^2/\mathbb{Z}_3\times\mathbb{C}$~\cite{Bianchi:2014qma, Bianchi:2020fuk}, dubbed from now on $\mathbb{C}^3/\mathbb{Z}'_3$, and thus for any orbifold $\mathbb{Z}_n$ we look for theories that result from mass deformations of non-chiral orbifold theories. These theories belong to the family of the $L^{a,b,a}$ theories~\cite{Cvetic:2005ft, Martelli:2005wy, Franco:2005sm}, and remarkably we find that orientifolds of $L^{\frac{3n}{2},\frac{3n}{2},\frac{3n}{2}}$ for $n$ even and $L^{\frac{3n-1}{2},\frac{3n+1}{2},\frac{3n-1}{2}}$ for $n$ odd have a superconformal fixed point with a value of the $a$ charge which coincides with that of $(\mathrm{SPP}/\mathbb{Z}'_n)^{\Omega}$. Besides, $\left( L^{\frac{3n}{2},\frac{3n}{2},\frac{3n}{2}}\right)^{\Omega}$ ($n$ even) realise the first scenario and therefore constitute an infinite class of models in which the mechanism described in~\cite{Antinucci:2020yki} is realised.

The plan of the paper is as follows. In Sec.~\ref{Sec:ConformalOmega} we give motivations for this work, describe the line of reasoning and summarize the results. In Sec.~\ref{Sec:SPPZn} we construct the non-chiral theories SPP$/\mathbb{Z}'_n$ and find their maximal central charge $a$. In Sec.~\ref{Sec:MassDef} we discuss different patterns of mass deformation of $\mathbb{C}^3/\mathbb{Z}'_{3n}$ to SPP$/\mathbb{Z}'_n$, $L^{\bar{n},\bar{n},\bar{n}}$ or $L^{\bar{n},\bar{n}+1,\bar{n}}$, and their Seiberg duals. In Sec.~\ref{Sec:SPPZnOmega} we find the conformal point of the orientifold projections of SPP$/\mathbb{Z}'_n$ and show that they belong to the third scenario. In Sec.~\ref{Sec:OmegaDeformation} we show that the unoriented theories obtained by mass deformation of the $\mathbb{C}^3/\mathbb{Z}'_{3n}$ share the same central charge, 't Hooft anomalies and superconformal index. In particular, for $n$ even this happens between models in third scenario and first scenario. In Sec.~\ref{sec:Elliptic} we construct the type IIA brane model related to our classes of theories. With the help of these elliptic models, we provide another evidence for the conformal point we find. Finally, Sec.~\ref{Sec:Discussion} contains a discussion of our results and perspectives on future work.

\section{Unoriented Conformal Theories}\label{Sec:ConformalOmega}

In this section we first review the main results of~\cite{Antinucci:2020yki}, which are the starting point for the present work. We then give motivations for this project and outline the rest of the paper.

\begin{figure}
\begin{center}
\begin{subfigure}{0.30\textwidth}
\centering{\begin{tikzpicture}[auto, scale= 0.34]
        \node [circle, draw=blue!50, fill=blue!20, inner sep=0pt, minimum size=5mm] (3) at (5,-2.5) {$2$};            
        \node [circle, draw=blue!50, fill=blue!20, inner sep=0pt, minimum size=5mm] (5) at (-5,-2.5) {$4$};
        \node [circle, draw=blue!50, fill=blue!20, inner sep=0pt, minimum size=5mm] (6) at (-5,2.5) {$5$}; 
        \node [circle, draw=blue!50, fill=blue!20, inner sep=0pt, minimum size=5mm] (2) at (5,2.5) {$1$}; 
        \node [circle, draw=blue!50, fill=blue!20, inner sep=0pt, minimum size=5mm] (1) at (0,6) {$0$};
        \node [circle, draw=blue!50, fill=blue!20, inner sep=0pt, minimum size=5mm] (4) at (0,-6) {$3$}; 
        \draw (1)  to node {} (2) [->, thick, ];
        \draw (2)  to node {} (3) [->, thick, ];
        \draw (3)  to node {} (4) [->, thick, ];     
        \draw (4)  to node {} (5) [->, thick, ];
        \draw (5)  to node {} (6) [->, thick, ];
        \draw (6)  to node {} (1) [->, thick, ];
        \draw (1)  to node {} (4) [<->, thick, ];
        \draw (6)  to node {} (2) [->, thick, ];
        \draw (2)  to node {} (4) [->, thick, ];
        \draw (4)  to node {} (6) [->, thick, ];
        \draw (3)  to node {} (5) [->, thick, ];
        \draw (5)  to node {} (1) [->, thick, ];
        \draw (1)  to node {} (3) [->, thick, ];

        \draw [thick, dashed, gray] (0, -8) to node [pos=0.01] {$\Omega$} (0,8) ;
        \end{tikzpicture}
        \subcap{The quiver of theories PdP$_{3b}$ and PdP$_{3c}$. The dashed gray line labelled as $\Omega$ represents the orientifold projection, which identifies the two sides of the quiver and projects fields and gauge groups that lie on top of it.} \label{fig:quiverPdP3bc}}
\end{subfigure}
\hfill
\begin{subfigure}{0.30\textwidth}
\centering{
\includegraphics[scale=0.28, trim={7.2cm 2cm 12.4cm 2.1cm}, clip]{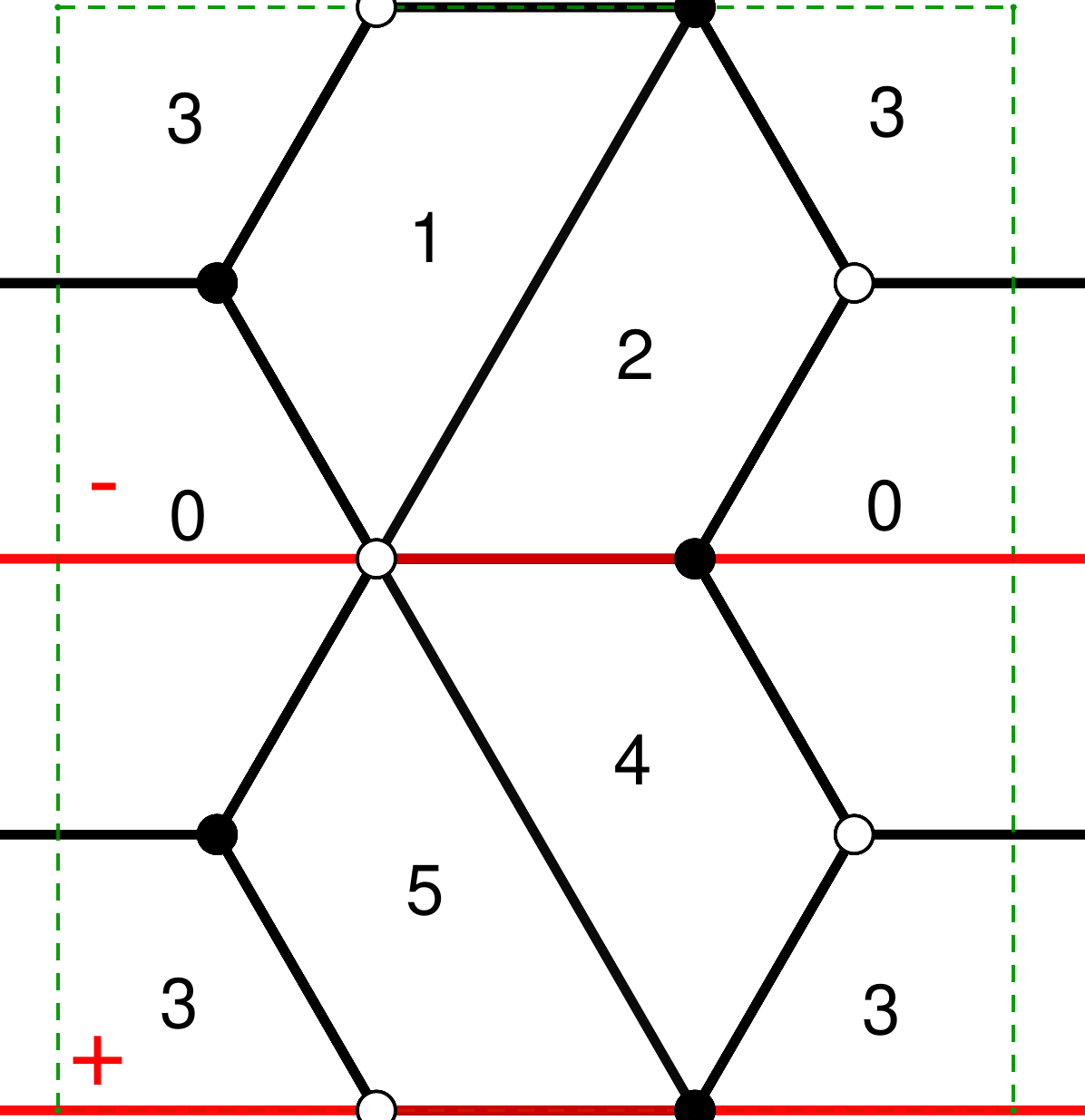}
\vspace{25pt}
\subcap{dimer of PdP$_{3b}$, where the dashed green line delimits the fundamental cell. The two red fixed lines and their signs represent the orientifold projection that yields the unoriented PdP$_{3b}^{\Omega}$.}\label{fig:PdP3bDimer}}
\end{subfigure}
\hfill
\begin{subfigure}{0.30\textwidth}
\centering{
\includegraphics[scale=0.32, trim={7.5cm 0.8cm 14.3cm 1.5cm}, clip]{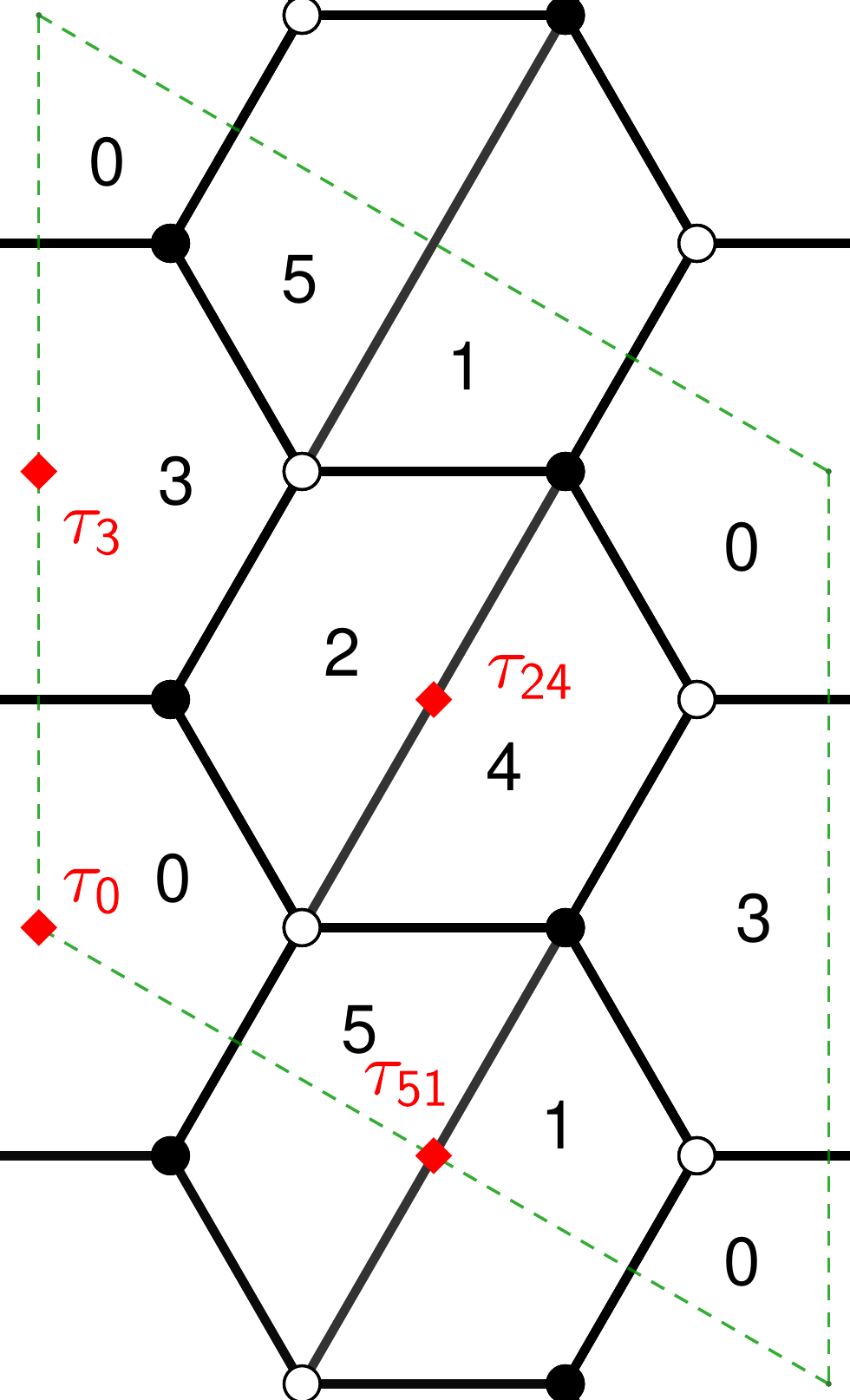}
\vspace{15pt}
\subcap{The dimer of PdP$_{3c}$, where the dashed green line delimits the fundamental cell. The four red fixed points $(\tau_0,\,\tau_3,\tau_{24},\,\tau_{51})$ represent the orientifold projection, where $(+,-,-,+)$ corresponds to PdP$_{3c}^{\Omega_1}$ and $(-,+,-,+)$ corresponds to PdP$_{3c}^{\Omega_2}$.}\label{fig:PdP3cDimer}}
\end{subfigure}
\end{center}
\end{figure}

In~\cite{Antinucci:2020yki} the orientifold projections $\Omega$ of theories over the surfaces Pseudo del Pezzo (PdP) $3b$ and $3c$ were analyzed, and in particular their superconformal central charges $a^{\Omega}$ were found and compared. The toric parent theories PdP$_{3b}$ and PdP$_{3c}$ can be represented via a  brane tiling of a torus and their gauge groups and matter content easily read. Both have the same gauge group $\prod_{i=0}^5 \; SU(N_i)$ and matter content, as showed in Fig.~\ref{fig:quiverPdP3bc}, and global symmetries $U(1)^2\times U(1)_R$ as mesonic ones and $U(1)^3$ as non-anomalous baryonic ones. The orientifold involution of toric models is discussed in~\cite{Bianchi:2007wy, Franco:2007ii, Argurio:2020dko, Garcia-Valdecasas:2021znu}: on the brane tiling, orientifold planes are represented as the fixed loci of the $\mathbb{Z}_2$ involution of the fundamental cell, either with fixed points or fixed lines, all carrying the charge of the orientifold plane, indicated by $\tau$. Gauge groups and bifundamental fields are identified with respect to them. In case of fixed points, toric isometries $U(1)^2\times U(1)_R$ are preserved, while fixed lines break the non-$R$ symmetries into a diagonal combination.

PdP$_{3b}$ is projected with fixed lines, see Fig.~\ref{fig:PdP3bDimer}, resulting in a theory with gauge group $SO/Sp(N_0) \times SU(N_1)\times SU(N_2) \times Sp/SO(N_3)$ and mesonic flavour symmetry $U(1) \times U(1)_R$. The signs of the fixed lines $(\pm,\mp)$ determines the nature of the orthosymplectic gauge factor, where a + is associated with $SO$ and a $-$ with $Sp$. These signs also determine the symmetry properties of the projected bifundamentals $X_{24}$ and $X_{51}$ which give rise to matter in the rank-two symmetric (with +) and antisymmetric (with $-$) tensors. Determining the conformal point of the unoriented PdP$_{3b}^{\Omega}$ via maximization of the two-variable function $a$, one finds that $a^{\Omega}/a = 1/2$, namely, the degrees of freedom are halved by the orientifold projection, which keeps the number of global abelian symmetry mixing with the $R$-symmetry as in the parent. This affects the $R$-charges and the central charge $a$ only at subleading order. This is the first scenario for an unoriented theory, as opposed to the second scenario in which the conformal point does not exist.

On the other hand, PdP$_{3c}$ is projected via fixed points, see Fig.~\ref{fig:PdP3cDimer}, preserving toric symmetries. Following the rules of~\cite{Franco:2007ii}, the signs of these fixed points must obey $\prod \tau = (-1)^{N_W/2}$, where $N_W$ is the number of terms in the superpotential of the parent theory. Two inequivalent choices of $(\tau_0,\,\tau_3,\tau_{24},\,\tau_{51})$ are allowed, $\Omega_1=(\pm,\, \mp ,\, \mp,\, \pm)$ and $\Omega_2=(\mp,\, \pm,\, \mp,\, \pm)$. The signs project, in order, gauge group 0 and 3 and bifundamental fields $X_{24}$ and $X_{51}$. The choice $\Omega_1$ leads to a conformal point that belongs to the first scenario. This is not the case for $\Omega_2$, in which the $R$-charges and the maximal central charge $a$ are affected by the orientifold already at leading order, due to the breaking of a $U(1)$ flavour symmetry. Consequently, $a^{\Omega}/a < 1/2$ and d.o.f. result to be more than halved at the fixed point, which, for the $a$-theorem, has been moved towards the infrared. This is the third scenario found in~\cite{Antinucci:2020yki}. 

Interestingly, the orientifolds PdP$_{3c}^{\Omega_2}$ and PdP$_{3b}^{\Omega}$ share the same gauge factors and matter content, and even more surprisingly the two models have exactly the same $R$-charges and central charge $a$. As a consequence, their 't Hooft anomalies and superconformal indices trivially match,
and since the two theories only differ because of superpotential terms one expects that they are connected by an exactly marginal deformation. The fact that both theories are orientifold projections of toric models suggests that they could be dual~\cite{Antinucci:2020yki}.

In this work we find other examples in which the third scenario occurs and, for a subset of them, their counterpart in the first scenario.  
One may ask whether for all these models the conformal point in the third scenario is physical. Since dimensions of gauge-invariant operators are modified and the fixed point is moved towards the IR, some operators may decouple before reaching the conformal point. As a consequence, accidental $U(1)$ flavour factors are generated. This is discussed in~\cite{Kutasov:2003iy}, where it is pointed out that one must correct the computation of the central charge $a$ by taking into account the fact that a chiral operator $\mathcal{C}$ has hit the bound $R\leq 2/3$ and become free. Only $\mathcal{C}$ is charged under the accidental $U(1)$ symmetry, which corrects its $R$ charge so that it remains 2/3. The $a$-charge must be corrected taking into account this accidental abelian factor as
\begin{align}\label{eq:CorrectionACharge}
\widetilde{a}^{\Omega} &= a^{\Omega} + \frac{3}{32} \left.\left( 3 \textrm{Tr } R^3 - \textrm{Tr } R \right)\right|_{R=2/3} - \frac{3}{32} \left( 3 \textrm{Tr } R_{\mathcal{C}}^3 - \textrm{Tr } R_{\mathcal{C}} \right) \nonumber \\[7pt]
&= a^{\Omega} + \frac{1}{96} \left(2 - 3 R_{\mathcal{C}} \right)^2\left( 5 - 3 R_{\mathcal{C}} \right) \; , 
\end{align}
where $\widetilde{a}^{\Omega}$ and $a^{\Omega}$ are the corrected and the uncorrected central charges, respectively. For each gauge-invariant chiral operator which crosses the free-field bound there is a correction term of this form. Note that, unless $R_{\mathcal{C}}=2/3$, when such decoupling occurs the overall $\textrm{Tr } R\neq 0$, which is harmless for the gauge field theory but spoils the holographic duality. This will occur in some cases in our analysis, and theories of this type will not be considered physical.


A further, and stronger, check that we will perform in this paper consists in finding the conformal point of a Seiberg dual theory. Agreement with the electric theory would provide solid evidence for the physical existence of the conformal fixed point. However, the context itself of unoriented models is subtle, due to the presence of tensorial matter and/or orthogonal or symplectic gauge groups. While several Seiberg dualities are known for such cases, in most of them the required superpotential terms are not allowed for toric theories. When this is the case, all one can say is that a fixed point seems to exist for the theory, but further analysis is needed.

In order to find more models that belong to the third scenario, we note that PdP$_{3c}$ is the chiral orbifold SPP$/\mathbb{Z}_2$. Thus, it is natural to investigate the orientifold projection of other orbifolds of SPP, both chiral and non-chiral theories. The former case is computationally complicated as $n$ grows, while $a$-maximization for unoriented theories in the latter case can be easily generalized, as we shall see. Indeed, their orientifold projections belong to the third scenario. For certain solutions, it happens that some operators decouple before the theory reaches the conformal point, then the central charge $a$ must be revisited, as discussed above. Finally, we apply Seiberg duality to find the magnetic dual and its conformal point. 

Furthermore, the class of SPP$/\mathbb{Z}'_n$ theories can be obtained by a certain mass deformation of $\mathbb{C}^3/\mathbb{Z}'_{3n}$. Different choices of mass deformation yield another class of theories, $L^{\bar{n},\bar{n},\bar{n}}$ for $\bar{n}=3n/2$ and $n$ even, and $L^{\bar{n},\bar{n}+1,\bar{n}}$ for $\bar{n}=(3n-1)/2$ and $n$ odd, not related by Seiberg duality to SPP$/\mathbb{Z}'_n$. These two classes provide a natural ground for searching signs of a relation as in~\cite{Antinucci:2020yki}, where fixed-point models in the third scenario share the same central charge $a$, 't Hooft anomalies and superconformal index with fixed-line models in the first scenario. This would be a stronger evidence of the existence of the conformal point in the third scenario. This happens between SPP$/\mathbb{Z}'_n$ and $L^{\bar{n},\bar{n},\bar{n}}$ for $n$ even, while for $n$ odd both theories belong to the third scenario. Interestingly, for $n$ even the orientifold projection of the theory $\mathbb{C}^3/\mathbb{Z}'_{3n}$ also belongs to the third scenario and for some choices of the ranks it again features the same central charge $a$, 't Hooft anomalies and superconformal index as the aforementioned pair of unoriented models. These results are summarized in the chart in Fig.~\ref{fig:Scheme}, which serves as a guide for the reader.

\begin{figure}
\begin{center}
\begin{tikzpicture}[node distance=1cm,auto, scale=0.45]
		\node (0) {};
		\node[punkt, above left=2.5cm of 0] (even) {$\prod \tau = +1$};
		\node [below=2.2cm of even] (1) {};
		\node [punkt, left=0.7cm of 1] (Om1) {\begin{tabular}{l}
		$\Omega_B$\\[4pt] $(\pm,\,\pm,\,\mp,\,\mp)$\\[4pt]
		$a^{\Omega_B}=\frac{81 n N^2 p^2 \tau_0}{16\left( p+2\tau_0\right)^3}$\\[4pt]
		$\frac{a^{\Omega_B}}{a}<\frac{1}{2}$
		\end{tabular}};
		\node [punkt, right=0.7cm of 1] (Om2) {\begin{tabular}{l}
		$\Omega_A$\\[4pt] 
		$(\pm,\,\mp,\,\pm,\,\mp)$\\[4pt]
		$a^{\Omega_A}=\frac{3}{8}nN^2$\\[4pt]$\frac{a^{\Omega_A}}{a}=\frac{1}{2}$
		\end{tabular}};
		\node[punkt, right=2cm of 0] (odd) {\begin{tabular}{l}
		$\prod \tau = -1$\\[4pt] 
		$(\pm,\,\mp,\,\mp,\,\mp)$\\[4pt]
		$a^{\Omega}=\frac{3}{8}nN^2$\\[4pt]
		$\frac{a^{\Omega}}{a}=\frac{1}{2}$
		\end{tabular}};
        \node[punkt, above=3cm of 0] (C3Z) {$\left( \mathbb{C}^3/\mathbb{Z}'_{3n}\right)^{\Omega}$};
        \node[below=1cm of odd, color=black!30!red, inner sep=0pt, minimum size=0] (2) {};
        \node[punkt2, below left=0.9cm of 2] (SPPZodd) {\begin{tabular}{l}
        $\left(\textrm{SPP}/\mathbb{Z}'_n\right)^{\Omega}$\\[4pt]
        $\prod \tau = +1$\\[4pt]
        $(\pm,\,\pm,\,\mp,\,\mp)$\\[4pt]        
        \end{tabular}};
        \node[punkt2, below right=0.9cm of 2] (Laba) {\begin{tabular}{l}
        $\left(L^{\bar{n},\bar{n}+1,\bar{n}}\right)^{\Omega}$\\[4pt]
        $\prod \tau = +1$\\[4pt]
        $(\pm,\,\pm,\,\mp,\,\mp)$        
        \end{tabular}};   
        \node[below=7cm of even] (3) {}; 
        \node[below=4.5cm of even, color=black!30!red] (33) {\begin{tabular}{c}Mass\\[1pt]Deformation\end{tabular}};   
        \node[punkt2, below left=0.5cm of 3] (SPPZeven) {\begin{tabular}{l}
        $\left(\textrm{SPP}/\mathbb{Z}'_n\right)^{\Omega}$\\[4pt]
        $\prod \tau = +1$\\[4pt]
        $(\pm,\,\pm,\,\mp,\,\mp)$\\[4pt]  
        $\frac{a^{\Omega}}{a}<\frac{1}{2}$\\[4pt]      
        \end{tabular}};
        \node[punkt2, below right=0.5cm of 3] (Laaa) {\begin{tabular}{l}
        $\left(L^{\bar{n},\bar{n},\bar{n}}\right)^{\Omega}$\\[4pt]
        Fixed lines\\[4pt]
        $(\pm,\mp)$\\[4pt]
        $\frac{a^{\Omega}}{a}=\frac{1}{2}$        
        \end{tabular}};

        \draw (C3Z.west) [pil,bend right=30,swap,pos=0.7] to node  {\begin{tabular}{l}$n$ even\\[4pt]$\bar{n}=3n/2$\\[4pt] $\left(\tau_{_0},\,\tau_{_{00}},\,\tau_{_{\bar{n}}},\,\tau_{_{\bar{n},\bar{n}}}\right)$ \end{tabular}} (even);
        \draw (C3Z.east) [pil,bend left=40,pos=0.4] to node  {\begin{tabular}{l}$n$ odd\\[4pt] $\bar{n}=(3n-1)/2$\\[4pt] $\left(\tau_{_0},\,\tau_{_{00}},\,\tau_{_{\bar{n},\bar{n}+1}},\,\tau_{_{\bar{n}+1,\bar{n}}}\right)$ \end{tabular}} (odd);
        \draw (even.south west) [pil] to node {} (Om1);
        \draw (even.south east) [pil] to node {} (Om2);
        \draw (odd) [-, thick, color=black!30!red] to node {Mass Deformation} (2);
        \draw (2) [->, thick, color=black!30!red, bend right=15] to node {} (SPPZodd.north);
        \draw (2) [->, thick, color=black!30!red, bend left=12.5] to node {} (Laba.north);
        \draw (SPPZodd.south) [pil2, color=black!60!green, bend right=35, swap] to node {\begin{tabular}{c}
        $a^{\Omega}=\frac{81}{256} n N^2$\\[4pt]
        $\frac{a^{\Omega}}{a} < \frac{1}{2}$        
\end{tabular}}
        (Laba.south);
        \draw (even) [pil, color=black!30!red, pos=0.8] to node {} (33);
        \draw (33) [pil, color=black!30!red, bend right=25] to node {} (SPPZeven);
        \draw (33) [pil, color=black!30!red, bend left=25] to node {} (Laaa);
        \draw (SPPZeven.south) [pil2, color=black!60!green, bend right=35, swap] to node {\begin{tabular}{c}
        $a^{\Omega}=\frac{81}{256} n N^2$\\[4pt]    
\end{tabular}}
        (Laaa.south);
        \draw (SPPZeven.north west) [pil2, color=black!60!green, bend left=15,pos=0.55, swap] to node {\begin{tabular}{l}
        $p=2\tau_0$
        \end{tabular}}
        (Om1.south); 
\end{tikzpicture}
\end{center}
\caption{The web of unoriented dualities found between $\mathbb{C}^3/\mathbb{Z}'_{3n}$, SPP$/\mathbb{Z}'_n$ and $L^{k,n-k,k}$.}\label{fig:Scheme}
\end{figure}

\section{SPP and its non-chiral orbifold $\mathbb{Z}'_n$}\label{Sec:SPPZn}

\begin{figure}
\centering{\includegraphics[scale=1, trim= 6cm 3cm 10cm 2.5cm, clip]{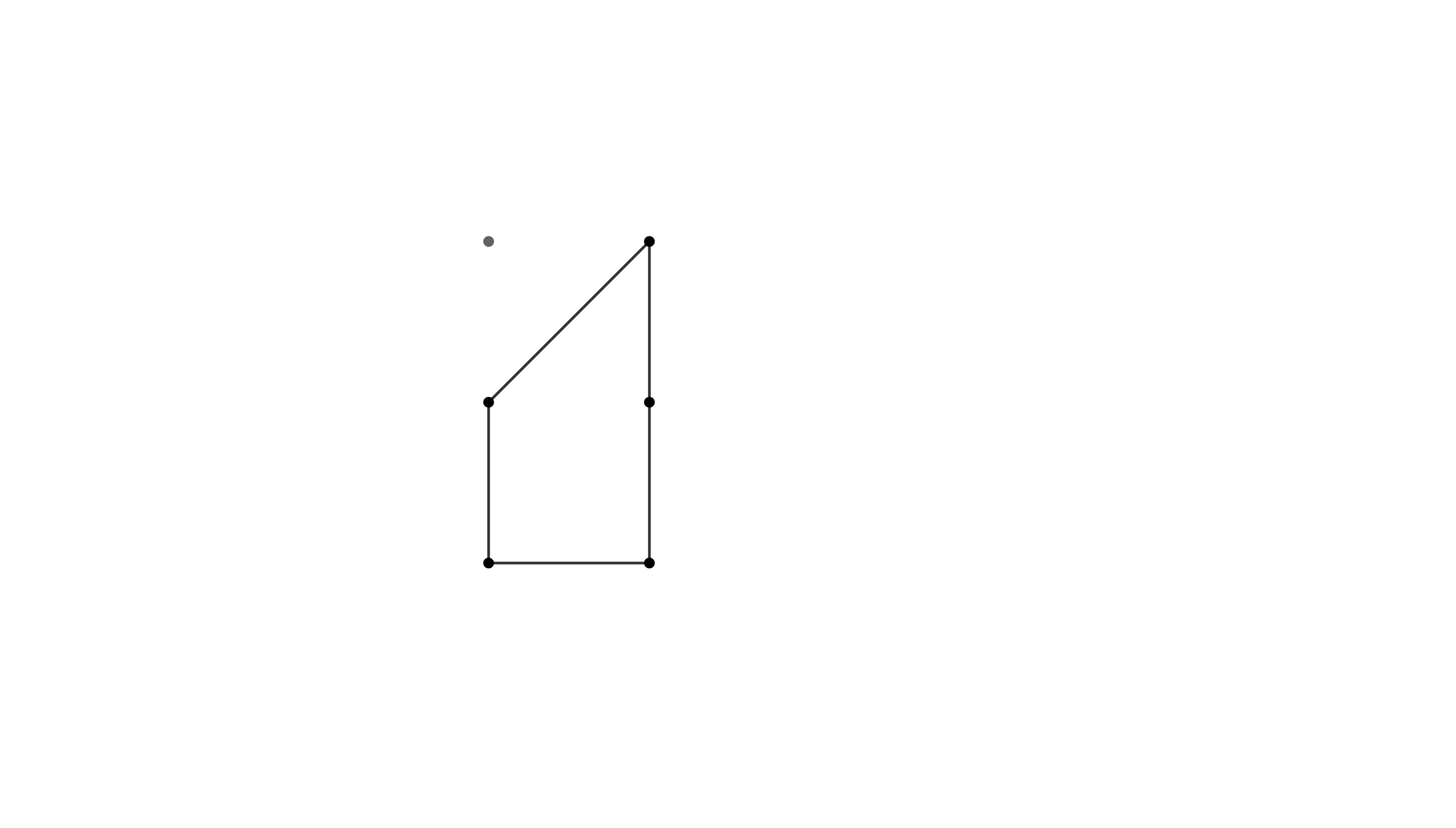}}
\caption{The toric diagram of the SPP singularity.}\label{fig:ToricSPP}
\end{figure}

\begin{figure}
\begin{center}
\begin{subfigure}{0.45\textwidth}
\centering{
\begin{tikzpicture}[auto, scale=0.45]
		\node [circle, draw=blue!50, fill=blue!20, inner sep=0pt, minimum size=5mm] (0) at (0,5) {$0$}; 
		\node [circle, draw=blue!50, fill=blue!20, inner sep=0pt, minimum size=5mm] (1) at (3,0) {$1$}; 
		\node [circle, draw=blue!50, fill=blue!20, inner sep=0pt, minimum size=5mm] (2) at (-3,0) {$2$};
        \draw (0) to node {} (1) [<->, thick];
		\draw (1) to node {} (2) [<->, thick];
		\draw (2) to node {} (0) [<->, thick];
		\draw (0) to [out=130, in=50, looseness=13] (0) [<->, thick] ;
		\draw (0,-1) to node [pos=0.01] {$\Omega$} (0,8) [dashed, draw=gray];
\end{tikzpicture}\subcap{The quiver diagram for SPP. The line $\Omega$ represents the orientifold projection.}\label{fig:SPP}}
\end{subfigure}
\hfill
\begin{subfigure}{0.45\textwidth}
\centering{
\includegraphics[scale=0.5, trim={5.7cm 5cm 10cm 4.1cm}, clip]{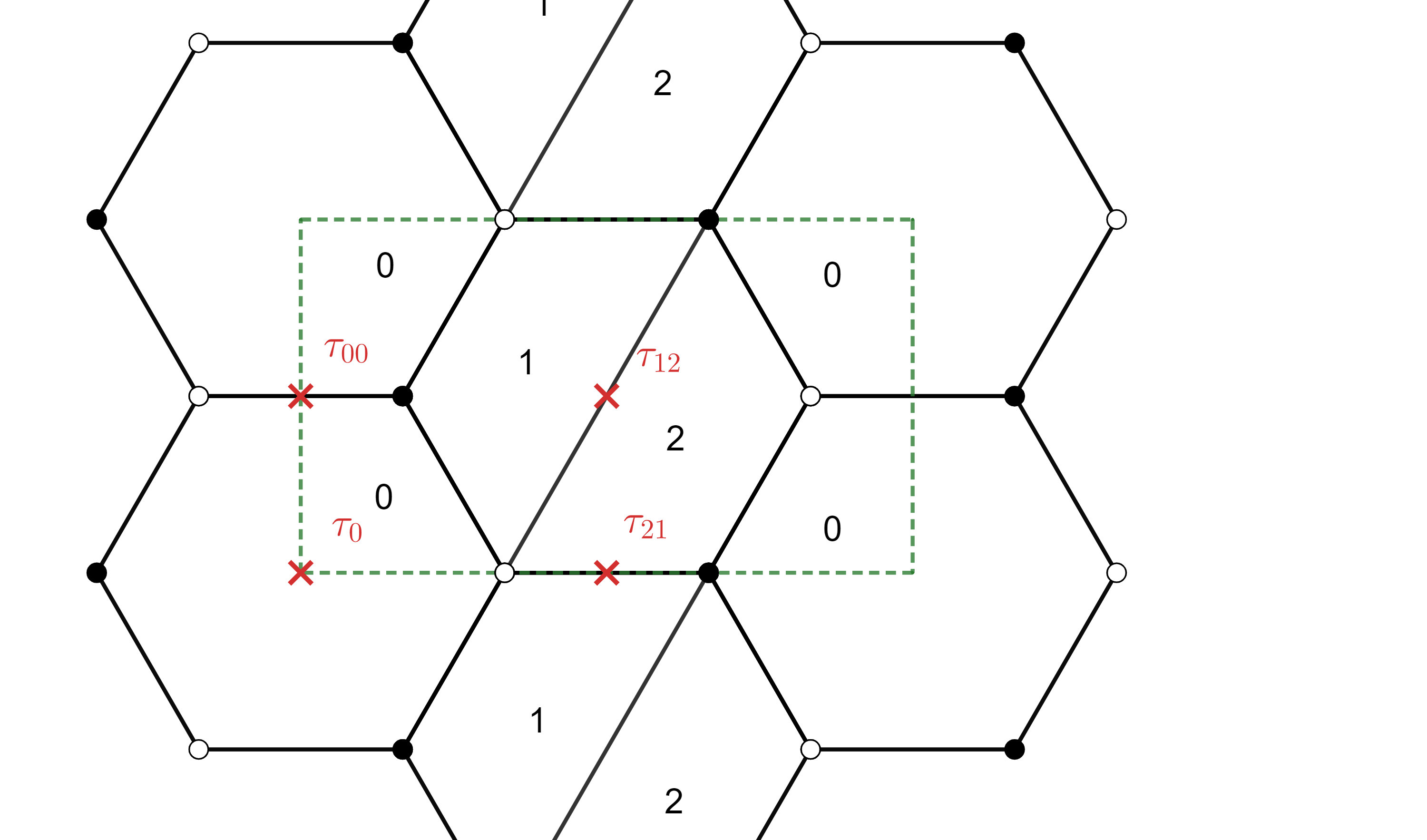}
\vspace{8pt}
\subcap{The dimer of the SPP theory, the four fixed points of the orientifold projection are drawn in red.}\label{fig:SPPdimer}}
\end{subfigure}
\end{center}
\end{figure}

The Suspended Pinch Point (SPP) is a (non-isolated) toric singularity, that can be realized as an affine variety in $\mathbb{C}^4$ with the relation
\begin{align}
x y = z^2 w \; ,
\end{align}
with $x,\,y,\,z,\,w\, \in \mathbb{C}^4$. The singularity is represented by the toric diagram in Fig.~\ref{fig:ToricSPP}, which has no internal points, signalling that the associated gauge theory is non-chiral. 
The gauge group is $U(N_0)\times U(N_1)\times U(N_2)$, while the matter content corresponds to six chiral fields denoted by $X_{ij}$ ($i\neq j$), transforming  under the fundamental representation of $U(N_i)$ and the anti-fundamental of $U(N_j)$, together with the chiral field $X_{00}$, that we denote by $\phi_0$,\footnote{In general we denote adjoint chiral fields $X_{ii}$ by $\phi_i$.} in the adjoint on the group $U(N_0)$.  We draw the quiver and the dimer of the theory in Figs.~\ref{fig:SPP} and \ref{fig:SPPdimer}.
The superpotential reads 
\begin{align}
W_{_\textrm{SPP}} = \phi_0 \left( X_{02}X_{20} - X_{01}X_{10} \right) + X_{12}X_{21}X_{10}X_{01} - X_{21}X_{12}X_{20}X_{02} \; ,\label{eq:SPPsuperpotential}
\end{align}
as can be deduced from the dimer. 

\begin{figure}
\begin{subfigure}{0.45\textwidth}
\begin{center}
\begin{tikzpicture}[auto, scale=0.45]
		\node [circle, draw=blue!50, fill=blue!20, inner sep=0pt, minimum size=5mm] (0) at (0,5) {$0$}; 
		\node [circle, draw=blue!50, fill=blue!20, inner sep=0pt, minimum size=5mm] (1) at (3,0) {$1$}; 
		\node [circle, draw=blue!50, fill=blue!20, inner sep=0pt, minimum size=5mm] (2) at (-3,0) {$2$};
        \draw (0) to node {} (1) [<->, thick];
		\draw (1) to node {} (2) [<->, thick];
		\draw (2) to node {} (0) [<->, thick];
		\draw (0) to [out=130, in=50, looseness=13] (0) [<->, thick];
		\draw (2) to [out=250, in=170, looseness=13] (2) [<->, thick, dashed];
		\draw (1) to [out=10, in=290, looseness=13] (1) [<->, thick, dashed];
\end{tikzpicture}
\subcap{The quiver diagram for $\mathbb{C}^3/\mathbb{Z}'_3$. Giving mass to the adjoint fields represented by dashed lines yields SPP.}\label{fig:C3Z3prime}
\end{center}
\end{subfigure}
\hfill
\begin{subfigure}{0.45\textwidth}
\vspace{-25pt}
\centering{\includegraphics[scale=0.25, trim=7.2cm 2.5cm 17.5cm 2.6cm, clip]{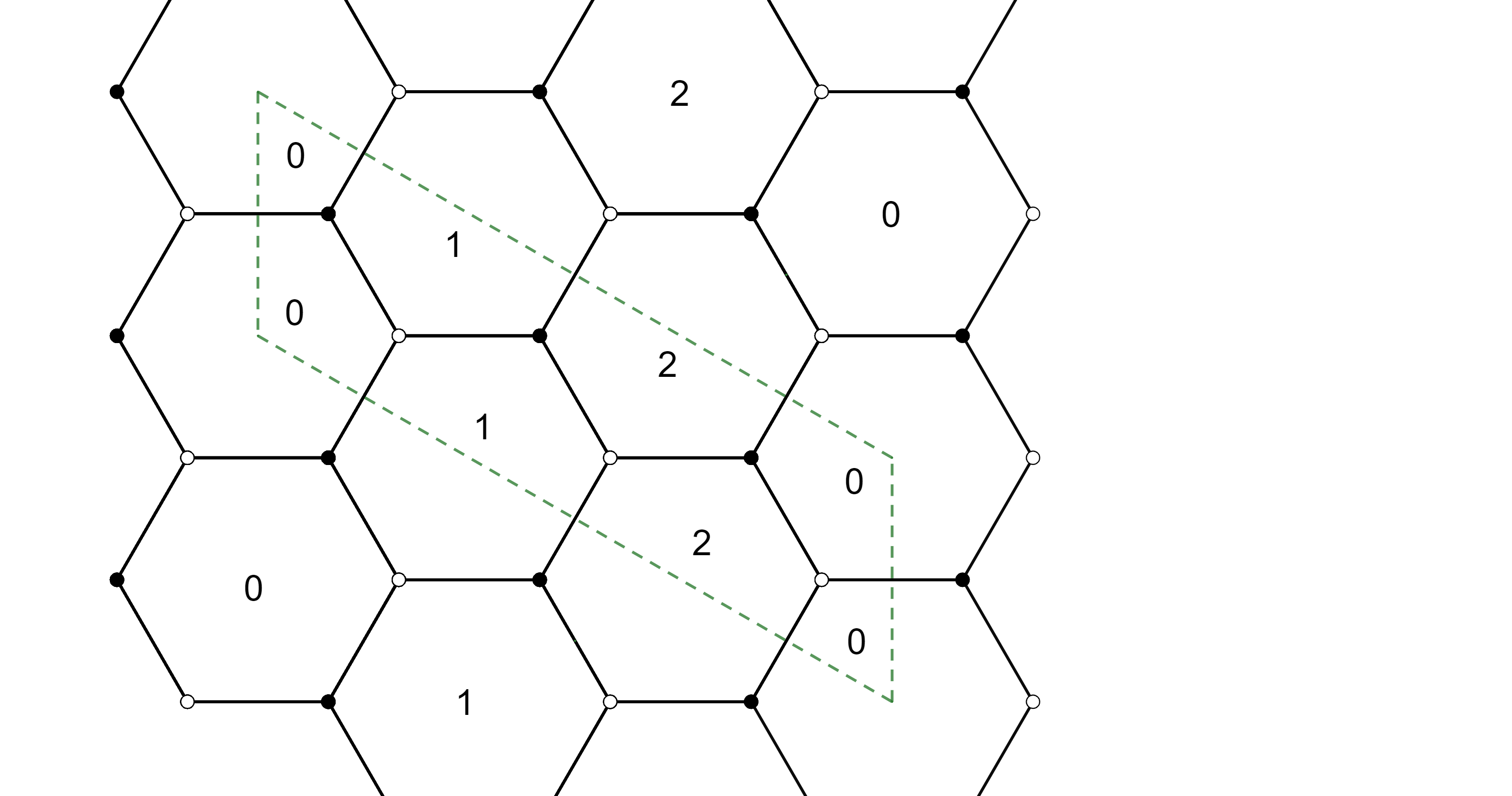}}
\vspace{20pt}
\subcap{The brane tiling for $\mathbb{C}^3/\mathbb{Z}'_3$.}\label{fig:TilingC3Z3}
\end{subfigure}
\end{figure}

The SPP theory can be obtained by mass deformation of another toric theory, the non-chiral orbifold of flat space $\mathbb{C}^3/\mathbb{Z}'_3$~\cite{Bianchi:2014qma}. Its graphical representation as dimer and quiver is shown in Fig.~\ref{fig:C3Z3prime}-\ref{fig:TilingC3Z3}.
The superpotential reads
\begin{align}
W_{_{\mathbb{C}^3/\mathbb{Z}'_3}} &= \phi_0 \left( X_{02}X_{20} - X_{01}X_{10} \right) + \phi_1 \left( X_{10}X_{01} - X_{12}X_{21} \right) + \phi_2 \left( X_{21}X_{12} - X_{20}X_{02} \right)
\end{align}
which reduces to the superpotential of SPP adding 
the mass term 
\begin{align}
\Delta W_{_{\mathbb{C}^3/\mathbb{Z}'_3}} = {M\over 2} \left( \phi_1^2 - \phi_2^2 \right)
\end{align}
and integrating the massive fields out. Plugging F-terms into the superpotential and redefining fields as 
\begin{align}
&X'_{21}X'_{12} = \frac{1}{M} X_{21}X_{12} \; , \nonumber \\[5pt]
&\phi_0' = \phi_0 + \frac{1}{2M} \left( X_{01}X_{10} - X_{02}X_{20} \right)
\end{align}
gives the superpotential of SPP.

We now review how the $R$-charges and the central charge $a$ of the conformal SPP theory are determined using $a$-maximization. 
We denote the $R$-charges of $X_{ij}$ by
\begin{align}
R_{ij} = r_{ij} + 1 \; , 
\end{align} 
where $r_{ij}$ is the $R$-charge of the fermionic field in the multiplet. First, we impose the constraint $R(W)=2$, that gives
\begin{align}\label{eq:RchargesSPP}
& r_{01} + r_{10} + r_{00}= -1 \; ,  \nonumber \\
& r_{12} + r_{21} - r_{00}= -1 \; , \nonumber \\
& r_{20} + r_{02} = r_{01} + r_{10} \; . 
\end{align}
This constraint implies that  all the $r$'s must  satisfy $-1<r<1$. Moreover, the $\mathbb{Z}_2$ symmetry of the quiver implies $r_{12} =r_{21}$, $r_{02}=r_{01}$ and $r_{20}=r_{10}$. The condition that the beta functions vanish (which in turn is equivalent to  the $R$-symmetry being anomaly-free) gives
\begin{align}\label{eq:RchargesAnomaly}
&r_{00} \left( 2N_0 - N_1 - N_2 \right) = - \left( 2N_0 - N_1 - N_2 \right) \; , \nonumber \\ 
&r_{00} \left( N_1 - N_2 \right) = \left( 2N_1 - N_0 - N_2 \right) \; , \nonumber \\ 
&r_{00} \left( N_0 - N_1 \right) = \left( 2N_2 - N_1 - N_0 \right) \; ,
\end{align}
where we have used Eq.~\eqref{eq:RchargesSPP}. 
We note that $2N_0 - N_1 - N_2 \neq 0$ would imply $r_{00}=-1$, violating unitarity.  We therefore impose $N_0 = N_1 = N_2 = N$, which leaves $r_{00}$  undetermined. Note that Eq.~\eqref{eq:RchargesSPP} is invariant under the exchange $r_{ji}\leftrightarrow r_{ij}$, then we put them equal. This is inherited from the $\mathcal{N}=2$ $\mathbb{C}^3/\mathbb{Z}'_3$ and its superpotential before mass deformation. Hence, we have a one-parameter family of solutions, corresponding to the fact that there is one non-anomalous $U(1)$ flavour symmetry that can in principle redefine the $R$-charge. The superconformal $R$-charge  is then determined by $a$-maximization. In particular, defining
$r_{01}=x$, the $a$-charge\footnote{Observe that $\textrm{Tr }R=0$ at leading order in $N$ for holographic theories.} at leading order in $N$
\begin{align}
a_{_{\textrm{SPP}}} = \frac{9}{32} \textrm{Tr }R^3 =\frac{9 N^2}{32} \left[ \left( -1-2x \right)^3 + 4 \left( x \right)^3 + 2(-1-x)^3 + 3 \right] \; 
\end{align}
has a local maximum at~\cite{Franco:2005sm}
\begin{align}
&r_{00} = 1 - \frac{2}{\sqrt{3}} \; , \qquad r_{12} = - \frac{1}{\sqrt{3}} \; , \qquad r_{01} = r_{10} =- 1 + \frac{1}{\sqrt{3}} \; , \label{eq:RchargesSPPmax}
\end{align}
which gives the superconformal $a$-charge
\begin{align}
a_{_{\textrm{SPP}}} = \frac{3 \sqrt{3}}{8} N^2 \; .
\end{align}

\subsection{Non-chiral orbifold SPP/$\mathbb{Z}'_n$}\label{Sec:Zn}

\begin{figure}
\begin{center}
\begin{subfigure}{0.4\textwidth}
\centering{\includegraphics[scale=0.6, trim={4cm 0 8cm 0}, clip]{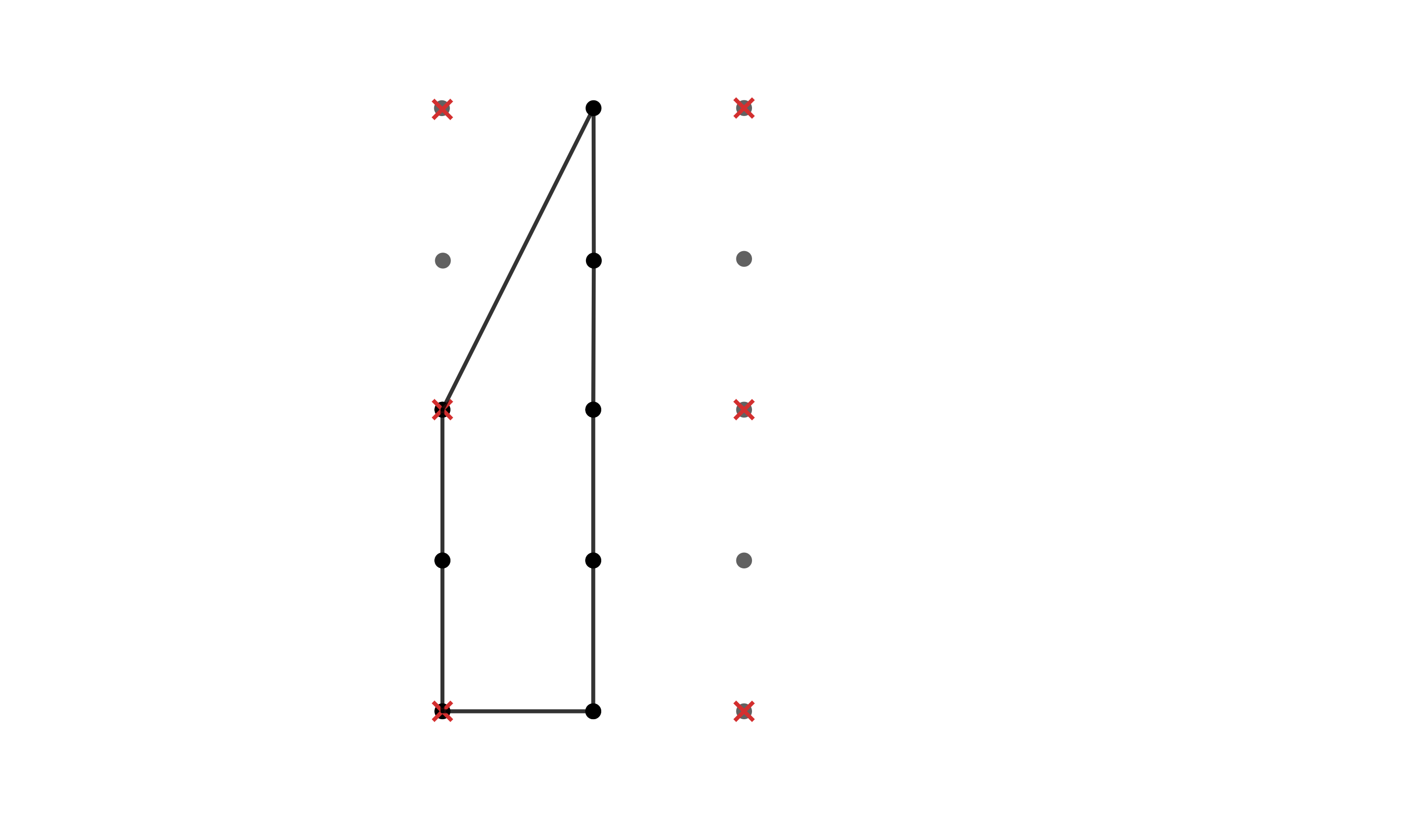}
\subcap{The toric diagram of SPP$/\mathbb{Z}'_2$, a.k.a. $L^{(2,4,2)}$, where fixed points of the orientifold projection are drawn in red.}\label{fig:SPPZ2toric}}
\end{subfigure}
\hfill
\begin{subfigure}{0.4\textwidth}
\centering{
\begin{tikzpicture}[auto, scale=0.4]
		\node [circle, draw=blue!50, fill=blue!20, inner sep=0pt, minimum size=5mm] (0) at (0,2.7) {$0$}; 
		\node [circle, draw=blue!50, fill=blue!20, inner sep=0pt, minimum size=5mm] (1) at (3.7,0) {$1$}; 
		\node [circle, draw=blue!50, fill=blue!20, inner sep=0pt, minimum size=5mm] (2) at (3.7,-4) {$2$}; 
		\node [circle, draw=blue!50, fill=blue!20, inner sep=0pt, minimum size=5mm] (3) at (0,-6.7) {$3$}; 		
		\node [circle, draw=blue!50, fill=blue!20, inner sep=0pt, minimum size=5mm] (4) at (-3.7,-4) {$4$};
		\node [circle, draw=blue!50, fill=blue!20, inner sep=0pt, minimum size=5mm] (5) at (-3.7,0) {$5$};
        \draw (0) to node {} (1) [<->, thick];
		\draw (1) to node {} (2) [<->, thick];
		\draw (2) to node {} (3) [<->, thick];
		\draw (3) to node {} (4) [<->, thick];
		\draw (4) to node {} (5) [<->, thick];
		\draw (5) to node {} (0) [<->, thick];
		\draw (0) to [out=130, in=50, looseness=11] (0) [<->, thick] ;
		\draw (3) to [out=310, in=230, looseness=11] (3)[<->, thick] ;
		\draw (0,-10) to node [pos=0.01, gray]{$\Omega$} (0,6) [dashed, gray];
\end{tikzpicture}
\subcap{The quiver of SPP$/\mathbb{Z}'_2$, a.k.a. $L^{(2,4,2)}$. The dashed line represents the orientifold projection.}\label{fig:SPPZ2}}
\end{subfigure}
\end{center}
\end{figure}

Starting from the SPP geometry, one can construct additional models by considering abelian orbifolds SPP$/\Gamma$. As we have already mentioned, a particular $\mathbb{Z}_2$ orbifold results in the PdP$_{3c}$ geometry, which is the model whose properties were the main motivation for the present work. There is another  $\mathbb{Z}_2$  involution that can be performed, resulting in the toric geometry denoted by $L^{2,4,2}$ in the literature, and whose toric diagram is given in Fig.~\ref{fig:SPPZ2toric}.  As will be discussed in the next section, this geometry leads to two different toric phases, and we are interested in particular in the one corresponding to the quiver in Fig.~\ref{fig:SPPZ2}, which can be seen as arising from a $\mathbb{Z}_2$ involution on the SPP gauge theory. The resulting gauge theory has six unitary gauge groups, it is non-chiral, and we denote it by SPP$/\mathbb{Z}'_2$. The superpotential is 
\begin{align}
W_{_{{\textrm{SPP}}/\mathbb{Z}'_2}} &= \phi_{0} \left( X_{05}X_{50} - X_{01}X_{10} \right) + 
\phi_{3} \left( X_{32}X_{23} - X_{34}X_{43} \right)
+  X_{10}X_{01}X_{12}X_{21} \nonumber \\[3pt] & -  X_{21}X_{12}X_{23}X_{32} 
+  X_{43}X_{34}X_{45}X_{54} -  X_{54}X_{45}X_{50}X_{05}  \; ,
\end{align}
and it can be explicitly obtained from the $\mathbb{Z}'_2$ action on the SPP superpotential in Eq.~\eqref{eq:SPPsuperpotential}.

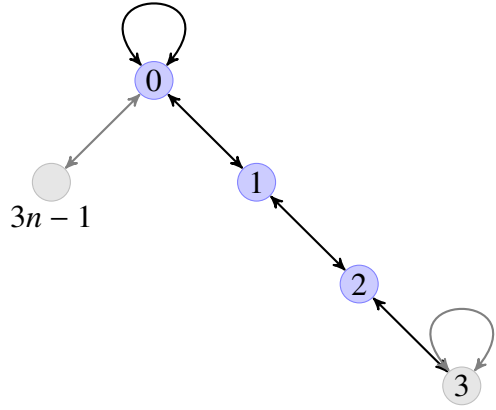
\begin{figure}
\begin{center}
\begin{tikzpicture}[auto, scale=0.45]
		\node [circle, draw=blue!50, fill=blue!20, inner sep=0pt, minimum size=5mm] (0) at (0,3) {$0$}; 
		\node [circle, draw=blue!50, fill=blue!20, inner sep=0pt, minimum size=5mm] (1) at (3,0) {$1$}; 
		\node [circle, draw=blue!50, fill=blue!20, inner sep=0pt, minimum size=5mm] (2) at (6,-3) {$2$}; 
		\node [circle, draw=gray!50, fill=gray!20, inner sep=0pt, minimum size=5mm] (3) at (9,-6) {$3$}; 		
		\node [circle, draw=gray!50, fill=gray!20, inner sep=0pt, minimum size=5mm] (n) at (-3,0) {};
		\node (nn) at (-3,-1) {$3n-1$};
        \draw (0) to node {} (1) [<->, thick];
		\draw (1) to node {} (2) [<->, thick];
		\draw (2) to node {} (3) [<->, thick];
		\draw (n) to node {} (0) [<->, thick, gray];
		\draw (0) to [out=130, in=50, looseness=11] (0) [<->, thick] ;
		\draw (3) to [out=130, in=50, looseness=11] (3) [<->, thick, gray] ;
\end{tikzpicture}\caption{Blue nodes form the recursive structure of the quiver SPP$/\mathbb{Z}'_n$.}\label{fig:SPPZnprime}
\end{center}
\end{figure}

The non-chiral $\mathbb{Z}'_2$ orbifold discussed above belongs to an infinite family of non-chiral models SPP$/\mathbb{Z}'_n$, whose quivers correspond to a sequence on $n$ copies of the structure of nodes and arrows in Fig.~\ref{fig:SPPZnprime}, giving in total $3n$ unitary gauge groups, $n$ of which have matter in the adjoint. The associated geometry is known as $L^{n,2n,n}$ in the literature. 
The superpotential reads
\begin{align}
W_{_{{\textrm{SPP}}/\mathbb{Z}'_n}} &= \sum_{i=0}^{n-1}\phi_{3i} \left( X_{3i,\,3i-1}X_{3i-1,\,3i} - X_{3i,\,3i+1}X_{3i+1,\,3i} \right) \nonumber \\[3pt]
&+\sum_{i=0}^{n-1} \left( X_{3i+1,\,3i}X_{3i,\,3i+1}X_{3i+1,\,3i+2}X_{3i+2,\,3i+1} \right. \nonumber \\[3pt]
& \left. - X_{3i+2,\,3i+1}X_{3i+1,\,3i+2}X_{3i+2,\,3i+3}X_{3i+3,\,3i+2} \right) \; ,
\end{align}
where it is understood that the group labels of the fields are defined modulo $3n$.


In order to determine the $R$ charges and $a$ charge of the SPP$/\mathbb{Z}'_n$ theory at the conformal fixed point, we impose the constraints coming from the condition that the $R$ charge of the superpotential be equal to 2, {\it viz.}
\begin{align}\label{eq:RchargesSPPZn}
& r_{3i,3i+1} + r_{3i+1, 3i} + r_{3i,3i}= -1 \; ,  \nonumber \\
& r_{3i, 3i-1} + r_{3i-1,3i} + r_{3i,3i}= -1 \; , \nonumber \\
& r_{3i+1,3i} + r_{3i,3i+i} + r_{3i+1,3i+2} + r_{3i+2,3i+1}=-2 \; , \nonumber \\
& r_{3i+2,3i+1} + r_{3i+1,3i+2} + r_{3i+2,3i+3} + r_{3i+3,3i+2}=-2
 \; . 
\end{align}
with $i=0,...,n-1$. As discussed in the previous subsection, $r_{3i,3i+1} = r_{3i+1,3i}$. The symmetry of the quiver also allows to impose various constraints on the charges. First of all, the charges are invariant under shifts in $i$. Besides, the $\mathbb{Z}_2$ symmetry around each adjoint node implies that $r_{3i-1,3i} = r_{3i+1,3i}$, $r_{3i,3i-1} = r_{3i,3i+1}$ and $r_{3i+1,3i+2} = r_{3i+2,3i+1}$. Finally, the condition that all the beta functions vanish is solved imposing that all the gauge groups have equal rank $N$. Putting all this together, one can show that the $a$ charge is simply $n$ times the $a$ charge of the SPP theory. In particular, performing $a$ maximization gives~\cite{Franco:2005sm} 
\begin{align}
&r_{3i,3i} = 1 - \frac{2}{\sqrt{3}} \; , \qquad r_{3i+1,3i+2} = - \frac{1}{\sqrt{3}} \; , \qquad r_{3i,3i+1} =- 1 + \frac{1}{\sqrt{3}} \;  
\end{align}
as in Eq. \eqref{eq:RchargesSPPmax}, and the corresponding  maximized $a$-charge reads
\begin{align}
a_{_{\textrm{SPP}/\mathbb{Z}'_n}} = n \, a_{_{\textrm{SPP}}} = n \frac{3\sqrt{3}}{8}N^2 \; .
\end{align}
As a consequence of the orbifold involution, we see that the d.o.f. of the field theory increase with $n$.


\section{Mass Deformation of $\mathbb{C}^3/\mathbb{Z}'_{3n}$}\label{Sec:MassDef}

As we have already seen, the SPP theory can be obtained via mass deformation of $\mathbb{C}^3/\mathbb{Z}'_3$, giving mass to two of the adjoints. This is more general and we can recover SPP$/\mathbb{Z}'_n$ via mass deformation of $\mathbb{C}^3/\mathbb{Z}_{3n}$, giving mass to more pairs of adjoints. In particular, starting with the superpotential 
\begin{align}
W_{_{\mathbb{C}^3/\mathbb{Z}'_{3n}}} = \sum_{i=0}^{3n-1} \phi_{i}\left( X_{i,\,i-1}X_{i-1,\,i} - X_{i,\,i+1}X_{i+1,\,i} \right) \; 
\end{align}
and deforming it with 
\begin{align}
\label{eq:massdeftoSPPZn}
\Delta W_{_{\mathbb{C}^3/\mathbb{Z}'_{3n}}} = \sum_{i=0}^{n-1} \frac{M}{2} \left( \phi_{3i+1}^2 - \phi_{3i+2}^2 \right) \; ,
\end{align}
below the scale $M$ the effective theory reads 
\begin{align}
W &= \sum_{i=0}^{n-1}\phi_{3i} \left( X_{3i,\,3i-1}X_{3i-1,\,3i} - X_{3i,\,3i+1}X_{3i+1,\,3i} \right) \nonumber \\[3pt]
&+ \sum_{i=0}^{n-1} \left( X_{3i+1,\,3i}X_{3i,\,3i+1}X_{3i+1,\,3i+2}X_{3i+2,\,3i+1} \right. \nonumber \\[3pt]
& \left. - X_{3i+2,\,3i+1}X_{3i+1,\,3i+2}X_{3i+2,\,3i+3}X_{3i+3,\,3i+2} \right) \; ,
\end{align}
which is the superpotential of SPP$/\mathbb{Z}'_n$. Recall that SPP is the toric geometry $L^{1,2,1}$. In~\cite{Bianchi:2014qma}, it is pointed out that giving mass to contiguous $k$ pair of adjoint fields in $\mathbb{C}^3/\mathbb{Z}'_{3n}$, one obtains the toric theory $L^{k,3n-k,k}$. 

\begin{figure}
\begin{center}
\begin{subfigure}{0.45\textwidth}
\centering{
\includegraphics[scale=0.65, trim={0 0 5cm 0}, clip]{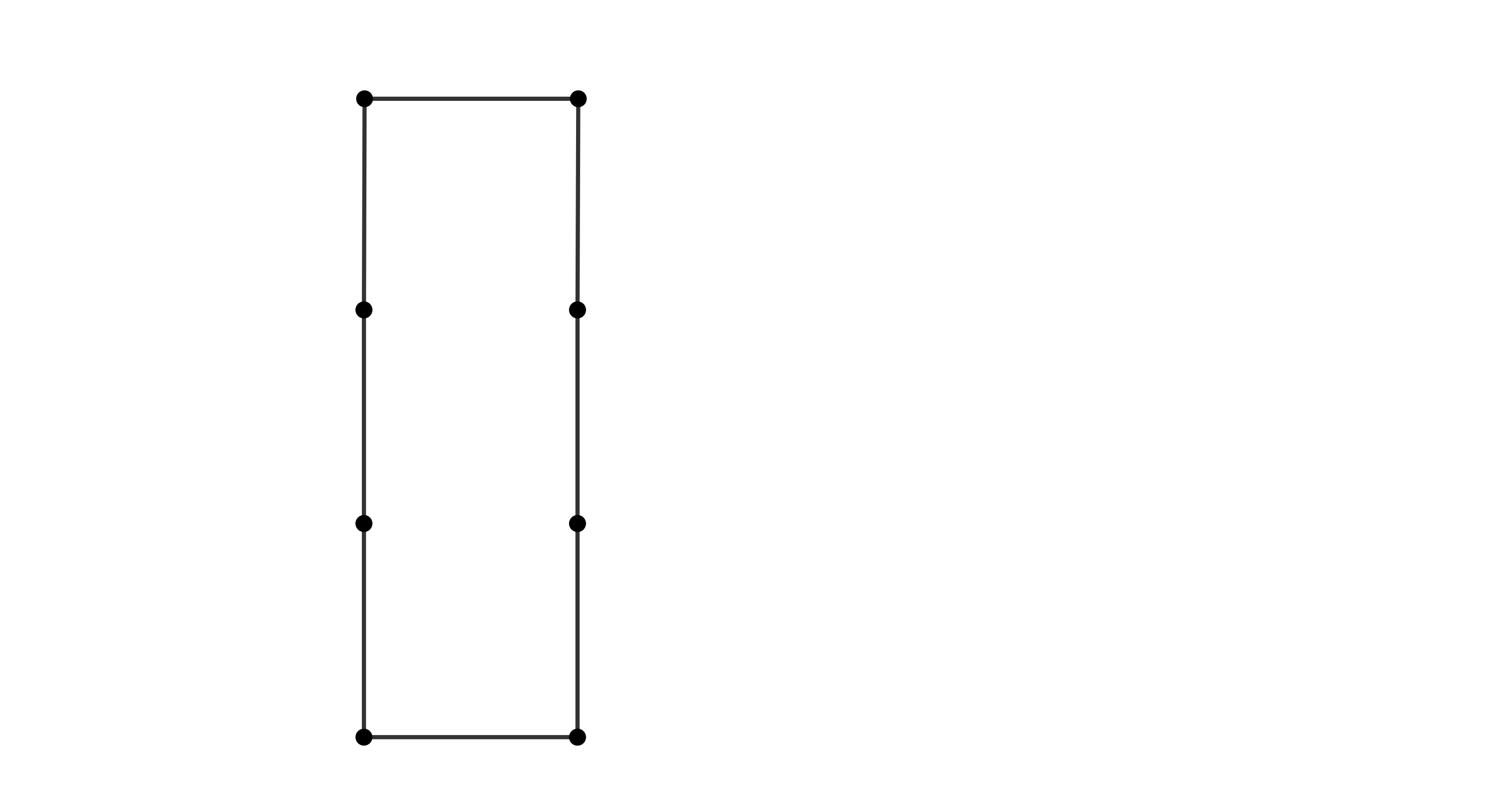}
\subcap{The toric diagram for $L^{3,3,3}$.}\label{fig:ToricL333}}
\end{subfigure}
\hfill
\begin{subfigure}{0.45\textwidth}
\centering{
\begin{tikzpicture}[auto, scale=0.38]
		\node [circle, draw=blue!50, fill=blue!20, inner sep=0pt, minimum size=5mm] (0) at (0,2.7) {$0$}; 
		\node [circle, draw=blue!50, fill=blue!20, inner sep=0pt, minimum size=5mm] (1) at (3.7,0) {$1$}; 
		\node [circle, draw=blue!50, fill=blue!20, inner sep=0pt, minimum size=5mm] (2) at (3.7,-4) {$2$}; 
		\node [circle, draw=blue!50, fill=blue!20, inner sep=0pt, minimum size=5mm] (3) at (0,-6.7) {$3$}; 		
		\node [circle, draw=blue!50, fill=blue!20, inner sep=0pt, minimum size=5mm] (4) at (-3.7,-4) {$4$};
		\node [circle, draw=blue!50, fill=blue!20, inner sep=0pt, minimum size=5mm] (5) at (-3.7,0) {$5$};
        \draw (0) to node {} (1) [<->, thick];
		\draw (1) to node {} (2) [<->, thick];
		\draw (2) to node {} (3) [<->, thick];
		\draw (3) to node {} (4) [<->, thick];
		\draw (4) to node {} (5) [<->, thick];
		\draw (5) to node {} (0) [<->, thick];
\end{tikzpicture}
\vspace{10pt}
\subcap{The quiver for $L^{3,3,3}$, obtained by mass deformation of all of the adjoints in $\mathbb{C}^3/\mathbb{Z}'_6$.}\label{fig:QuiverL333}}
\end{subfigure}
\end{center}
\end{figure}

\begin{figure}
\begin{center}
\begin{subfigure}{0.45\textwidth}
\centering{
\includegraphics[scale=0.5, trim={5cm 0.5cm 9cm 0}, clip]{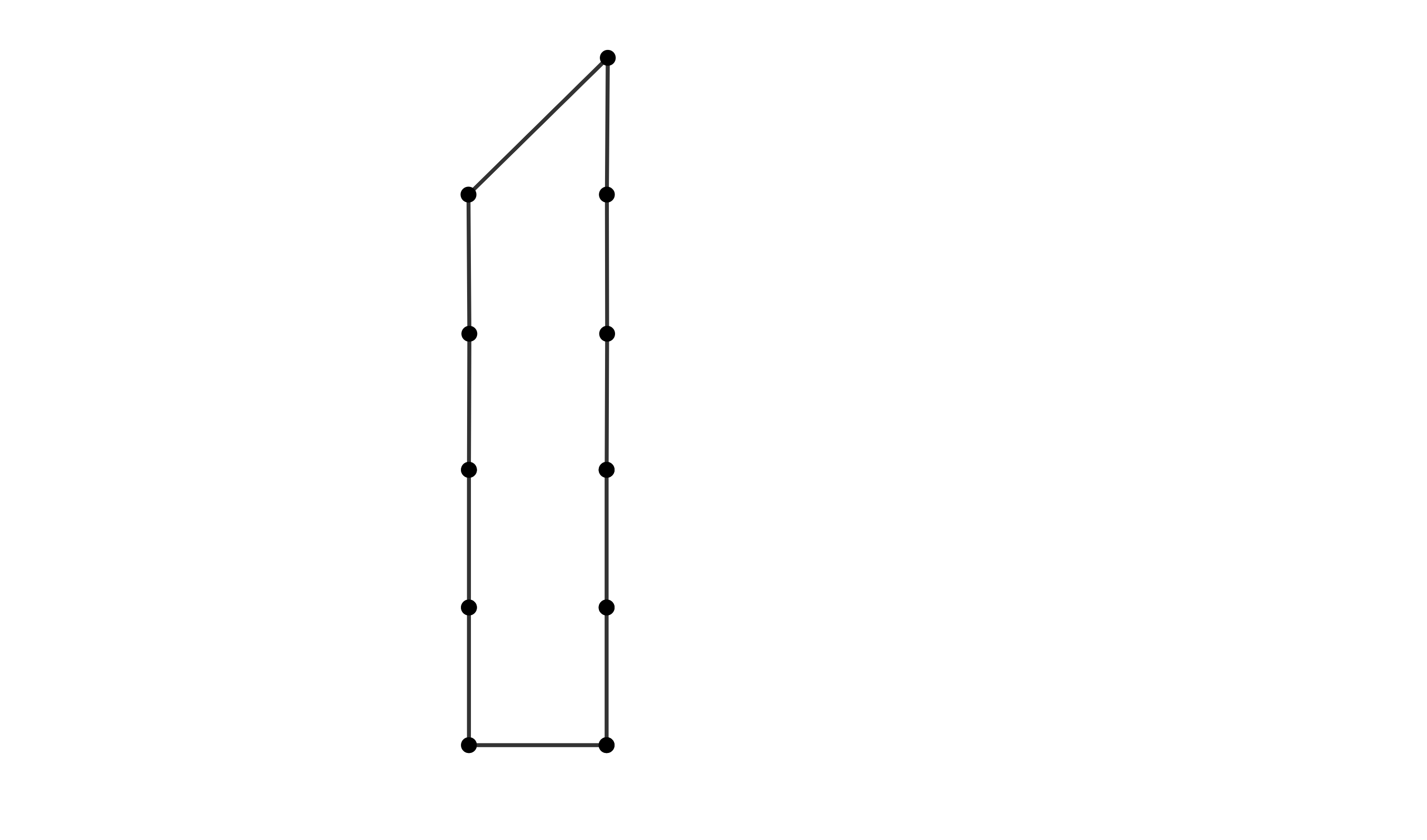}
\subcap{The toric diagram for $L^{4,5,4}$.}\label{fig:ToricL454}}
\end{subfigure}
\hfill
\begin{subfigure}{0.45\textwidth}
\centering{
\begin{tikzpicture}[auto, scale=0.32]
		\node [circle, draw=blue!50, fill=blue!20, inner sep=0pt, minimum size=5mm] (0) at (0,2.7) {$0$}; 
		\node [circle, draw=blue!50, fill=blue!20, inner sep=0pt, minimum size=5mm] (1) at (3.7,0) {$1$}; 
		\node [circle, draw=blue!50, fill=blue!20, inner sep=0pt, minimum size=5mm] (2) at (3.7,-4) {$2$}; 
		\node [circle, draw=blue!50, fill=blue!20, inner sep=0pt, minimum size=5mm] (3) at (3.7,-8) {$3$}; 		
		\node [circle, draw=blue!50, fill=blue!20, inner sep=0pt, minimum size=5mm] (4) at (3.7,-12) {$4$};
		\node [circle, draw=blue!50, fill=blue!20, inner sep=0pt, minimum size=5mm] (5) at (-3.7,-12) {$5$};
		\node [circle, draw=blue!50, fill=blue!20, inner sep=0pt, minimum size=5mm] (6) at (-3.7,-8) {$6$};
		\node [circle, draw=blue!50, fill=blue!20, inner sep=0pt, minimum size=5mm] (7) at (-3.7,-4) {$7$};
		\node [circle, draw=blue!50, fill=blue!20, inner sep=0pt, minimum size=5mm] (8) at (-3.7,0) {$8$};		
        \draw (0) to node {} (1) [<->, thick];
		\draw (1) to node {} (2) [<->, thick];
		\draw (2) to node {} (3) [<->, thick];
		\draw (3) to node {} (4) [<->, thick];
		\draw (4) to node {} (5) [<->, thick];
		\draw (5) to node {} (6) [<->, thick];
		\draw (6) to node {} (7) [<->, thick];
		\draw (7) to node {} (8) [<->, thick];
		\draw (8) to node {} (0) [<->, thick];
		\draw (0) to [out=130, in=50, looseness=11] (0) [<->, thick] ;
\end{tikzpicture}
\vspace{3pt}
\subcap{The quiver for $L^{4,5,4}$, obtained by mass deformation of all but one of the adjoints in $\mathbb{C}^3/\mathbb{Z}'_9$.}\label{fig:QuiverL454}}
\end{subfigure}
\end{center}
\end{figure}

If we perform a mass deformation such that the highest number of pairs of adjoints are integrated out, the resulting theory depends on whether $n$ is even or odd. In fact, for $n$ even we can integrate out all the adjoint fields, with $k=3n/2$, to obtain $L^{\frac{3n}{2},\frac{3n}{2},\frac{3n}{2}}$, whose toric diagram is a rectangle. In Fig.~\ref{fig:QuiverL333}-\ref{fig:ToricL333} we show an example for $n=2$. The final superpotential reads
\begin{align}
W_{_{{\frac{3n}{2},\frac{3n}{2},\frac{3n}{2}}}} & = \sum_{i=0}^{3n-1} \left( X_{i+1,\,i}X_{i,\,i+1}X_{i+1,\,i+2}X_{i+2,\,i+1} \right. \nonumber \\[3pt]
&\left.- X_{i+2,\,i+1}X_{i+1,\,i+2}X_{i+2,\,i+3}X_{i+3,\,i+2} \right) \; .
\end{align}
On the other hand, for $n$ odd at most we can integrate out $\frac{n-1}{2}$ pair of adjoints and we are left with a single adjoint field, which we can choose to be on node 0 without loss of generality. In this case we are left with $L^{\frac{3n-1}{2},\frac{3n+1}{2},\frac{3n-1}{2}}$, whose toric diagram is a trapezoid, see Fig.~\ref{fig:ToricL454}. The resulting superpotential reads
\begin{align}
W_{_{{\frac{3n-1}{2},\frac{3n+1}{2},\frac{3n-1}{2}}}} &= \phi_0 \left( X_{0, \, 3n-1}X_{3n-1,\,0} - X_{01}X_{10} \right)  \nonumber \\[3pt]
& +\sum_{i=0}^{3n-3} \left( X_{i+1,\,i}X_{i,\,i+1}X_{i+1,\,i+2}X_{i+2,\,i+1} \right. \nonumber \\[3pt]
&\left.- X_{i+2,\,i+1}X_{i+1,\,i+2}X_{i+2,\,i+3}X_{i+3,\,i+2} \right) \; .
\end{align}

\subsection{Web of Seiberg dualities}\label{sec:Seiberg}

As we have just shown, from $\mathbb{C}^3/\mathbb{Z}'_{3n}$ we can reach SPP$/\mathbb{Z}'_n$ by mass deforming pairs of adjoint fields in a particular pattern, which is the one given in Eq.~\eqref{eq:massdeftoSPPZn}. On the other hand, toricity is preserved as long as we give mass to an adjacent pair of adjoint fields, that is two adjoint fields whose gauge groups are connected in the quiver. As an example, for the case of $\mathbb{C}^3/\mathbb{Z}'_{6}$ ({\it i.e.} $n=2$) if we give mass to two adjacent pairs of adjoints we have two possibilities, up to symmetries: we can either give mass to the adjoints of the groups 1,2,4 and 5 or to the adjoints of 1,2,3 and 4. The resulting theories are two different toric phases of $L^{2,4,2}$, with the former being SPP$/\mathbb{Z}'_2$. If instead we give mass to a single pair or to all the adjoints, there is clearly only one possibility in each case, corresponding to $L^{1,5,1}$ and $L^{3,3,3}$ respectively.

This can be generalized to any $n$.  Starting from $\mathbb{C}^3/\mathbb{Z}'_{3n}$, we have one possibility if we give mass to a single pair, which corresponds to $L^{1,3n-1,1}$, while if we give mass to two pairs we have $\left[ \frac{3n}{2} \right] -1$ different $L^{2,3n-2,2}$ gauge theories. It is a combinatorial exercise to determine all possible theories that one obtains giving mass to $k$ pairs. If $k=n$ one gets 
$L^{n,2n,n}$, which contains SPP$/\mathbb{Z}'_n$.  If $n$ is even, one can remove all adjoint fields giving mass to $\frac{3n}{2}$ pairs, which gives
the  $L^{\frac{3n}{2},\frac{3n}{2},\frac{3n}{2}}$ theory. If $n$ is odd, one reaches the 
 $L^{\frac{3n-1}{2},\frac{3n+1}{2},\frac{3n-1}{2}}$ theory giving mass to $\frac{3n-1}{2}$ adjacent pairs.

\begin{figure}
\centering{
\begin{tikzpicture}[auto, scale=0.45]

		\node [circle, draw=blue!50, fill=blue!20, inner sep=0pt, minimum size=5mm] (0) at (0,4) {}; 
		\node (00) at (-2,4) {$a-1$}; 
		\node [circle, draw=blue!50, fill=blue!20, inner sep=0pt, minimum size=5mm] (1) at (0,0) {}; 
		\node (11) at (-1.5,0) {$a$}; 
		\node [circle, draw=blue!50, fill=blue!20, inner sep=0pt, minimum size=5mm] (2) at (0,-4) {}; 
		\node (22) at (-2,-4) {$a+1$}; 
		
		\node (a) at (2,0) {};
		\node (b) at (8,0) {};
		\node (c) at (-3,0) {};
		\node (d) at (-9,0) {};
		
		\node [circle, draw=blue!50, fill=blue!20, inner sep=0pt, minimum size=5mm] (0a) at (11,4) {}; 
		\node (00a) at (9,4) {$a-1$}; 
		\node [circle, draw=blue!50, fill=blue!20, inner sep=0pt, minimum size=5mm] (1a) at (11,0) {}; 
		\node (11a) at (9.5,0) {$a$}; 
		\node [circle, draw=blue!50, fill=blue!20, inner sep=0pt, minimum size=5mm] (2a) at (11,-4) {}; 
		\node (22a) at (9,-4) {$a+1$}; 
		
		\node [circle, draw=blue!50, fill=blue!20, inner sep=0pt, minimum size=5mm] (aa) at (-11,4) {}; 
		\node (aa0) at (-13,4) {$a-1$}; 
		\node [circle, draw=blue!50, fill=blue!20, inner sep=0pt, minimum size=5mm] (bb) at (-11,0) {}; 
		\node (bb1) at (-12.5,0) {$a$}; 
		\node [circle, draw=blue!50, fill=blue!20, inner sep=0pt, minimum size=5mm] (cc) at (-11,-4) {}; 
		\node (cc2) at (-13,-4) {$a+1$}; 
		
        \draw (0) to node {} (1) [<->, thick];
		\draw (1) to node {} (2) [<->, thick];
		\draw (0) to [out=40, in=320, looseness=9] (0) [<->, thick, gray] ;
		\draw (0) to [out=40, in=320, looseness=13] (0) [<->, thick, dashed, gray] ;
		\draw (2) to [out=60, in=295, looseness=1] (0) [<->, thick, dashed, gray];
		\draw (2) to [out=40, in=320, looseness=13] (2) [<->, thick] ;
				
		\draw (a) [pil] to node {} (b);
		\draw (d) [pil] to node {} (c);
		
		\draw (0a) to node {} (1a) [<->, thick];
		\draw (1a) to node {} (2a) [<->, thick];
		\draw (2a) to [out=40, in=320, looseness=10] (2a) [<->, thick] ;
		
		\draw (aa) to node {} (bb) [<->, thick];
		\draw (bb) to node {} (cc) [<->, thick];
		\draw (aa) to [out=40, in=320, looseness=10] (aa) [<->, thick] ;
\end{tikzpicture}}
\caption{Performing Seiberg duality on node $(a)$, while node $(a-1)$ has an adjoint, results in moving the adjoint from $(a-1)$ to $(a+1)$. Dashed lines are mesons, while gray lines represent fields that have been integrated out in the process.}\label{fig:SeibergDualAdj}
\end{figure}
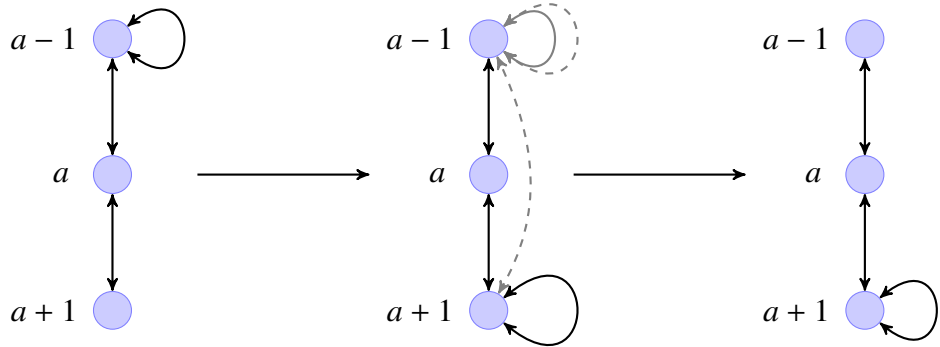
\begin{figure}
\begin{center}
\begin{subfigure}{0.35\textwidth}
\centering{\includegraphics[scale=0.24, trim=15cm 3cm 9cm 3.4cm, clip]{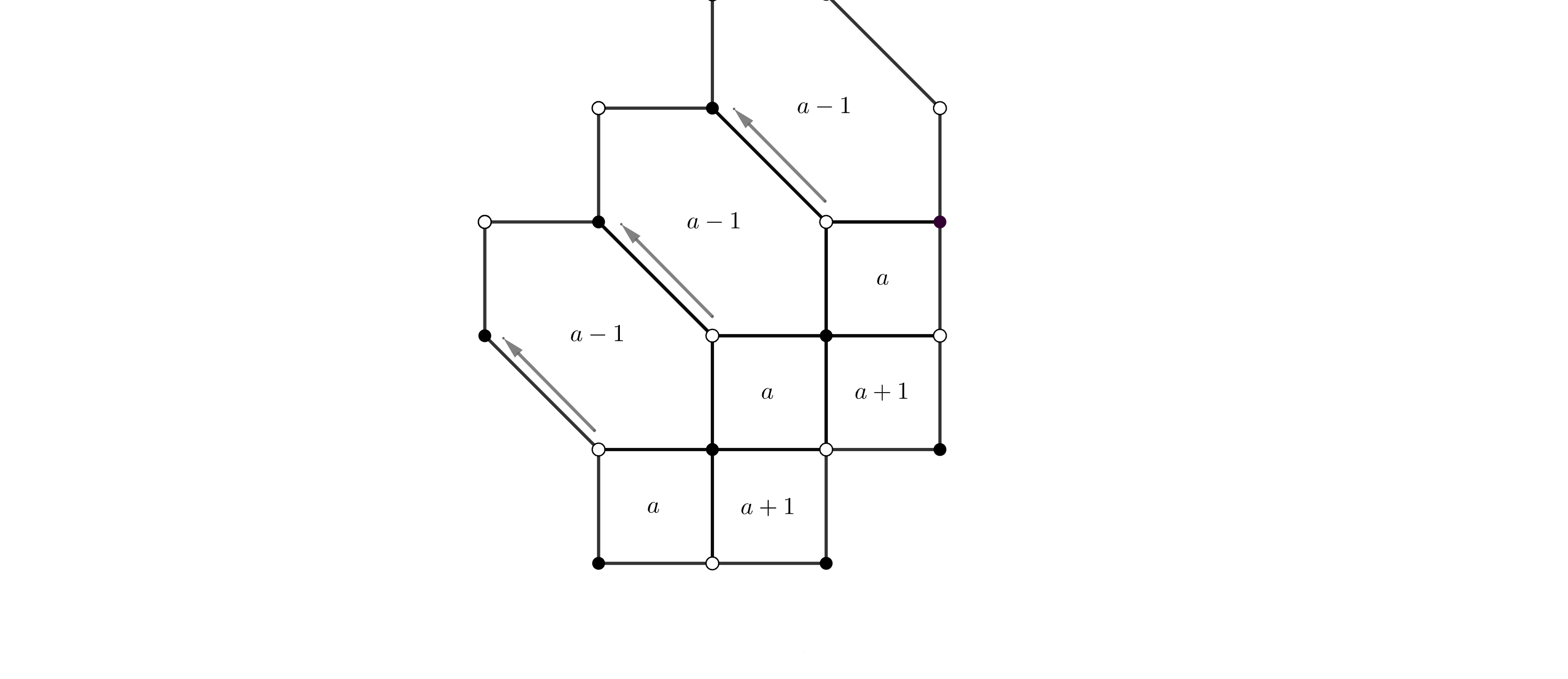}}
\end{subfigure}
\begin{subfigure}{0.1\textwidth}
\centering{
\begin{tikzpicture}[auto, scale=0.43]
       \node (0) at (0,0) {};
       \node (1) at (4,0) {};
       \draw (0) [pil] to node {} (1);
\end{tikzpicture}}
\end{subfigure}
\begin{subfigure}{0.35\textwidth}
\centering{\includegraphics[scale=0.9, trim=0 1cm 3cm 0, clip]{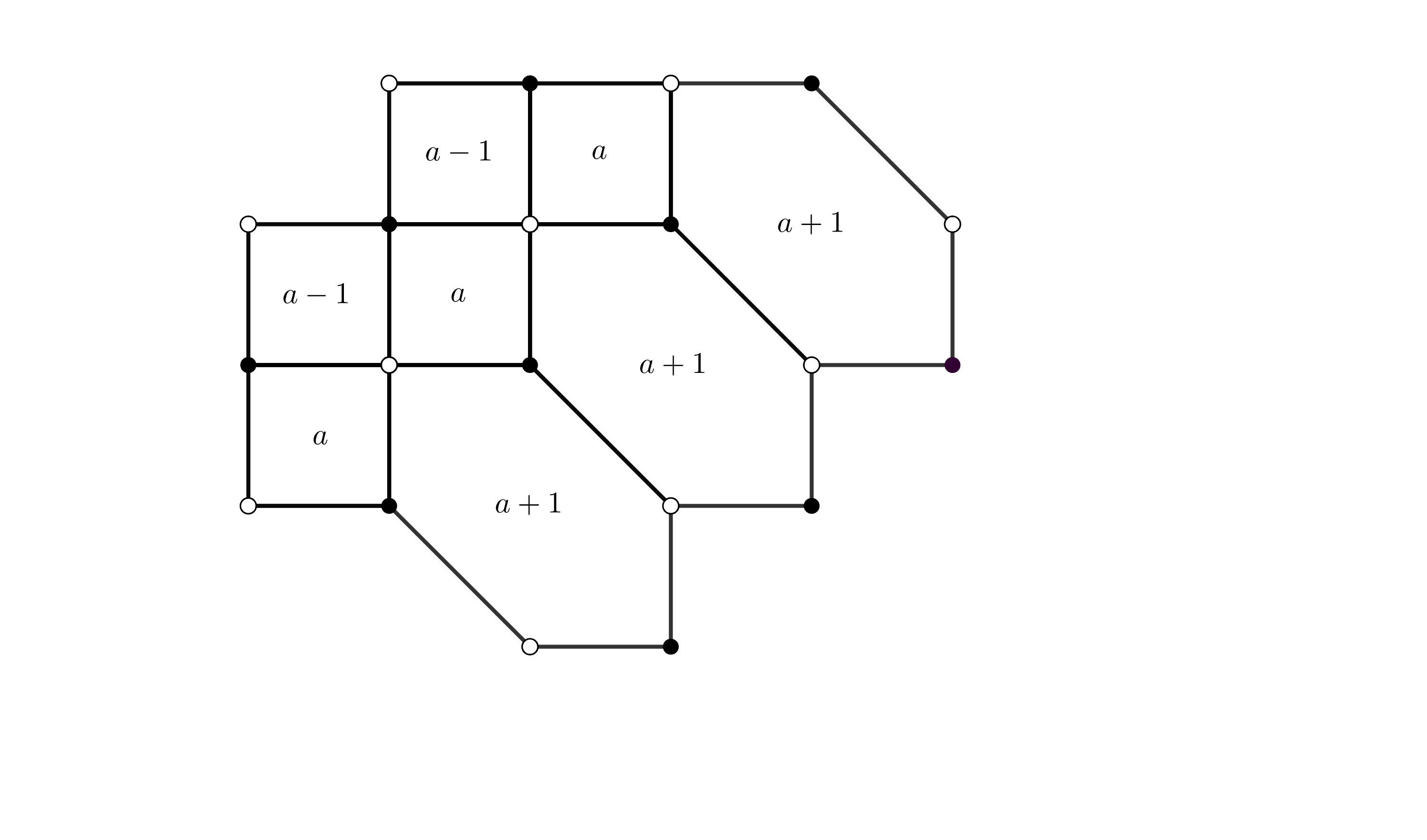}}
\end{subfigure}
\end{center}
\caption{Seiberg duality on node $(a)$, whereas node $(a-1)$ has an adjoint field, from the perspective of the dimer. As a result of integration of massive fields, the exagon $(a-1)$ collapses into a square, while the a pair of extra edges transform the square $(a+1)$ into an exagon, generating the adjoint fields.}\label{fig:SeibergDualAdjDimer}
\end{figure}

We now explicitly show that these $L^{k,3n-k,k}$ gauge theories for a given $k$ are related by Seiberg duality, which in turn means that they are dual phases of the same toric diagram. In particular, we perform Seiberg duality on a gauge  group with no adjoint fields. Suppose that $(a)$ is such a node. There are two possibilities: either both nodes $(a-1)$ and $(a+1)$ have no adjoint, or one of the two, say $(a-1)$, has an adjoint. In the former case, Seiberg duality gives a theory with an adjoint on both the $(a-1)$ and the $(a+1)$ node. The latter case is the interesting one. Suppose the group at node $a$ has rank $N_a$. The usual rules for the Seiberg dual give $\widetilde{N}_a=N_{a+1}+N_{a-1}-N_a$ and if $N_a=N $ for all $a$ then $\widetilde{N}_a=N$. The matter  content includes dual bifundamental fields and mesons. Integrating the massive fields out one obtains the dual magnetic theory. Note that the net result is to move an adjoint field from node $(a-1)$ to $(a+1)$, as displayed in Fig.~\ref{fig:SeibergDualAdj}. This is represented as the operation in Fig.~\ref{fig:SeibergDualAdjDimer} from the dimer perspective.  Repeating the process, one can construct all possible theories with $3n-2k$ adjoints and non-adjoint nodes all in pairs.  In particular, one can choose to dualise both nodes $(a)$ and $(n-a)$, realizing a theory that is $\mathbb{Z}_2$-symmetric. This will be useful in the case of orientifold projections.


\section{Orientifold of SSP$/\mathbb{Z}'_n$}\label{Sec:SPPZnOmega}

In this section we study the orientifold projection $\Omega$ of the non-chiral orbifold SSP$/\mathbb{Z}'_n$ and in particular we seek the conformal point of the unoriented theory. We first discuss the cases with $n=1$ and $n=2$ and then general $n$.

\subsection{Unoriented SPP}\label{sec:SPPOm}

Let us perform the orientifold projection $\Omega$ with four fixed points, see Fig.~\ref{fig:SPPdimer}, whose signs are denoted by $\tau_0$ and $\tau_{00}$ for the gauge group and the adjoint field, and  $\tau_{12}$ and  $\tau_{21}$ for the projected bifundamental fields. The anomaly cancellation condition imposes $\tau_{12} = \tau_{21}$. Since half the number of terms in the supepotential is even, the sign rule requires $\prod \tau = +1$. 
The superpotential of the unoriented theory reads
\begin{align}
W^{\Omega}_{\textrm{SPP}} = - \phi_0 X_{01}X_{10} + X_{12}X_{21}X_{10}X_{01} \; .
\end{align}
The condition $R(W)=2$ remains as in Eq.~\eqref{eq:RchargesSPP}, while the cancellation of the $R$-symmetry anomaly gives 
\begin{align}\label{eq:RchargeAnomalyOmega}
&r_{00} \left( N_0 - N_1 + 2 \tau_{0} \right) = - \left( N_0 - N_1 - 2 \tau_{0} \right) \; , \\ \nonumber
&r_{00} \left( N_0 - N_1 - 2 \tau_{12} \right) = - \left( N_0 - N_1 + 2 \tau_{12} \right) \; .
\end{align}
At the conformal point of the parent theory, $N_0=N_1=N$, the orientifold projection gives $r_{00}=+1$, violating unitarity. On the other hand, imposing $N_1 = N_0 - 2 \tau_0 = N_0 + 2 \tau_{12}$ we obtain 
\begin{align}
r_{00} = 0 \; , \qquad \tau_{0} = - \tau_{12} \; , 
\end{align}
fixing $r_{01}=r_{12}=-1/2$. The superconformal $R$-charges are already determined and the $a$-charge at large $N = N_0 \simeq N_1$ reads
\begin{align}
&a^{\Omega}_{_{\textrm{SPP}}} = \frac{81}{256} N^2 \; .
\end{align}

\begin{center}
\vspace{15pt}
\begin{tabular*}{0.6\textwidth}{@{\extracolsep{\fill}}cccc}
\toprule
 & $Sp/SO(N_0)$ & $SU(N_1)$ &  $U(1)_R$\\
\midrule
 $\phi_0$ & $\tiny{\yng(1,1)}$/$\tiny{\yng(2)}$ & $\bf{1}$ & 1  \\[3pt]
 $X_{01}$ & $\tiny{\yng(1)}$ & $\overline{\tiny{\yng(1)}}$ &   $\frac{1}{2}$  \\[3pt]
 $X_{10}$ & $\overline{\tiny{\yng(1)}}$ & $\tiny{\yng(1)}$  & $\frac{1}{2}$ \\[3pt]
 $X_{12}^{S/A}$ & $\bf{1}$ & $\tiny{\yng(2)}$/$\tiny{\yng(1,1)}$ & $\frac{1}{2}$\\[3pt]
 $X_{21}^{S/A}$ & $\bf{1}$ & $\overline{\tiny{\yng(2)}}$/$\overline{\tiny{\yng(1,1)}}$ & $\frac{1}{2}$\\[3pt]
\bottomrule
\end{tabular*}
\captionof{table}{The matter content and the superconformal $R$-charges of $(\textrm{SPP})^{\Omega}$.}\label{tab:SPP}
\end{center}

The ratio between the $a$-charge of the parent and of the unoriented theory 
\begin{align}\label{eq:aChargeRatio}
\frac{a^{\Omega}_{_{\textrm{SPP}}}}{a_{_{\textrm{SPP}}}} = \frac{9 \sqrt{3}}{32} \simeq 0.4871 \; 
\end{align}
is more than halved. From a geometrical perspective, the volume of the horizon is less than halved, if compared with the parent one with the same radius\footnote{The (fourth power of the) radius of the horizon is proportional to the unit of five-form flux $N$. In unoriented theories, the radius is then proportional to $N/2$, then we must rescale $N \to N/2$ in $V^{\Omega}$ in order to compare it with the parent volume.}
\begin{align}\label{eq:VolumeRatio}
\frac{V^{\Omega}_{_{\textrm{SPP}}}}{V_{_{\textrm{SPP}}} } = \frac{8 \sqrt{3}}{27} \simeq 0.5132 \; .
\end{align}
The third scenario occurs here, where the $R$-charges after the orientifold projection are different from those of the parent theory already at leading order, in contrast with the first scenario. This can be traced back to the fact that the number of abelian symmetries that mix with the $R$-symmetry is less than in the parent theory, for the $r_{00}$ being already fixed, which in turn fixes all the $r_{ij}$ before $a$-maximization. In contrast, in the parent theory the $r_{ij}$ are determined by $a$-maximization. The same mechanism, the breaking of an abelian symmetry, is discussed in~\cite{Antinucci:2020yki}. This is the reason behind the values in Eqs.~\eqref{eq:aChargeRatio} and~\eqref{eq:VolumeRatio}: since the $R$-charges are related to the Reeb vector, the consequence is that the geometry of the horizon is different between the first and the third scenario. 

One last solution is allowed for the $R$-charges. Imposing in Eq.~\eqref{eq:RchargeAnomalyOmega} that $N_0 - N_1 + 2 \tau_0 \neq 0$ and $N_0 - N_1 - 2 \tau_{12} \neq 0$ and $N_0 - N_1 = p \neq 0$, we have
\begin{align}
r_{00} = - \frac{p - 2 \tau_0}{p + 2 \tau_0} = - \frac{p + 2 \tau_{12}}{p - 2 \tau_{12}} \; , \qquad r_{01}=- \frac{2 \tau_0}{p+2\tau_0} \; , \qquad r_{12} = - \frac{p}{p+2\tau_0} \; .
\end{align} 
which requires $\tau_0 = - \tau_{12}$. They yields
\begin{align}
a_{_{\textrm{SPP}}}^{\Omega} = \frac{27}{8}N^2\frac{p\tau_0}{\left( p + 2 \tau_0 \right)^3}\left( p + \tau_0 \right) \; ,
\end{align}
for $\tau_{0}=-1$ and $-N_1<p<0$, or $\tau_{0}=+1$ and $N_0>p>0$, considering unitarity and $a^{\Omega}>0$, $N_0>0$, $N_1>0$. Note that if $p=2\tau_0$ we recover the previous case, then this is a more general solution.

To sum up, for the unoriented SPP the result would seem to naively suggest the existence of a whole family of conformal theories with $\tau_0=\tau_{00}=-\tau_{12}=-\tau_{21}$ and parametrized by $p$, the shift between ranks $N_0$ and $N_1$, i.e. the number of fractional branes. They would all belong to the third scenario, since a $U(1)$ is anomalous and at the fixed point the $R$-charges differ from those of the parent already at leading order. The fact that any value of $p$ could in principle yield a conformal point is somewhat surprising, so the existence of this family of solution must be investiged more. We are going to discuss this point further.

We should worry about operators that may become free and decouple before the theory reaches the conformal point and correct the central charge $a$ as in Eq.~\eqref{eq:CorrectionACharge}. In applying this analysis, we look at several operators potentially dangerous. Operators which contain mesons in the superpotential never become free, however other can be constructed. Since $\tau_0 = \tau_{00}$
\begin{align}
& \mathcal{O}_{0,j}=\textrm{Tr }\phi_0^j \; , \quad j > 1 \; , \nonumber \\[5pt]
& \mathcal{M}_m = \left( X_{12}X_{21} \right)^m \; , \quad m \geq 1 \; , \nonumber \\[5pt]
& \widetilde{\mathcal{M}}_{0,lk} = \phi_0^l \left( X_{01}X_{10} \right)^k \; , \quad l \geq 0 \; , \quad k \geq 1 \; 
\end{align}
are allowed and their $R$-charge reads 
\begin{align}
& R_{\mathcal{O}}^{(j)} = j \frac{4 \tau_0}{p + 2 \tau_0} \; , \nonumber \\[5pt]
& R_{\mathcal{M}}^{(m)} = m \frac{4 \tau_0}{p + 2 \tau_0} \; , \nonumber \\[5pt]
& R_{\widetilde{\mathcal{M}}}^{(lk)} = \frac{4 \tau_0}{p + 2 \tau_0} (l-k) + 2k \; .
\end{align}
The singlet with $j=1$ vanishes in the unoriented theory. In the parent theory this parametrizes the movement of fractional branes along a curve of singularity, but this mode is projected out by the orientifold plane  since fractional branes are stuck at the orientifold singularity. The configuration will be explicit in the elliptic model (see Sec.~\ref{sec:Elliptic}).

\begin{figure}
\centering{\includegraphics[scale=0.8]{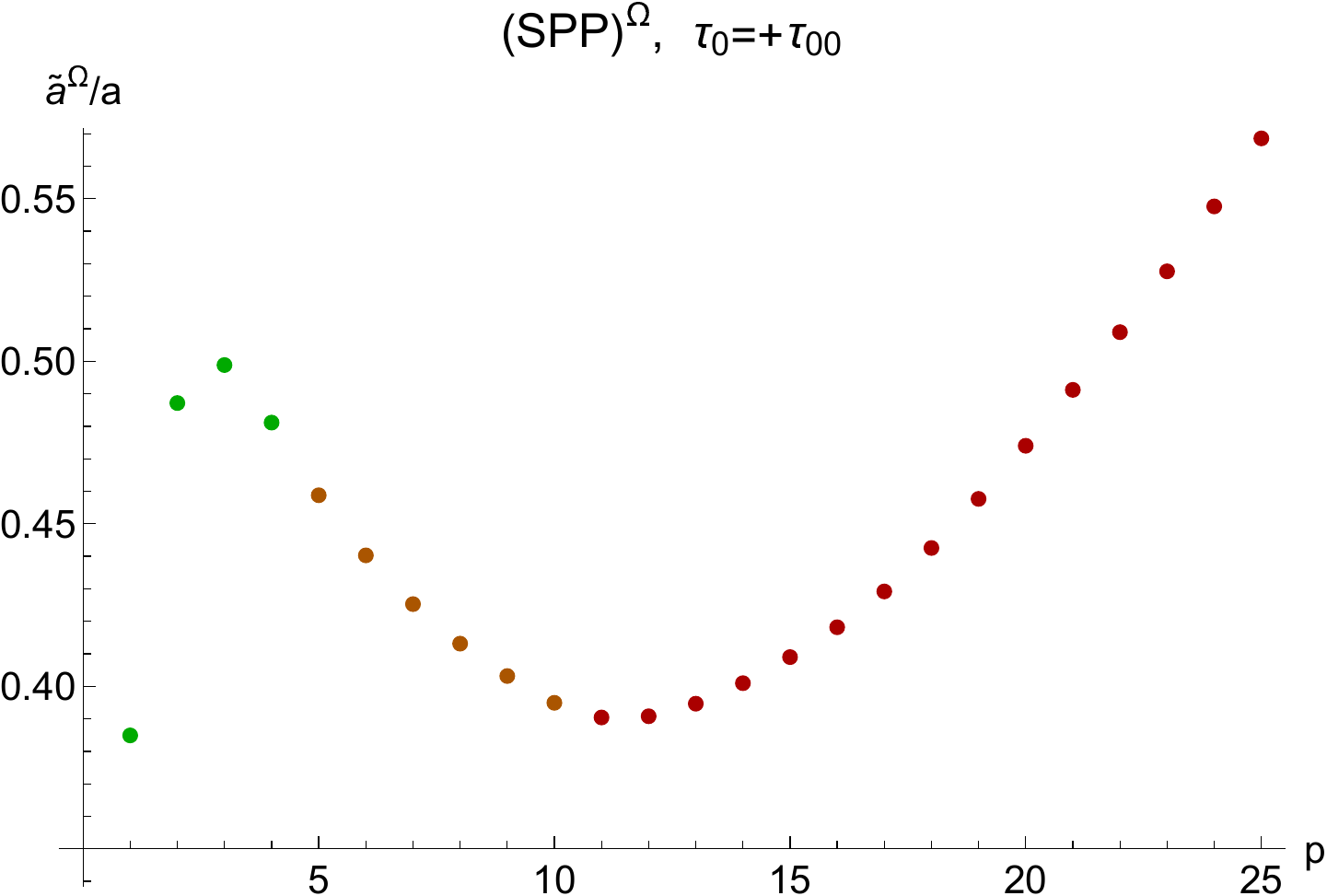}}
\caption{The ratio $\widetilde{a}_{_{\textrm{SPP}}}^{\Omega}/a_{_{\textrm{SPP}}}$ vs $p=N_0 - N_1$. The green points signal that there are no correction to the central charge, on orange points $\left( X_{12}X_{21} \right)^m$ becomes free, while on red ones operators $\textrm{Tr }\phi_0^j$ start to decouple.}\label{fig:SPPOmegaCorrected}
\end{figure}

Clearly, these operators may decouple depending on the value of $p$. We stress that each value of $p$ defines an independent theory and we are not describing an RG-flow parametrized by $p$.\footnote{In principle, one could perform a duality cascade, under which the theory is self-similar and after a number of Seiberg dualitites it goes back to its original structure. This is not possible in this case, as one eventually needs to dualize a gauge group with tensorial matter and the superpotential does not meet the known dualities.} Let us focus on the case with $\tau_0 = \tau_{00} = +1$ and $0<p<N_0$, the opposite choice is similar. The operator $\mathcal{O}_j$ becomes free for 
\begin{align}
j \frac{4}{p + 2} \leq \frac{2}{3} \; ,
\end{align} 
so, for all integers $j \leq \bar{j}=(p+2)/6$ an operator decouples and the $a$-charge gets corrected. For example, $\textrm{Tr }\phi_0^2$ decouples for $p=10$ (at which the correction is zero though), while for $p=16$ both $\textrm{Tr }\phi_0^2$ and $\textrm{Tr }\phi_0^3$ decouple. As for $\widetilde{\mathcal{M}}_{0,lk}$ hits $R_{\widetilde{\mathcal{M}}}^{(lk)}=2/3$ only for $p=1$ and the correction to the $a$-charge is zero. On the other hand, $\mathcal{M}_m$ is free for $m \leq (p+2)/6$ and the first correction enters for $p=4$, where it is zero. 
In Fig.~\ref{fig:SPPOmegaCorrected} we can see that the corrected ratio $\widetilde{a}^{\Omega}_{_{\textrm{SPP}}}/a_{_{\textrm{SPP}}}$. It increases and approaches the value 0.5, beyond which we doubt the existence of the conformal theory at all: it is the $\mathbb{Z}_2$ projection of SPP.  From Eq.~\eqref{eq:CorrectionACharge}, we also note that for $p>4$ $\textrm{Tr }R\neq0$ at leading order, due to the correction itself. Hence, beyond this point the holographic duality should not hold in its simple form and we are not allowed to think of the field theory as the gauge dual of a gravity theory. The existence of the conformal point can be bound to $p \leq 4$, for which the third scenario always occurs. Moreover, applying the analysis of~\cite{Benvenuti:2017lle} we find that for $p=1$ all terms in the superpotential should be removed, posing doubts on the existence of the conformal point. We exclude $p=1$ from the allowed range.

A hypothetical magnetic theory could confirm the existence of the conformal point and maybe select only one value of $p$, number of fractional branes. Unfortunately, for this case there is no known Seiberg duality compatible with the superpotential of the unoriented toric theory. As we shall see, this is not the case when $n>1$.

\subsection{Unoriented SPP$/\mathbb{Z}'_2$}

Let us now focus on the case with $n=2$, the first with unitary nodes with no tensorial matter, as can be seen from the quiver in Fig.~\ref{fig:SPPZ2}. Depending on the $\tau$'s, gauge groups at node zero are orthogonal or symplectic, with a bifundamental hypermultiplet and tensorial matter. The superpotential reads
\begin{align}
W^{\Omega}_{\textrm{SPP}/\mathbb{Z}'_2} = - \phi_0 X_{01}X_{10} + X_{12}X_{21}X_{10}X_{01} - X_{21}X_{12}X_{23}X_{32} + \phi_3 X_{32}X_{23} \; .
\end{align}

Proceeding as before, we solve the constraints for the $R$-charges. Along with Eq.~\eqref{eq:RchargesSPP}, $r_{00}=r_{33}$ and $2r_{01}= 2r_{23}=-1-r_{00}$, we have
\begin{align}
& r_{00} \left( N_0 - N_1 + 2 \tau_{00} \right) = - \left( N_0 - N_1 - 2 \tau_0 \right) \; , \nonumber \\[3pt]
& r_{00} \left( N_0 - N_2 \right) = 2 N_1 - N_0 - N_2  \; , \nonumber \\[3pt]
& r_{00} \left( N_3 - N_1 \right) =  2 N_2 - N_3 - N_1  \; , \nonumber \\[3pt]
& r_{00} \left( N_3 - N_2 + 2 \tau_{33} \right) = - \left( N_3 - N_2 - 2 \tau_{3} \right) \; .
\end{align}

From the projected nodes we may either have $\tau_0 = - \tau_{00}$ and $\tau_{3} = - \tau_{33}$ and shift between the first and last pair of ranks determined, or $\tau_0 = \tau_{00}$ and $\tau_{3} = \tau_{33}$. We denote them as solution A and B, respectively.

\subsubsection{Solution A}

\begin{center}
\vspace{15pt}
\begin{tabular*}{0.9\textwidth}{@{\extracolsep{\fill}}cccccc}
\toprule
 & $Sp/SO(N_0)$ & $SU(N_1)$ & $SU(N_2)$ & $SO/Sp(N_3)$ & $U(1)_R$\\
\midrule
 $\phi_0$ & $\tiny{\yng(2)}$/$\tiny{\yng(1,1)}$ & $\bf{1}$ & $\bf{1}$ & $\bf{1}$ &  1  \\[3pt]
 $X_{01}$ & $\tiny{\yng(1)}$ & $\overline{\tiny{\yng(1)}}$ & $\bf{1}$ & $\bf{1}$ &  $\frac{1}{2}$  \\[3pt]
 $X_{10}$ & $\overline{\tiny{\yng(1)}}$ & $\tiny{\yng(1)}$ & $\bf{1}$ & $\bf{1}$ & $\frac{1}{2}$ \\[3pt]
 $X_{12}$ & $\bf{1}$ & $\tiny{\yng(1)}$ & $\overline{\tiny{\yng(1)}}$ & $\bf{1}$ & $\frac{1}{2}$\\[3pt]
 $X_{21}$ & $\bf{1}$ & $\overline{\tiny{\yng(1)}}$ & $\tiny{\yng(1)}$ & $\bf{1}$ & $\frac{1}{2}$\\[3pt]
 $X_{23}$ & $\bf{1}$ & $\bf{1}$ & $\tiny{\yng(1)}$ & $\overline{\tiny{\yng(1)}}$ &   $\frac{1}{2}$  \\[3pt]
 $X_{32}$ & $\bf{1}$ & $\bf{1}$ & $\overline{\tiny{\yng(1)}}$ & $\tiny{\yng(1)}$  & $\frac{1}{2}$ \\[3pt]
  $\phi_3$ & $\bf{1}$ & $\bf{1}$ & $\bf{1}$ & $\tiny{\yng(1,1)}$/$\tiny{\yng(2)}$ & 1  \\[3pt]
\bottomrule
\end{tabular*}
\captionof{table}{The matter content and the superconformal $R$-charges of $(\textrm{SPP}/\mathbb{Z}'_2)^{\Omega}$ solution A.}\label{tab:SPPZ2A}
\end{center}

Consider the case $\tau_0 = - \tau_{00}$ and $\tau_{3} = - \tau_{33}$. Denoting rank shifts as $N_0 - N_2 = p$, $N_1 - N_2 = q$ requires that $q = p - 2 \tau_0$ and $ \tau_0 = - \tau_3 $. Then one gets
\begin{align}\label{eq:RankZ2A}
& N_1 = N_0 - 2\tau_0 \; , \nonumber \\
& N_2 = N_0 - p \; , \nonumber \\
& N_3 = N_0 - p - 2 \tau_0 \; , 
\end{align}
along with
\begin{align}
r_{00} = 1 - 4\frac{\tau_0}{p} \; , \qquad r_{01} = - \frac{p - 2 \tau_0}{p} \; ,\qquad r_{12} = - 2 \frac{\tau_0}{p} \; .
\end{align}
Thus, at large $N$
\begin{align}
&a_{_{\textrm{SPP}/\mathbb{Z}'_2}}^{\Omega} = \frac{27}{8}N^2 \left(- \frac{\tau_0}{p^3} \right) \left( 4 - p^2 \right) 
\end{align}
and the ratio w.r.t. the parent reads
\begin{align}
&\frac{a_{_{\textrm{SPP}/\mathbb{Z}'_2}}^{\Omega}}{a_{_{\textrm{SPP}/\mathbb{Z}'_2}}} = \frac{3 \sqrt{3}}{2} \left( - \frac{\tau_0}{p^3} \right) \left( 4 - p^2 \right) \; ,
\end{align}
with $\tau_0 = +1$ and $ 2 < p <  N_0-2$, or $\tau_0 = -1$ and $ 2 < -p < N_3 - 2 $, from unitarity and positivity of $N_a$.

\begin{figure}
\centering{\includegraphics[scale=0.8]{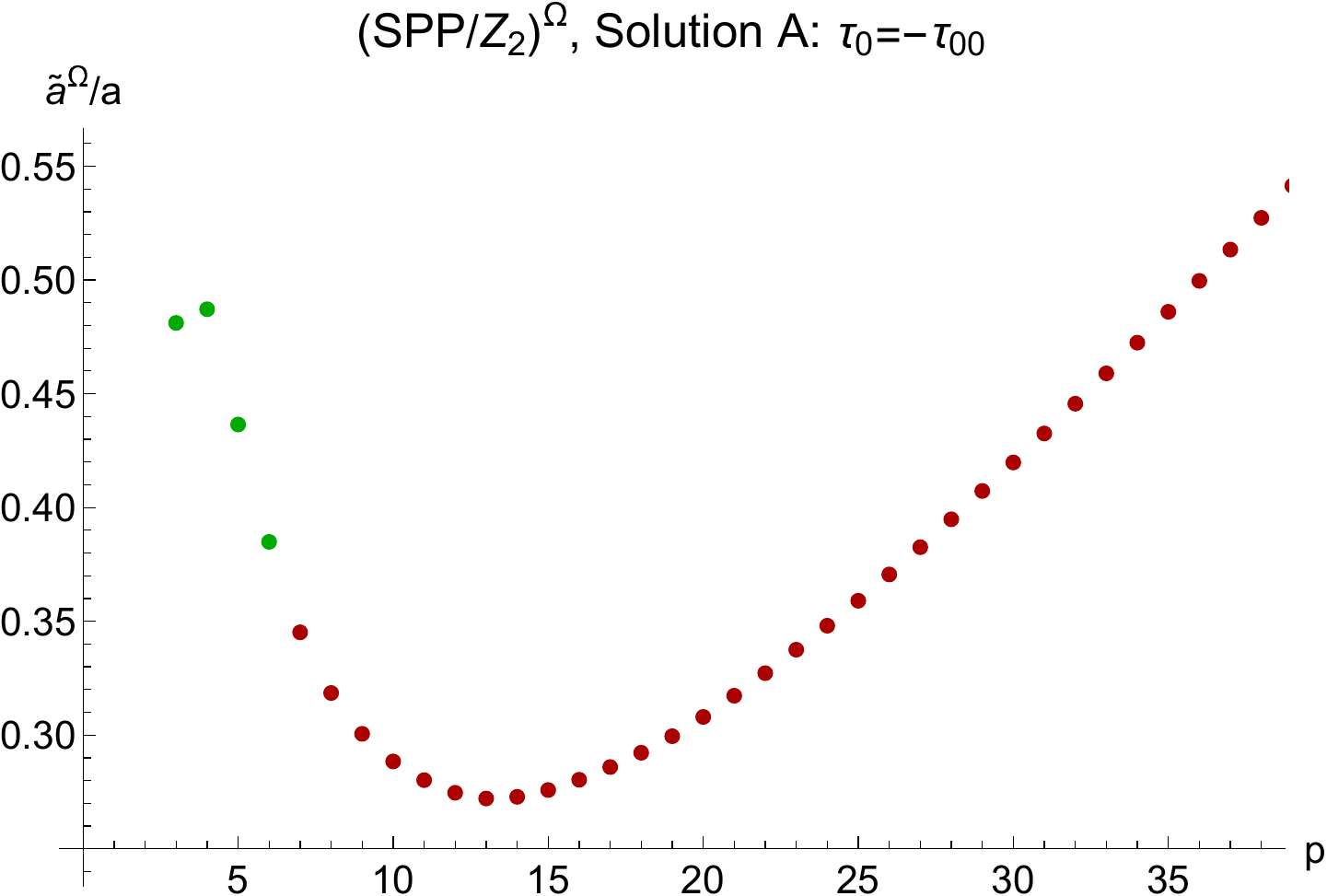}}
\caption{The ratio $\widetilde{a}_{_{\mathrm{SPP}/\mathbb{Z}'_2}}^{\Omega}/a_{_{\textrm{SPP}/\mathbb{Z}'_2}}$ vs $p=N_0 - N_2$ in Solution A. The green points signal that there are no correction to the central charge,  while on red points operators $\phi_0^l \left( X_{01}X_{10} \right)^k$ and $\phi_3^i \left( X_{32}X_{23} \right)^k$ start to decouple.}\label{fig:SPPZ2RatioChargeA}
\end{figure}

Some operators are dangerous, in the sense that may decouple and correct the computation of the central charge $a$. Since $\tau_0 = - \tau_{00}$ there are no operators of the form $\textrm{Tr }\phi_{0}^j$ or $\textrm{Tr }\phi_{3}^j$. However, the following gauge-invariant operators  
\begin{align}
& \mathcal{M}_m = \left( X_{12}X_{21} \right)^m \; , \quad m \geq 1 \; , \nonumber \\[5pt]
& \widetilde{\mathcal{M}}_{0,lk} = \phi_0^l \left( X_{01}X_{10} \right)^k \; , \quad l \geq 0 \; , \quad k \geq 1 \; \nonumber \\[5pt]
& \widetilde{\mathcal{M}}_{3,lk} = \phi_3^l \left( X_{32}X_{23} \right)^k \; , \quad l \geq 0 \; , \quad k \geq 1 \; 
\end{align}
with $R$-charges 
\begin{align}
& R_{\mathcal{M}}^{(m)} = 2 m \frac{p - 2 \tau_0}{p} \; , \nonumber \\[5pt]
& R_{\widetilde{\mathcal{M}}}^{(lk)} = 2\frac{p - 2 \tau_0}{p} (l-k) + 2k \; 
\end{align}
may decouple. Operator $\mathcal{M}_m$ becomes free only for $m=1$ and $p=3$, where the correction to $a$ vanish. Instead, operators $\widetilde{\mathcal{M}}_{0,lk}$ and $\widetilde{\mathcal{M}}_{3,lk}$ become free for $l=0$, $k\leq p/6$ and $p \geq 6$ and the $a$-charge gets corrected for $p>6$. The final ratio $a_{_{\mathrm{SPP}/\mathbb{Z}'_2}}^{\Omega}/a_{_{\textrm{SPP}/\mathbb{Z}'_2}}$ is displayed in Fig.~\ref{fig:SPPZ2RatioChargeA}. As before, the existence of the conformal point is bound to $p\leq 6$, where holography still holds and the third scenario occurs.

\subsubsection{Solution B}

\begin{center}
\vspace{15pt}
\begin{tabular*}{0.9\textwidth}{@{\extracolsep{\fill}}cccccc}
\toprule
 & $Sp/SO(N_0)$ & $SU(N_1)$ & $SU(N_2)$ & $SO/Sp(N_3)$ & $U(1)_R$\\
\midrule
 $\phi_0$ & $\tiny{\yng(1,1)}$/$\tiny{\yng(2)}$ & $\bf{1}$ & $\bf{1}$ & $\bf{1}$ &  1  \\[3pt]
 $X_{01}$ & $\tiny{\yng(1)}$ & $\overline{\tiny{\yng(1)}}$ & $\bf{1}$ & $\bf{1}$ &  $\frac{1}{2}$  \\[3pt]
 $X_{10}$ & $\overline{\tiny{\yng(1)}}$ & $\tiny{\yng(1)}$ & $\bf{1}$ & $\bf{1}$ & $\frac{1}{2}$ \\[3pt]
 $X_{12}$ & $\bf{1}$ & $\tiny{\yng(1)}$ & $\overline{\tiny{\yng(1)}}$ & $\bf{1}$ & $\frac{1}{2}$\\[3pt]
 $X_{21}$ & $\bf{1}$ & $\overline{\tiny{\yng(1)}}$ & $\tiny{\yng(1)}$ & $\bf{1}$ & $\frac{1}{2}$\\[3pt]
 $X_{23}$ & $\bf{1}$ & $\bf{1}$ & $\tiny{\yng(1)}$ & $\overline{\tiny{\yng(1)}}$ &   $\frac{1}{2}$  \\[3pt]
 $X_{32}$ & $\bf{1}$ & $\bf{1}$ & $\overline{\tiny{\yng(1)}}$ & $\tiny{\yng(1)}$  & $\frac{1}{2}$ \\[3pt]
  $\phi_3$ & $\bf{1}$ & $\bf{1}$ & $\bf{1}$ & $\tiny{\yng(2)}$/$\tiny{\yng(1,1)}$ & 1  \\[3pt]
\bottomrule
\end{tabular*}
\captionof{table}{The matter content and the superconformal $R$-charges of $(\textrm{SPP}/\mathbb{Z}'_2)^{\Omega}$ solution B.}\label{tab:SPPZ2B}
\end{center}

This solution is obtained for $\tau_0 = \tau_{00}$ and $\tau_{3} = \tau_{33}$ and $N_0 \neq N_1 - 2 \tau_0$, $N_3 \neq N_2 - 2 \tau_{3}$. Denoting the shifts as $N_0 - N_1 = p$, $N_1 - N_2 = q$, $N_2 - N_3 = s$ leads to $q=2\tau_0$, $p=s$, $\tau_0 = -\tau_3$ and 
\begin{align}\label{eq:RanksZ2B}
& N_1 = N_0 - p \; , \nonumber \\
& N_2 = N_0 - p - 2 \tau_0 \; , \nonumber \\
& N_3 = N_0 - 2p - 2 \tau_0 \; ,
\end{align}
along with
\begin{align}
r_{00} =  - \frac{p-2\tau_0}{p+2\tau_0} \; , \qquad r_{01} = - 2 \frac{\tau_0}{p + 2 \tau_0} \quad , \qquad r_{12} = - \frac{p}{p+2\tau_0} \; .
\end{align}
This family of solutions generalizes the one discussed for SPP and for $p=2\tau_0$ it gives $r_{00}=0$, and it will appear again for general $\mathbb{Z}'_n$. In order to impose unitarity, $-1<r_{00}<1$ holds for $\tau_0=+1$ and $0<p<(N_0-2)/2$, or for $\tau_0=-1$ and $-(N_3-2)/2 < p < 0$. 

\begin{figure}
\centering{\includegraphics[scale=0.8]{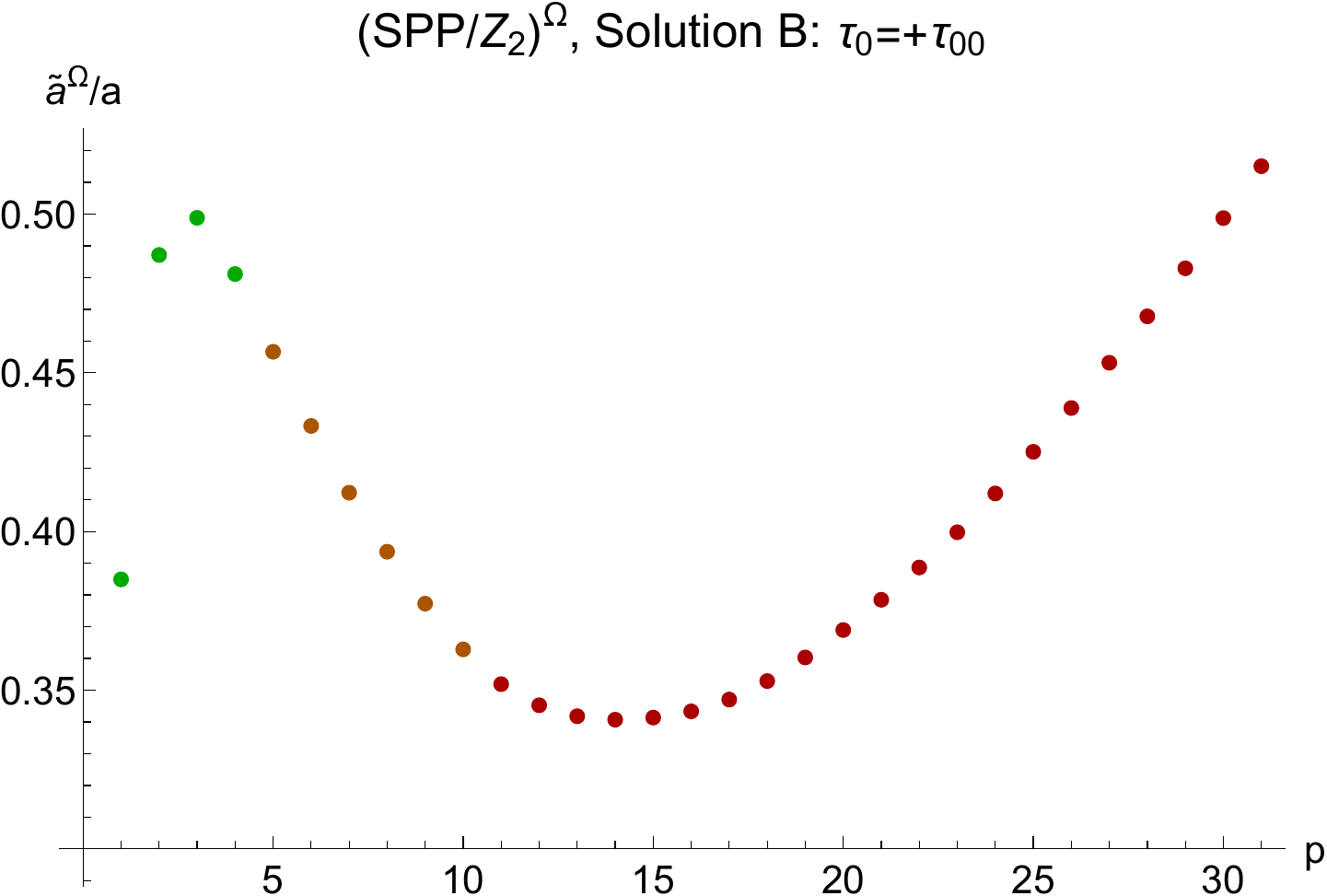}}
\caption{The ratio $\widetilde{a}_{_{\textrm{SPP}/\mathbb{Z}_2}}^{\Omega}/a_{_{\textrm{SPP}/\mathbb{Z}_2}}$ vs $p=N_0 - N_1$ in solution B. The green points signal that there are no correction to the central charge, on orange points $\left( X_{12}X_{21} \right)^m$ becomes free, while on red points operators $\textrm{Tr }\phi_0^j$ and $\textrm{Tr }\phi_3^j$ start to decouple.}\label{fig:SPPZ2OmegaCorrected}
\end{figure}

The $a$-charge and the ratio read
\begin{align}
&a_{_{\textrm{SPP}/\mathbb{Z}'_2}}^{\Omega} = \frac{27}{4}N^2 \frac{p \tau_0}{\left( p + 2 \tau_0 \right)^3} \left( p + \tau_0 \right) \; , \nonumber \\[5pt] 
&\frac{a_{_{\textrm{SPP}/\mathbb{Z}'_2}}^{\Omega}}{a_{_{\textrm{SPP}/\mathbb{Z}'_2}}} = \frac{3\sqrt{3}}{\left( p + 2 \tau_0 \right)^3}p\tau_0 \left( p + \tau_0 \right) \; .
\end{align}
In this case operators of the form
\begin{equation}
\begin{array}{lcl}
\mathcal{O}_{0,j} = \textrm{Tr }\phi_0^j \; , & j \geq 1 \; , &  \\[7pt]
\mathcal{O}_{3,j} = \textrm{Tr }\phi_3^j \; , & j \geq 1 \; , & R_{\mathcal{O}}^{(j)} =  j \frac{4\tau_0}{p+2\tau_0} \; , \\[7pt]
\mathcal{M}_m = \left( X_{12}X_{21} \right)^m \; , & m \geq 1 \;  , & R_{\mathcal{M}}^{(m)} = m \frac{4 \tau_0}{p + 2 \tau_0} \; ,  \\[7pt]
\widetilde{\mathcal{M}}_{0,lk} = \phi_0^l \left( X_{01}X_{10} \right)^k \; , & l \geq 0 \; , \quad k \geq 1 \; , &  \\[7pt]
\widetilde{\mathcal{M}}_{3,lk} = \phi_3^l \left( X_{32}X_{23} \right)^k \; , & l \geq 0 \; , \quad k \geq 1 \; , & R_{\widetilde{\mathcal{M}}}^{(lk)} = \frac{4 \tau_0}{p + 2 \tau_0} (l-k) + 2k \; , \\
\end{array} 
\end{equation}
may decouple and the central charge must be corrected as in Eq.~\eqref{eq:CorrectionACharge}. The corrections are the same as those discussed in SPP$/\Omega$ and the corrected central charge is displayed in Fig.~\ref{fig:SPPZ2OmegaCorrected}. As in the $\Omega$ projection of SPP and its solution with $r_{00} \neq 0$, for $p>4$ it turns out that $\textrm{Tr }R \neq 0$ and the conformal theory may exist only for $p\leq 4$, where it realises the third scenario, as can be seen from Fig.~\ref{fig:SPPZ2OmegaCorrected}.

\subsubsection{Seiberg duality for SPP$/\mathbb{Z}'_2$}

The solutions with $r_{00}\neq 0$ discussed in the previous subsections are somewhat difficult to interpret. The central charge $a$ must be corrected by the contribution of those operators which decouple along the flow towards the IR, where the conformal point in the third scenario stays. A side-effect of these corrections is that $\textrm{Tr }R \neq 0$ at leading order, then once they contribute, the holographic duality does not hold anymore. All one can say is that this reasoning bounds the number of fractional branes $p$, up to 6 in solution A and up to 4 in solution B (as in the SPP case), for the theory to have a gravity dual. Beyond this limiting value, the ratio between the corrected $a$-charge and the parent one in no more significative for the existence of the conformal point. As a consequence, the distinction between first and third scenario no longer holds. 

It is puzzling that the conformal point exists only for a range of number of fractional branes. Seiberg duality may help in finding a clear evidence for the very existence of the conformal point. From this point of view, we perform Seiberg duality on an $SU$ node in SPP$/\mathbb{Z}'_2$ and look for the conformal point in the magnetic theory, then compare it with the electric theory. The two must be the same.

\begin{figure}
\begin{center}
\begin{tikzpicture}[auto, scale=0.45]
		\node [circle, draw=blue!50, fill=blue!20, inner sep=0pt, minimum size=5mm] (0) at (0,2.7) {$0$}; 
		\node [circle, draw=blue!50, fill=blue!20, inner sep=0pt, minimum size=5mm] (1) at (3.7,0) {$\widetilde{1}$}; 
		\node [circle, draw=blue!50, fill=blue!20, inner sep=0pt, minimum size=5mm] (2) at (3.7,-4) {$2$}; 
		\node [circle, draw=blue!50, fill=blue!20, inner sep=0pt, minimum size=5mm] (3) at (0,-6.7) {$3$}; 		
		\node [circle, draw=blue!50, fill=blue!20, inner sep=0pt, minimum size=5mm] (4) at (-3.7,-4) {$4$};
		\node [circle, draw=blue!50, fill=blue!20, inner sep=0pt, minimum size=5mm] (5) at (-3.7,0) {$\widetilde{5}$};
        \draw (0) to node {} (1) [<->, thick];
		\draw (1) to node {} (2) [<->, thick];
		\draw (2) to node {} (3) [<->, thick];
		\draw (3) to node {} (4) [<->, thick];
		\draw (4) to node {} (5) [<->, thick];
		\draw (5) to node {} (0) [<->, thick];
		\draw (3) to [out=310, in=230, looseness=11] (3) [<->, thick];
		\draw (2) to [out=10, in=-70, looseness=11] (2) [<->, thick] ;
		\draw (4) to [out=170, in=250, looseness=11] (4)[<->, thick] ;
		\draw (0,-10) to node [pos=0.01, gray]{$\Omega$} (0,4.5) [dashed, gray];
\end{tikzpicture}\caption{The quiver for the orientifold projection of magnetic SPP$/\mathbb{Z}'_2$.}\label{fig:SPPZ2Dual}
\end{center}
\end{figure}
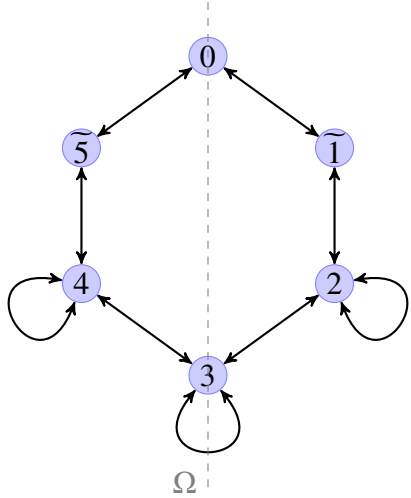

First, let us focus on the quiver theory in Fig.~\ref{fig:SPPZ2} with solutions A for the $r$-charges, $\tau_0 = - \tau_{00} = - \tau_3 = \tau_{33}$, with ranks given in Eq.~\eqref{eq:RankZ2A}. Performing Seiberg duality on gauge group $SU(N_1)$, the resulting magnetic node has rank 
\begin{align}
\widetilde{N}_1 = N_0 + N_2 - N_1 = N_0 - p + 2 \tau_0 \; ,
\end{align}
while mesons and dual quarks are constructed as discussed in Sec.~\ref{sec:Seiberg}. The final quiver is shown in Fig.~\ref{fig:SPPZ2Dual} and the superpotential reads
\begin{align}
\widetilde{W}^{\Omega}_{_{\textrm{SPP}/\mathbb{Z}'_2}} &=  \phi_3 X_{32}X_{23} - \widetilde{X}_{12}\widetilde{X}_{21}\widetilde{X}_{10}\widetilde{X}_{01} + M_2 \left(  \widetilde{X}_{21}\widetilde{X}_{12} - X_{23}X_{32} \right)  \; .
\end{align}
The conformal point is given by $r_{22}=0$ and $p=4\tau_0$. At these values, none of the gauge invariant operators decouples, neither in the electric nor in the magnetic theory. 
The $a$-charge does not change in the magnetic, hence, there is only one value for the number of fractional branes so that the conformal point exists and it features the third scenario.

The matter content and superpotential for solution B remains unchanged, while orientifold signs are $\tau_0 = \tau_{00} = - \tau_3 = - \tau_{33}$ and ranks given in Eq.~\eqref{eq:RanksZ2B} and the dual gauge node has rank $\widetilde{N}_1 = N_0 - 2 \tau_0$. For the fixed point, it must be $r_{22}=0$ and $p=2\tau_0$ and, again, the central charge $a$ gets no correction both in the electric and magnetic theory. We conclude that this is the conformal point, in third scenario, we looked for. We notice that in both cases the fixed point exists only for $r_{00}=0$.

\subsection{Unoriented SPP$/\mathbb{Z}'_n$}

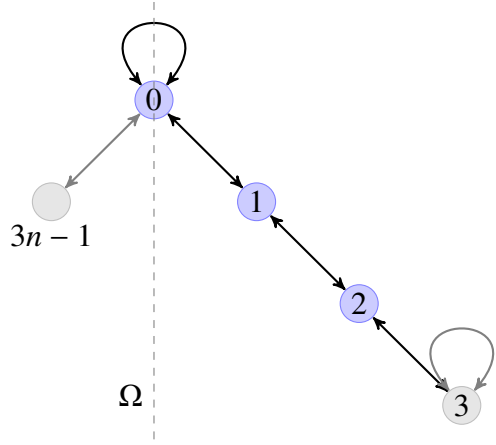
\begin{figure}
\begin{center}
\begin{tikzpicture}[auto, scale=0.45]
		\node [circle, draw=blue!50, fill=blue!20, inner sep=0pt, minimum size=5mm] (0) at (0,3) {$0$}; 
		\node [circle, draw=blue!50, fill=blue!20, inner sep=0pt, minimum size=5mm] (1) at (3,0) {$1$}; 
		\node [circle, draw=blue!50, fill=blue!20, inner sep=0pt, minimum size=5mm] (2) at (6,-3) {$2$}; 
		\node [circle, draw=gray!50, fill=gray!20, inner sep=0pt, minimum size=5mm] (3) at (9,-6) {$3$}; 		
		\node [circle, draw=gray!50, fill=gray!20, inner sep=0pt, minimum size=5mm] (n) at (-3,0) {};
		\node (nn) at (-3,-1) {$3n-1$};
        \draw (0) to node {} (1) [<->, thick];
		\draw (1) to node {} (2) [<->, thick];
		\draw (2) to node {} (3) [<->, thick];
		\draw (n) to node {} (0) [<->, thick, gray];
		\draw (0) to [out=130, in=50, looseness=11] (0) [<->, thick] ;
		\draw (3) to [out=130, in=50, looseness=11] (3) [<->, thick, gray] ;
		\draw (0,-7) to node [pos=0.1]{$\Omega$} (0,6) [dashed, gray] ;
\end{tikzpicture}\caption{Blue nodes form the recursive structure of the quiver SPP$/\mathbb{Z}'_n$.}\label{fig:SPPZnOmega}
\end{center}
\end{figure}

As we have seen in Sec.~\ref{Sec:Zn}, the parent gauge theory SPP$/\mathbb{Z}'_n$ has a recursive structure that allows us to solve the set of equation for the $R$-charges. The computation for the unoriented theory is similar, with some modifications due to the $\Omega$ projection. The $\mathbb{Z}_2$ maps two sides of the quiver and we keep nodes from 0 to $\bar{n}$, the latter being
\begin{align}
&\bar{n} = \frac{3}{2} n \; \quad n\textrm{ even} \; ,\nonumber \\[5pt]
&\bar{n} = \frac{3n - 1}{2} \; \quad n\textrm{ odd} \; .
\end{align}
Half the superpotential is projected out and it reads 
\begin{align}
W_{_{SPP/\mathbb{Z}'_n}}^{\Omega} &= - \phi_0 X_{01}X_{10} + \sum_{i=1}^{\lfloor\frac{n}{2}\rfloor}\phi_{3i} \left( X_{3i,\,3i-1}X_{3i-1,\,3i} - X_{3i,\,3i+1}X_{3i+1,\,3i} \right) \nonumber \\[3pt]
&+ \sum_{i=0}^{\bar{n}-1} \left( X_{3i+1,\,3i}X_{3i,\,3i+1}X_{3i+1,\,3i+2}X_{3i+2,\,3i+1} \right. \nonumber \\[3pt]
& \quad \left. - X_{3i+2,\,3i+1}X_{3i+1,\,3i+2}X_{3i+2,\,3i+3}X_{3i+3,\,3i+2} \right) \nonumber \\[3pt]
&+ \left\{ \begin{array}{lcr}
\phi_{\bar{n}}X_{\bar{n},\bar{n}-1}X_{\bar{n}-1,\bar{n}} & , & n \textrm{ even} \\[5pt]
X_{\bar{n},\bar{n}+1}^{S/A}X_{\bar{n}+1,\bar{n}}^{S/A}X_{\bar{n},\bar{n}-1}X_{\bar{n}-1,\bar{n}} & , & n \textrm{ odd}
\end{array}\right.
\end{align}
The gauge group at node 0 and its adjoint field are projected by the orientifold involution, with signs $\tau_{0}$ and $\tau_{00}$, respectively. Depending on the parity of $n$, the other projected elements are the gauge group at node $\bar{n}$ and its adjoint if $n$ is even, with signs $\tau_{\bar{n}}$ and $\tau_{\bar{n},\bar{n}}$. On the other hand, if $n$ is odd, fields $X_{\bar{n},\bar{n}+1}$ and $X_{\bar{n}+1,\bar{n}}$ are projected onto symmetric or anti-symmetric representations by $\tau_{\bar{n},\bar{n}+1}$ and $\tau_{\bar{n}+1,\bar{n}}$. In this case, the anomaly-cancellation condition is not trivial and requires that $\tau_{\bar{n},\bar{n}+1}=\tau_{\bar{n}+1,\bar{n}}$. From the sign rule, this means that $\tau_{0} = \tau_{00}$. 

Let us look at the constraints on the $R$-charges. Eq.~\eqref{eq:RchargesSPP} still holds and recursion yields
\begin{align}\label{eq:RecursiveR}
&r_{00} = r_{3i,3i} \; , \nonumber \\
&2r_{3i,3i+1} + r_{00}= -1  \; , \nonumber \\
&2r_{3i+1,3i+2} - r_{00} = -1 \; . 
\end{align}
Using Eq.~\eqref{eq:RecursiveR}, anomaly-free $R$-symmetry gives
\begin{align}
&r_{00} \left( N_0 - N_1 + 2 \tau_{00} \right) = - \left( N_0 - N_1 - 2 \tau_0 \right) \; , \\[3pt]
&r_{00} \left( 2 N_{3i} - N_{3i-1} - N_{3i+1} \right) = - \left( 2N_{3i} - N_{3i-1} - N_{3i+1} \right) \; , \label{eq:RZeroZeroAdj} \\[3pt]
&r_{00} \left( N_{3i} - N_{3i+2} \right) = 2N_{3i+1} - N_{3i} - N_{3i+2} \; , \\[3pt]
&r_{00} \left( N_{3i+3} - N_{3i+1} \right) = 2N_{3i+2} - N_{3i+3} - N_{3i+1} \; , \\[3pt]
&r_{00} \left( N_{\bar{n}} - N_{\bar{n}-1} + 2 \tau_{\bar{n}\bar{n}} \right) = - \left( N_{\bar{n}} - N_{\bar{n}-1} - 2 \tau_{\bar{n}} \right) \; , \\[3pt]
&r_{00} \left( N_{\bar{n}} - N_{\bar{n}-1} + 2 \tau_{\bar{n},\bar{n}+1} \right) = - \left( N_{\bar{n}} - N_{\bar{n}-1} - 2 \tau_{\bar{n},\bar{n}+1} \right) \; . 
\end{align}
If we impose that $N_{i}=N$ for all $i$, we obtain
\begin{align}
r_{00} = \frac{\tau_0}{\tau_{00}} = \pm 1 \; ,
\end{align}
but both choices of signs would violate unitarity. Thus, the conformal point of the parent theory is excluded. Allowing for different ranks gives, from the second equation, 
\begin{align}
r_{00} = -1 \; , \quad \textrm{if } \quad 2N_{\bar{n}} - N_{\bar{n}-1} + N_{\bar{n}+1} \neq 0 \; 
\end{align}
and again $R_{00}=0$, violating unitarity. On the other hand, if we impose the right hand sides to vanish, we get 
\begin{align}\label{eq:RanksSPPZn}
& r_{00} = 0 \; , \\[3pt]
& N_{i} = N_0 - i\, 2 \tau_0 \; , \quad 0 \leq i \leq \bar{n} \; , \nonumber \\[3pt]
&\tau_0 = - \tau_{\bar{n}} \; , \quad n \textrm{ even} \; , \nonumber \\
&\tau_0 = - \tau_{\bar{n},\bar{n}+1} \; , \quad n \textrm{ odd} \; .
\end{align}
Note that the sign rule restricts the possible choices for the $\tau$ signs, such that only $(\tau_0 ,\, \tau_{00} ,\, \tau_{\bar{n}} ,\, \tau_{\bar{n},\bar{n}})=(\pm,\,\pm,\,\mp,\,\mp)$ are allowed, both for $n$ even and, {\it mutatis mutandis}, for $n$ odd.

Before the orientifold projection there are $3n$ nodes, while all $R$-charges can be expressed in terms of one of them, say $r_{01}$, due to the recursive structure. In case of $n$ even (the odd case is similar), after the $\mathbb{Z}_2$ involution there are 2 projected nodes and $3n/2-1$ not projected. The condition $r_{00}=0$ ensures that $r_{01} = r_{12} = -1/2$, then there are $3n$ fields carrying $r$-charge $(-1/2)$. We obtain, at large $N$, 
\begin{align}
a^{\Omega}_{_{\textrm{SPP}/{\mathbb{Z}'_n}}} = \frac{81}{256}n N^2 \; ,
\end{align}
which holds also for $n$ odd. The volume of the horizon is
\begin{align}
V^{\Omega}_{_{\textrm{SPP}/{\mathbb{Z}'_n}}} = \frac{\pi}{n}\frac{64}{81} \; .
\end{align}
The ratio with the $a$-charge at the conformal point before the orientifold projection is
\begin{align}
\frac{a^{\Omega}_{_{\textrm{SPP}/{\mathbb{Z}'_n}}}}{a_{_{\textrm{SPP}/{\mathbb{Z}'_n}}}} = \frac{9\sqrt{3}}{32} \simeq 0.4871 \; .
\end{align}
This solution exists for all non-chiral orbifold $\mathbb{Z}'_n$ with $n\geq 1$, where for $n=1$ and $n=2$ it is part of the more general family of solution we found. Note that a solution with $\tau_{0}=-\tau_{00}$ is allowed only for $n=2$, because for $n>2$ nodes with adjoint fields prevent the solution to exist.

\subsection{Seiberg duality for Unoriented SPP$/\mathbb{Z}'_n$}\label{sec:DualSPPZn}

The $a$-maximization procedure for unoriented SPP/$\mathbb{Z}'_n$ shows that there is a conformal point for $r_{00}=0$. For this value of the $R$-charges, none of the gauge-invariant operators decouples. Then, the conformal point is determined without any ambiguities in the electric theory. As a further check, we study the dual magnetic theory and look for the conformal point. Due to the recursive structure, we have two options: first, for $n>1$, we can dualize only the first unitary gauge group and compute the maximal central charge $a$. Second,  we can dualize the first node of each fundamental structure, along all the quiver for even $n$ or all but the last unitary group for odd $n$. This is because the last unitary group has tensorial matter and we do not know how Seiberg duality works in this case with the toric superpotential, the same problem as in SPP. However, in this second method we just obtain the first one recursively repeated. Note that for all dualized node the rank remains the same, as
\begin{align}
\widetilde{N}_i = N_{i-1} + N_{i+1} - N_i = N_i \; ,
\end{align}
where we have used Eq.\eqref{eq:RanksSPPZn}.

Then, proceeding as in the first case, the resulting gauge groups, matter and superpotential is the same as for the magnetic SPP$/\mathbb{Z}'_2$, with the remaining quiver and superpotential unchanged:
\begin{align}
\widetilde{W}^{\Omega}_{_{\textrm{SPP}/\mathbb{Z}'_n}} &=  M_2 \left(  \widetilde{X}_{21}\widetilde{X}_{12} - X_{23}X_{32} \right) - \widetilde{X}_{12}\widetilde{X}_{21}\widetilde{X}_{10}\widetilde{X}_{01} + \phi_3 \left( X_{32}X_{23} - X_{34}X_{43} \right) + \ldots \; .
\end{align}
Solving the constraints for the $R$-charges, we find that for those bifudamental fields transforming under gauge groups which have also an adjoint field
\begin{align}
r_{i,\,i+1} + r_{i+1,\,i} = -1 - r_{22} \; , 
\end{align}
for example $r_{12}+r_{21}=-1-r_{22}$, since in the magnetic theory the second node has an adjoint. For the adjoints along the quiver $r_{i,i} = r_{22}$. On the other hand, for those bifundamental transforming under gauge groups which do not have an adjoint 
\begin{align}
r_{i,\,i+1} + r_{i+1,\,i} = -1 + r_{22} \; ,
\end{align}
for example $r_{01}+r_{10}= -1 + r_{22}\,$ in the magnetic theory. For the $R$-symmetry to be anomaly-free, the only solution is $r_{22}=0$, as for the electric theory. The central charge $a$ gets no correction, since no operators decouple. We conclude that the conformal point of SPP$/\mathbb{Z}'_n$ is the one with $r_{22}=0$, $\tau_0=\tau_{00}=-\tau_3=-\tau_{33}$ and central charge, in the third scenario,
\begin{align}
& a^{\Omega}_{_{\textrm{SPP}/{\mathbb{Z}'_n}}} = \frac{81}{256}n N^2 \; , \nonumber \\[5pt]
&\frac{a^{\Omega}_{_{\textrm{SPP}/{\mathbb{Z}'_n}}}}{a_{_{\textrm{SPP}/{\mathbb{Z}'_n}}}} = \frac{9\sqrt{3}}{32} \simeq 0.4871 \; ,
\end{align}
as for the electric theory.

\section{Orientifold projection of  $\mathbb{C}^3/\mathbb{Z}'_{3n}$ and deformations}\label{Sec:OmegaDeformation}

All parent theories we study in this work can be obtained from $\mathbb{C}^3/\mathbb{Z}'_{3n}$ by mass deformation. As we are interested in the conformal point after the orientifold projection and to compare it with the parent one, we should analyze the orientifold of $\mathbb{C}^3/\mathbb{Z}'_{3n}$ to get insights on the RG flow to the unoriented $\left(\mathrm{SPP}/\mathbb{Z}'_{n}\right)^{\Omega}$ and the unoriented $\left(L^{k,n-k,k}\right)^{\Omega}$. The strategy for the computation of the fixed point follows closely the one adopted in the previous sections. The only difference is that in this case adjoint fields are present at all nodes and the superpotential has only cubic interactions. Then one finds
\begin{align}
    r_{00} = r_{ii} \; , \qquad 2r_{01} = 2r_{i,i+1} = -1 - r_{00} \; , \qquad \forall i \; . 
\end{align}
For the parent theory, the constraints on the $R$-charges give $N_i=N$ for all $i$ and
\begin{align}
    a_{_{\mathbb{C}^3/\mathbb{Z}'_{3n}}} = \frac{3}{4}nN^2 \;.
\end{align}

The orientifold projection we want to study is given by four fixed points, but their signs depend on $n$ being even or odd. If $n$ is even $\prod \tau = (-1)^{N_W/2}=+1$, whereas if $n$ is odd $\prod \tau = -1$. Let us first focus on the case with $n$ odd. The anomaly cancellation condition imposes $\tau_{\bar{n},\bar{n}+1}=\tau_{\bar{n}+1,\bar{n}}$, where $\bar{n}=(3n-1)/2$ is the last node in the orientifolded quiver. Then, it follows that $\tau_0 = - \tau_{00}$. One finds a solution for the $R$-charges at $N_i=N_0- i\, 2\tau_0$, $0\leq i \leq \bar{n}$, $\tau_0=-\tau_{\bar{n},\bar{n}+1}$, with
\begin{align}\label{eq:C3Z3noddCharges}
    &r_{00} = r_{01} = - \frac{1}{3} \; , \nonumber \\[3pt]
    &a^{\Omega}_{_{\mathbb{C}^3/\mathbb{Z}'_{3n}}} = \frac{3}{8} n N^2 \; ,
\end{align}
so the theory belongs to the first scenario. 

We turn to $n$ even, in which case the last node $\bar{n}=3n/2$ is projected by the orientifold plane. There is no condition for anomaly cancellation, thus two distinct choices are allowed: solution A with $\tau_0=-\tau_{00}$ and solution B with $\tau_0=\tau_{00}$.\footnote{Note that the same choices are allowed for $\left(\textrm{SPP}/\mathbb{Z}'_2\right)^{\Omega}$.} For solution A, one finds that $N_i=N_0-i\, 2\tau_0$, $0\leq i \leq \bar{n}$, $\tau_0=\tau_{\bar{n}}$ and the same values of Eq.~\eqref{eq:C3Z3noddCharges}, so again the theory realises the first scenario. 

\begin{figure}
    \centering
    \includegraphics[scale=0.8]{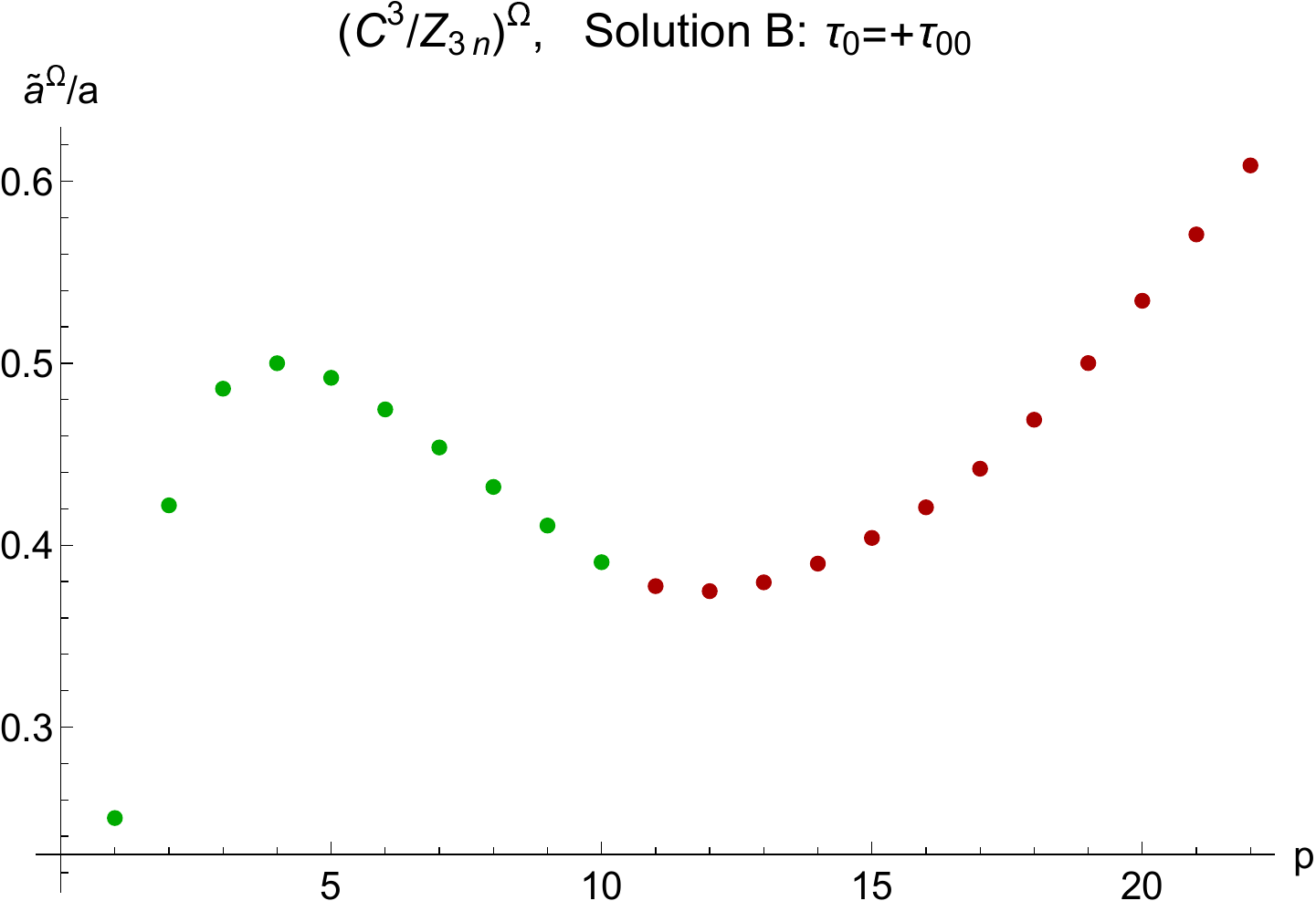}
    \caption{The corrected central charge $a$ for the unoriented $\mathbb{C}^3/\mathbb{Z}'_{12}$ solution B, where operators $\textrm{Tr }\phi_i^j$ start to decouple at $p=10$.}
    \label{fig:C3Z3nevenCorrectedCharge}
\end{figure}

On the other hand, for solution B one gets $N_i = N_0 - i \, p$, $0\leq i \leq \bar{n}$, $\tau_0 = - \tau_{\bar{n}}$, $p\tau_0 > 0$ and  
\begin{align}
    & a^{\Omega}_{_{\mathbb{C}^3/\mathbb{Z}'_{3n}}} = \frac{81}{16}nN^2 \frac{p^2 \tau_0}{\left( p + 2 \tau_0\right)^3} \; , \nonumber \\[5pt]
    & \left( \frac{a^{\Omega}}{a} \right)_{n \textrm{ even}} = \frac{27}{4} \frac{p^2 \tau_0}{\left( p + 2 \tau_0\right)^3} \; ,
\end{align}
which is always less than 1/2, hence the theory belongs to the third scenario. For some values of $p$ there are gauge-invariant operators that decouple before reaching the conformal point. In this case, they are 
\begin{align}
    & \mathcal{O}_{i,j} = \textrm{Tr }\phi_i^j \; , \qquad j>1 \; , \nonumber \\[5pt]
    & \widetilde{\mathcal{M}}_{i,lk} = \phi_i^l \left( X_{i,i+1}X_{i+1,i} \right)^k \; , \qquad l\geq 0 \; , k\geq 1 \; ,
\end{align}
whose $R$-charges read
\begin{align}
    &R_{\mathcal{O}}^{(j)}=j \frac{4\tau_0}{p + 2 \tau_0} \; , \nonumber \\[5pt]
    &R_{\widetilde{\mathcal{M}}}^{(lk)}=\frac{4\tau_0}{p+2\tau_0}(l-k) + 2k \; .
\end{align}
While the second operator hits the unitary bound only for $p=1$, $l=0$ and $k=1$ with vanishing correction to the central charge, operators $\mathcal{O}_{i,j}$ start to decouple at $p=10$. The corrected central charge is shown in Fig.~\ref{fig:C3Z3nevenCorrectedCharge}.
Interestingly, at $p=2\tau_0$, none of the gauge-invariant operators decouple and the central charge $a^{\Omega}$ results to be equal to that of SPP$/\mathbb{Z}'_n$, $n$ even. Moreover, the pattern of mass deformation needed to flow from $\mathbb{C}^3/\mathbb{Z}'_{3n}$ to SPP$/\mathbb{Z}'_n$ enjoys the $\mathbb{Z}'_2$ required for the orientifold projection, so this flow is preserved under the orientifold involution. That is not the case for the Seiberg dual phases. In fact, the phases that are not $\mathbb{Z}_2$ symmetric are projected out by the orientifold. 

\begin{figure}
\centering{
\includegraphics[scale=0.6, trim= 2.15cm 5cm 6.35cm 5.3cm, clip]{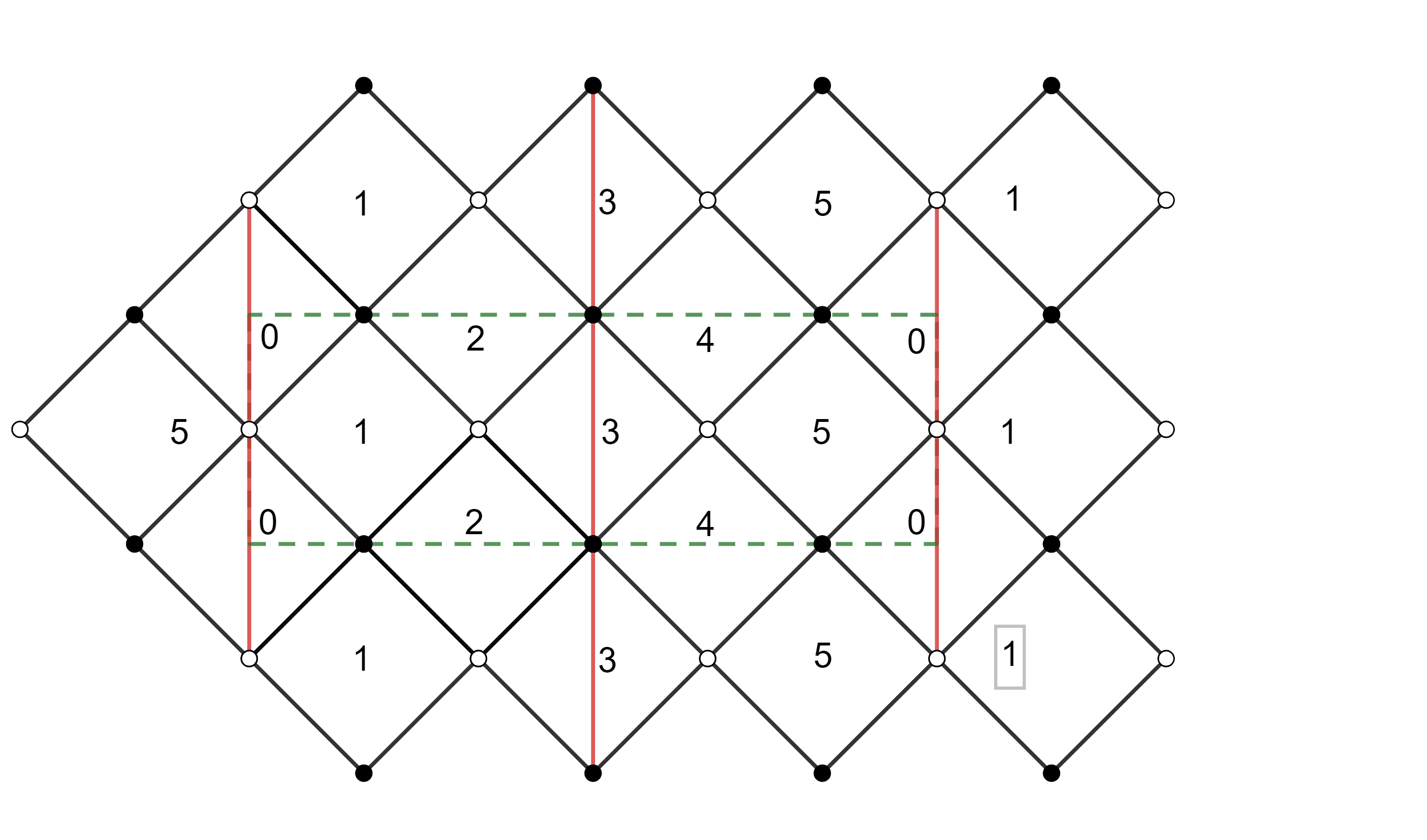}}
\caption{The brane tiling for $L^{3,3,3}$, where fixed lines represent the orientifold projection.}\label{fig:TilingL333}
\end{figure}

The deformation where the highest number of pairs of adjoints become massive allow for the orientifold involution, so the class of theories $\left(L^{\frac{3n}{2},\frac{3n}{2},\frac{3n}{2}}\right)$, $\left(  L^{\frac{3n-1}{2},\frac{3n+1}{2},\frac{3n-1}{2}}\right)$ can be reached in presence of orientifold planes. Their difference in the toric diagrams is crucial, since for $n$ even we can perform the orientifold projection we are interested in only with fixed lines\footnote{The related brane tiling is made only of squares and a fixed point can not lie on a square, because it must map nodes with different colors.}. See Fig.~\ref{fig:TilingL333} for the brane tiling of $n=2$, namely $L^{3,3,3}$. On the contrary, for $n$ odd fixed lines are not allowed, as can be seen from the toric diagram. An example of such a case on the brane tiling is $n=1$, namely SPP, in Fig.~\ref{fig:SPPdimer}.

Let us find and compare the conformal point both for the parent and the unoriented theories. Consider first the case with $n$ even, where all adjoints have been integrated out. The constraints for the $R$-charges~\cite{Franco:2005sm} read
\begin{align}
&r_{i,\,i+1} = -\frac{1}{2} \; , \nonumber \\
&N_i = N \; , \quad i = 0,\,\ldots, n-1 \; ,
\end{align} 
and the central charge $a$ is
\begin{align}
a_{_{n \textrm{ even}}} = \frac{81}{128} n N^2 \; .
\end{align}
Performing the orientifold projection, with fixed lines, we find $N_i = N_0 - i\, 2 \tau_0$, $0\leq i \leq \bar{n}$, $\tau_0 = -\tau_3$, where $\tau_0$ and $\tau_3$ are the sign of the two fixed lines. At large $N$, the $a$-charge reads
\begin{align}\label{eq:Aneven}
a^{\Omega}_{_{n \textrm{ even}}} = \frac{81}{256} n N^2 \; .
\end{align}
Clearly, for $n$ even the theory realizes the first scenario, since $a^{\Omega}/a = 0.5$ and the $R$-charges are the same both for parent and unoriented ones, at leading order. 

\begin{center}
\vspace{15pt}
\begin{tabular*}{0.9\textwidth}{@{\extracolsep{\fill}}cccccc}
\toprule
 & $Sp/SO(N_0)$ & $SU(N_1)$ & $SU(N_2)$ & $SO/Sp(N_3)$ & $U(1)_R$\\
\midrule
 $X_{01}$ & $\tiny{\yng(1)}$ & $\overline{\tiny{\yng(1)}}$ & $\bf{1}$ & $\bf{1}$ &  $\frac{1}{2}$  \\[3pt]
 $X_{10}$ & $\overline{\tiny{\yng(1)}}$ & $\tiny{\yng(1)}$ & $\bf{1}$ & $\bf{1}$ & $\frac{1}{2}$ \\[3pt]
 $X_{12}$ & $\bf{1}$ & $\tiny{\yng(1)}$ & $\overline{\tiny{\yng(1)}}$ & $\bf{1}$ & $\frac{1}{2}$\\[3pt]
 $X_{21}$ & $\bf{1}$ & $\overline{\tiny{\yng(1)}}$ & $\tiny{\yng(1)}$ & $\bf{1}$ & $\frac{1}{2}$\\[3pt]
 $X_{23}$ & $\bf{1}$ & $\bf{1}$ & $\tiny{\yng(1)}$ & $\overline{\tiny{\yng(1)}}$ &   $\frac{1}{2}$  \\[3pt]
 $X_{32}$ & $\bf{1}$ & $\bf{1}$ & $\overline{\tiny{\yng(1)}}$ & $\tiny{\yng(1)}$  & $\frac{1}{2}$ \\[3pt]
\bottomrule
\end{tabular*}
\captionof{table}{The matter content and the superconformal $R$-charges of $(L^{3,3,3})^{\Omega}$, dual to $(\textrm{SPP}/\mathbb{Z}'_2)^{\Omega}$.}\label{tab:L333}
\end{center}
\vspace{15pt}

\begin{figure}
\centering{
\includegraphics[scale=0.8]{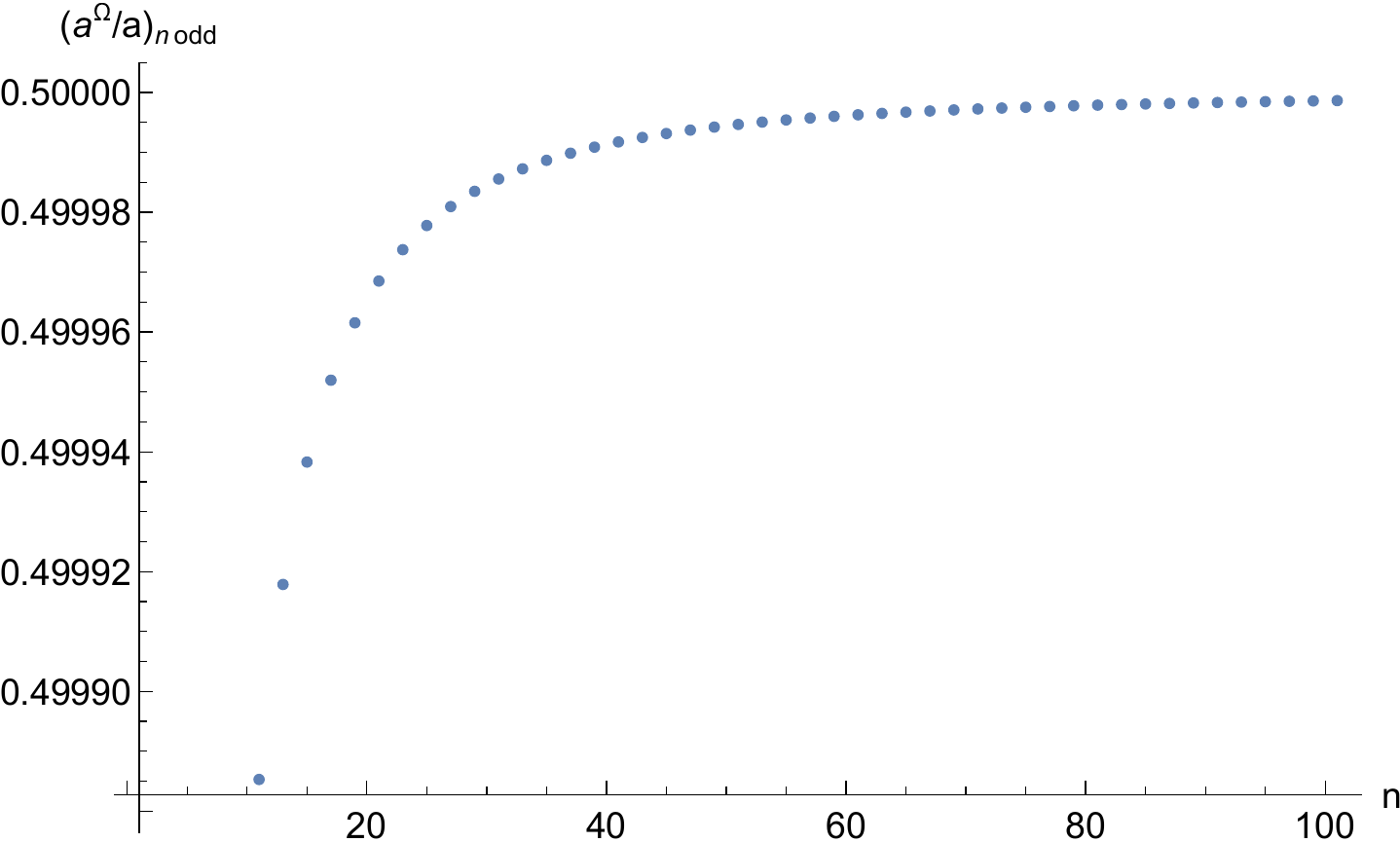}}
\caption{The ratio $\left(\frac{a^{\Omega}}{a}\right)$ for $n$ odd, which asymptotically approaches 0.5 .}\label{fig:ChargeAnodd}
\end{figure}

However, for $\mathbb{C}^3/\mathbb{Z}'_{3n}$ with $n$ odd, after mass deformation the presence of the adjoint field $\phi_0$ in the parent theory gives~\cite{Franco:2005sm}
\begin{align}
&2r_{2i,\,2i+1} = -1 - r_{00} \; , \nonumber \\
&2r_{2i+1,\,2i+2} = -1 + r_{00} \; , \nonumber \\
&N_i = N \quad , \quad i = 0,\ldots, n-1 \; .
\end{align}
Maximization of the central charge yields
\begin{align}
&r_{00} = n - \sqrt{\frac{1+3n^2}{3}} \; , \nonumber \\[3pt]
&a_{_{n \textrm{ odd}}} = \frac{9}{64} \left[ 3n\left(1-n^2\right) + \left( 1 + 3 n^2 \right) \sqrt{\frac{1+3n^2}{3}} \right] \; .
\end{align}
For the orientifold projection, $N_i = N_0 - i\, 2 \tau_0$, $0\leq i \leq \bar{n}$, $\tau_0=\tau_{00}=-\tau_{\bar{n},\bar{n}+1}=-\tau_{\bar{n}+1,\bar{n}}$. Since it is required that $r_{00}=0$, the $R$-charges and the central charge $a$ are the same of $n$ even in Eq.~\eqref{eq:Aneven}. The ratio between the central charges reads
\begin{align}
\left(\frac{a^{\Omega}}{a}\right)_{n \textrm{ odd}} = \frac{9}{4}n \left[ 3 n \left( 1 - n^2 \right) + \left( 1 + 3 n^2 \right) \sqrt{\frac{1+3n^2}{3}} \right]^{-1} \; ,
\end{align}
which asymptotically tends to 0.5, as shown in Fig.~\ref{fig:ChargeAnodd}. This case belongs to the third scenario.

Hence, for the unoriented case there is no distinction between $n$ even or odd and both show the same central charge $a$ of the unoriented SPP$/\mathbb{Z}'_n$. But, for $n$ odd $L^{\frac{3n-1}{2},\frac{3n+1}{2},\frac{3n-1}{2}}$ the third scenario occurs and the orientifold projection is performed with fixed points. For $n$ even $L^{\frac{3n}{2},\frac{3n}{2},\frac{3n}{2}}$ features the first scenario and its orientifold projection is performed with fixed lines. In $L^{\frac{3n}{2},\frac{3n}{2},\frac{3n}{2}}$ the fact that fixed lines are needed for the orientifold projection is crucial: fixed lines breaks a $U(1)$ mesonic symmetry, part of the toric $U(1)^2\times U(1)_R$. On the contrary, the orientifold projection with fixed points does not break toricity, but at the conformal point $r_{00}=0$ and, as a consequence, $r_{01}=r_{122} = -1/2$. Thus, the number of nodes and flavour symmetries matches, and 't Hooft anomalies do as well.

The superpotential and the matter content are however different and equivalence of the superconformal index must be checked. Contributions to the index come from matter fields $X_{ij}$ and vector multiplets $V_i$ as\footnote{Recall that $R_{ij}=r_{ij} +1$ in our notation.}
\begin{align}
& i_{X}(t,{{s}}) = \sum \frac{t^{R_{ij}}\chi_{_{X_{ij}}} - t^{2-R_{ij}}\chi_{_{\overline{X}_{ij}}}}{\left( 1- t{{s}} \right)\left( 1 - t{{s}}^{-1}\right)} \; , \nonumber \\[3pt]
& i_{V}(t,{{s}}) = \sum_{i=0}^{3n-1} \frac{\left[ 2t^2 - t\left({{s}} + {{s}}^{-1} \right)  \right]}{\left( 1 - t{{s}} \right)\left( 1 - t{{s}}^{-1} \right)} \chi_{\textrm{adj}_i} \; ,
\end{align}
where the first sum runs over all matter fields, $\chi_{_{X_{ij}}}$, $\chi_{_{\overline{X}_{ij}}}$ are the characters of the representation of $X_{ij}$ and its conjugate, $t$ and ${{s}}$ are the fugacities for $R$-charge and (twice the) spin, respectively. When matter fields $\phi_i$ are present, either in the adjoint or in the anti/symmetric representation, in the unoriented models $R_{ii}=1$ at the conformal fixed point and their contributions to the superconformal index vanish. The remaining contributions are equal for unoriented model with the same number of gauge groups. 

\begin{figure}
\begin{subfigure}{0.45\textwidth}
\centering{\includegraphics[scale=0.5, trim={0 0 8cm 0}, clip]{ToricSPPZ2Or.png}}
\end{subfigure}
\begin{subfigure}{0.45\textwidth}
\centering{\includegraphics[scale=0.525, trim={3cm 1cm 8cm 2cm}, clip]{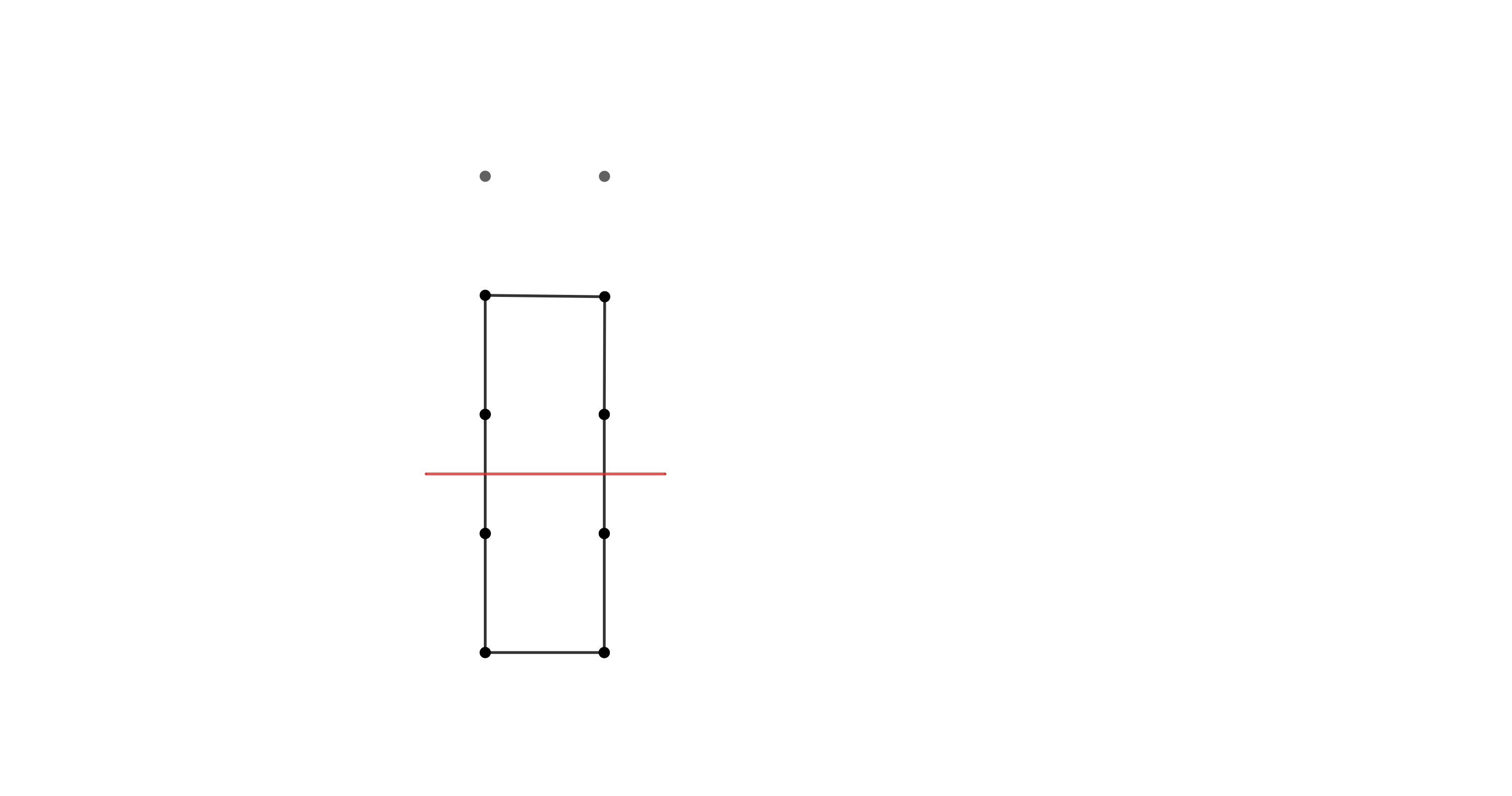}}
\end{subfigure}
\caption{Two dual theories: on the left, SPP$/\mathbb{Z}'_2$ with fixed points and, on the right, $L^{3,3,3}$ with fixed lines.}\label{fig:DualEven}
\end{figure}

Hence, the central charge $a^{\Omega}$, 't Hooft anomalies and the superconformal index of $\left({SPP}/\mathbb{Z}'_n\right)^{\Omega}$ match that of $\left(L^{\frac{3n}{2},\frac{3n}{2},\frac{3n}{2}}\right)^{\Omega}$ and of $\left(\mathbb{C}^3/\mathbb{Z}'_{3n}\right)^{\Omega}$ with $p=2\tau_0$ for $n$ even, while for $n$ odd the same quantities match between $\left({SPP}/\mathbb{Z}'_n\right)^{\Omega}$ and $\left(L^{\frac{3n-1}{2},\frac{3n+1}{2},\frac{3n-1}{2}}\right)^{\Omega}$. We want to stress that, for $n$ even, the former theory is an orientifold with fixed points in the third scenario, while the latter is an orientifold with fixed lines in the first scenario, as shown in Fig.~\ref{fig:DualEven}, exactly as the PdP$_{3c}^{\Omega_2}$ and PdP$_{3b}^{\Omega}$  theories in~\cite{Antinucci:2020yki}. The reader is invited to go back to Fig.~\ref{fig:Scheme}, where the full web of relations that we find is summarized.

\begin{figure}
\begin{center}
\begin{subfigure}{0.45\textwidth}
\centering{\includegraphics[scale=0.46, trim={0 0 10cm 0}, clip]{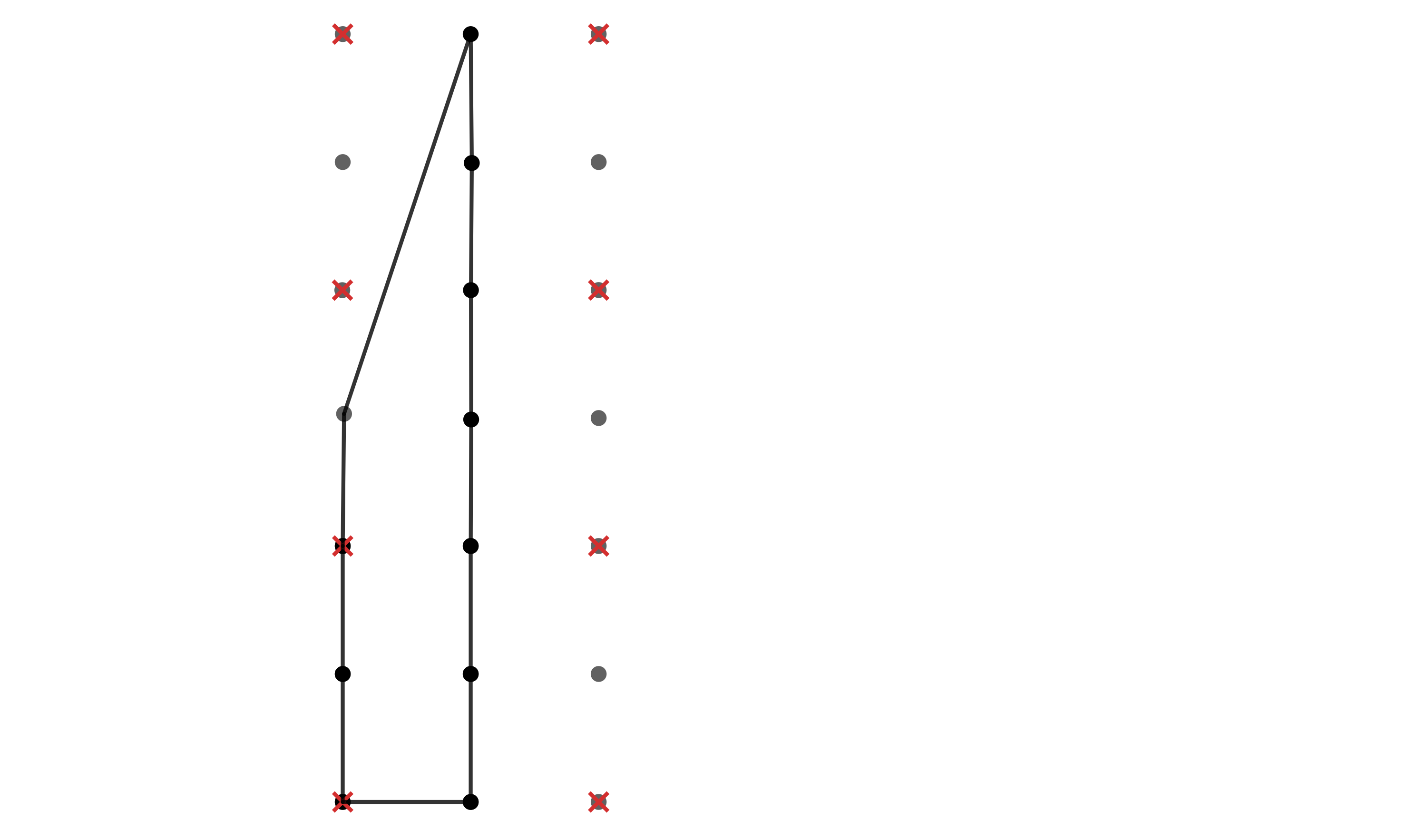}}\end{subfigure}
\begin{subfigure}{0.45\textwidth}
\centering{\includegraphics[scale=0.47, trim={1.5cm 0 0 0}, clip]{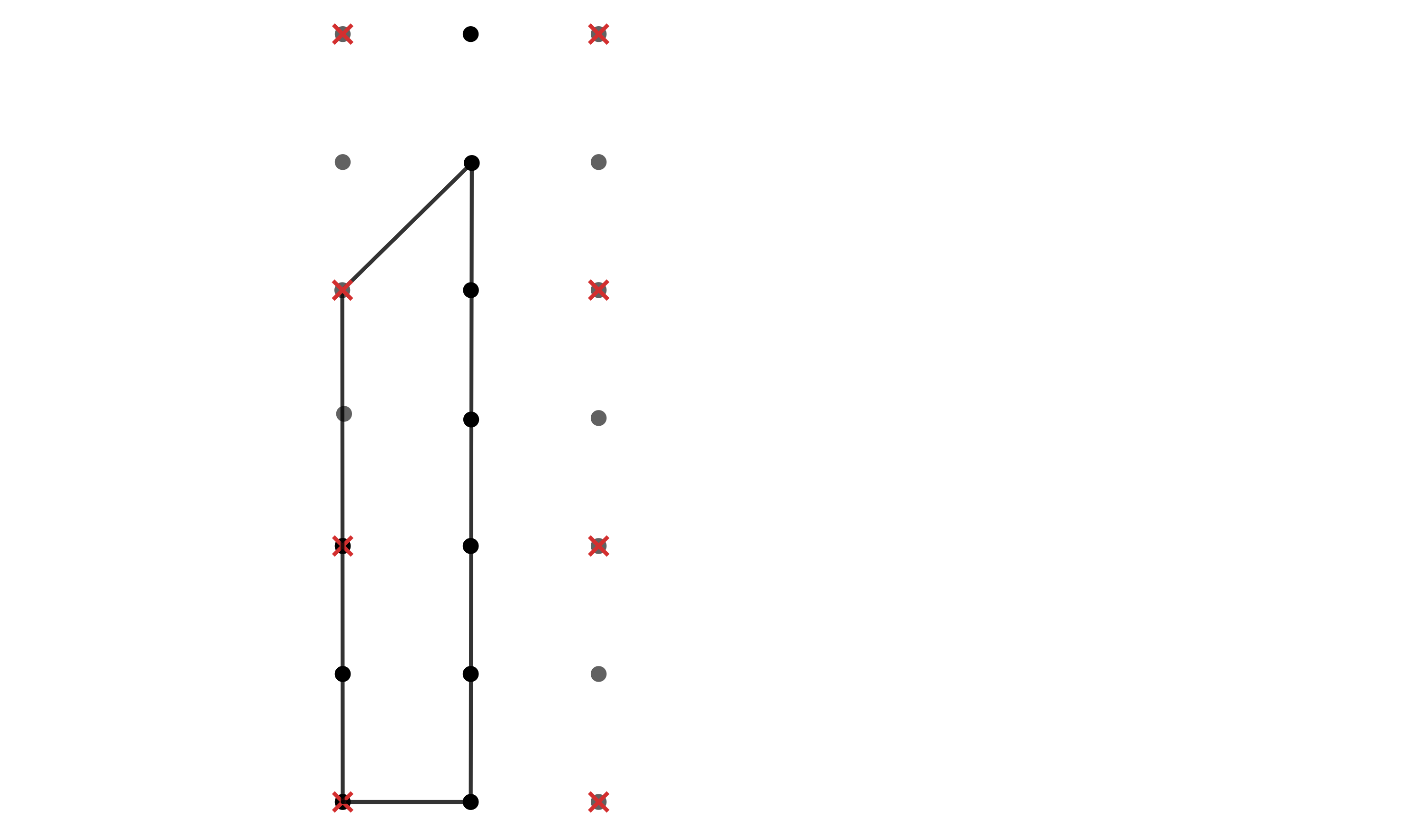}}
\end{subfigure}
\caption{Two dual theories: on the left, SPP$/\mathbb{Z}'_3$ and, on the right, $L^{4,5,4}$, both with fixed points.}\label{fig:DualOdd}
\end{center}
\end{figure}

\section{Elliptic models}\label{sec:Elliptic}

The models described above arise as the supersymmetric gauge field theory living on the world-volume (WV) of D3-branes at a toric CY singularity. We mentioned that, after two T-dualities, one may recover a system of NS5-branes and D5-branes wrapped along a torus and generating the brane tiling. Both descriptions are defined in type IIB string theory. However, performing one T-duality one can construct another useful description of the brane system in type IIA, as D4-branes suspended between NS5 branes. In particular, the D4-branes are wrapped along a compact direction, say $x^6$, while NS5s can extend along directions $x^4$, $x^5$ or $x^8$, $x^9$. In the former case they are called simply NS5, in the latter NS5$^\prime$. Both types split the WV of D4-branes. See~\cite{Uranga:1998vf} for the construction of such theories. The configuration is summarized in Tab.~\ref{tab:D4NS5}, where also $\Omega$6-planes are included, yielding the $\Omega$ projection.

\begin{center}
	\begin{tabular*}{0.5\textwidth}{@{\extracolsep{\fill}}cc|cccccc}
		\toprule
		& 0 1 2 3 & 4 & 5 & 6 & 7 & 8 & 9 \\
		\midrule
		D4 & $-$ $-$ $-$ $-$ & $\cdot$ & $\cdot$ & $-$ & $\cdot$ & $\cdot$ & $\cdot$ \\
		NS5 & $-$ $-$ $-$ $-$ & $-$ & $-$ & $\cdot$ & $\cdot$ & $\cdot$ & $\cdot$ \\  
		NS5$^\prime$ & $-$ $-$ $-$ $-$ & $\cdot$ & $\cdot$ & $\cdot$ & $\cdot$ & $-$ & $-$ \\
		${\Omega}$6$^{\pm}$ & $-$ $-$ $-$ $-$ & $\cdot$ & $\cdot$ & $\cdot$ & $-$ & $-$ & $-$ \\
		\bottomrule
	\end{tabular*}
	\captionof{table}{The T-dual picture with D4-branes, NS5 and NS5$^\prime$-branes. The direction $x^6$ is compact.}
	\label{tab:D4NS5}
\end{center} 

\begin{figure}
\centering{\includegraphics[scale=0.5, trim=2cm 2cm 3cm 3cm, clip]{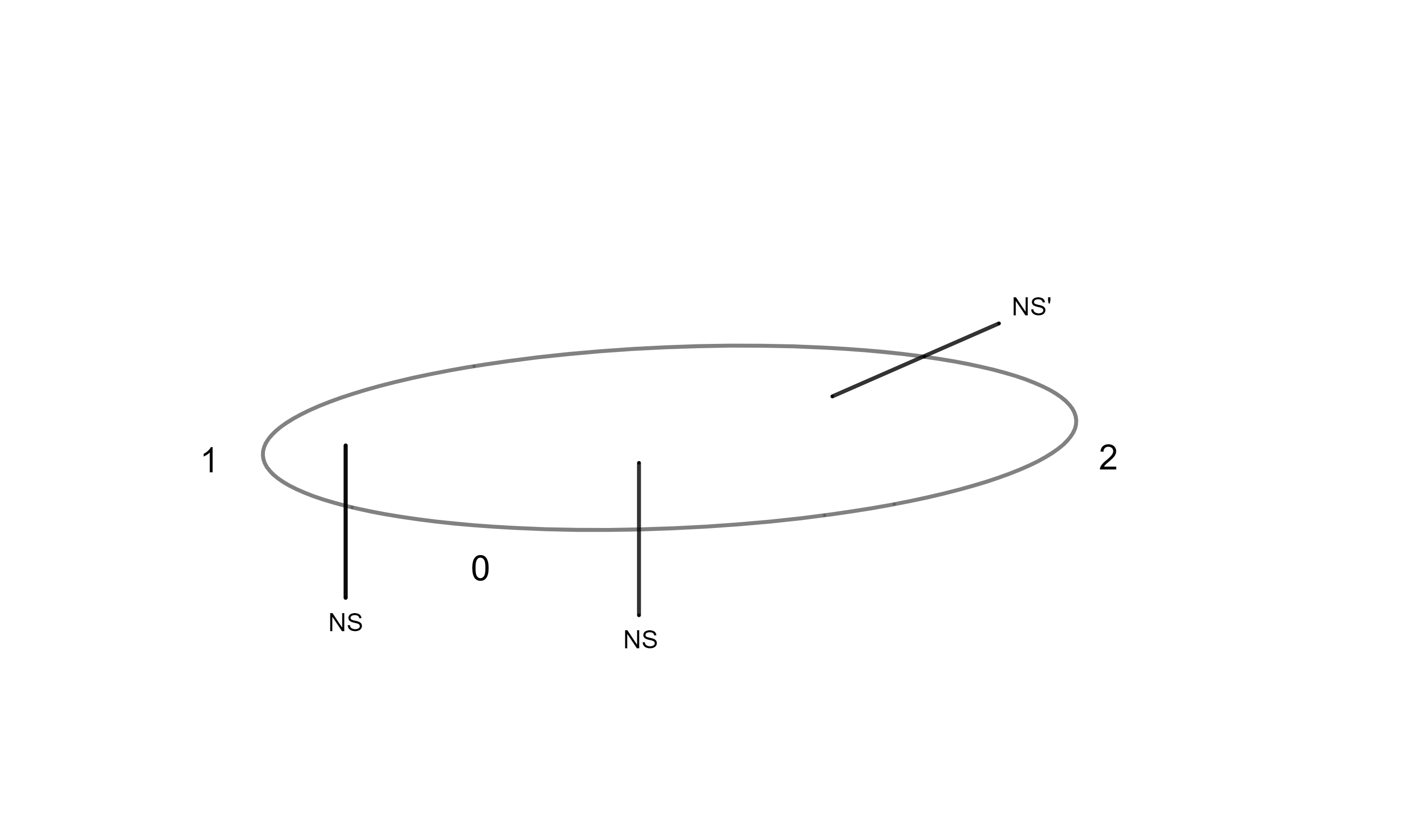}}
\caption{The brane system in type IIA which corresponds to the SPP singularity. The circular direction is $x^6$.}\label{fig:D4SPP}
\end{figure}

A simple example that can be obtained from this configuration is the SPP, where two NS and one NS$'$ 5-branes, located at different positions on $x^6$, divide the WV of D4-branes into three stacks, which we label by 0,1 and 2, see Fig.~\ref{fig:D4SPP}. Bifundamental fields $X_{01}$, $X_{01}$, $X_{12}$, $X_{12}$ and $X_{20}$, $X_{02}$ arise at the intersections between D4s and a 5-brane. The stack denoted by 0 is suspended between two parallel NS5 branes and D4-branes can be moved along them. This generates an adjoint field $X_{00}=\phi_0$, which completes an $\mathcal{N}=2$ vector multiplet. In fact, locally the physics resembles $\mathcal{N}=2$. D4-branes between two orthogonal 5-branes generate a quartic superpotential term of the form $\pm X_{ij}X_{ji}X_{ij'}X_{j'i}$, taken with positive sign if an NS5 lies at the left of the D4. When D4s are suspended between two parallel 5-branes, a cubic term as $\phi_i\left( X_{ij}X_{ji} - X_{ij'}X_{j'i}\right)$ is generated.

\begin{figure}
\centering{\includegraphics[scale=0.5, trim=2cm 3cm 3cm 1cm, clip]{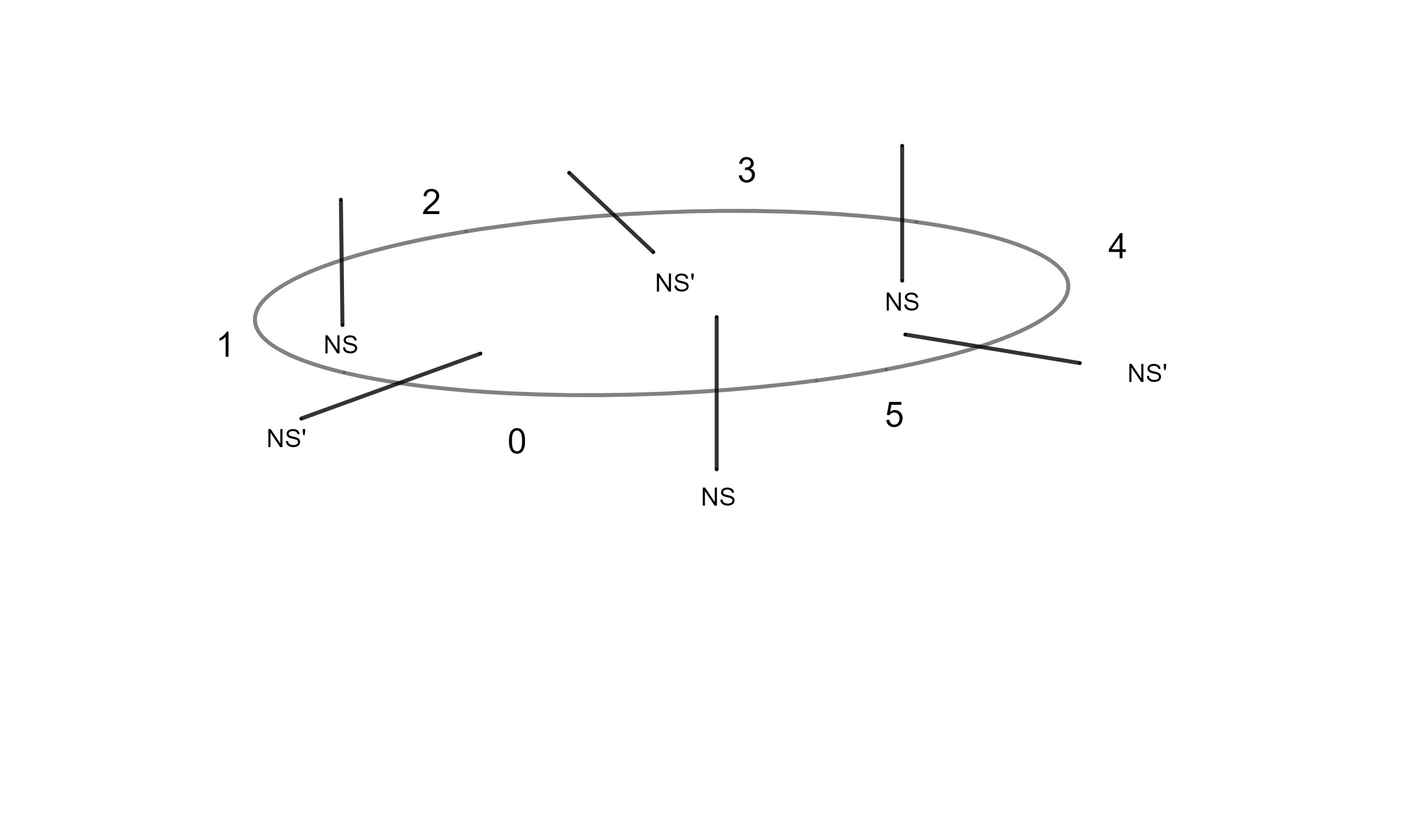}}
\caption{The brane system in type IIA which corresponds to the $L^{3,3,3}$ singularity. The circular direction is $x^6$.}\label{fig:D4L333}
\end{figure}

The geometry of the singularity can be described in $\mathbb{C}^4$ as the locus
\begin{align}
x y = z w^2 \; .
\end{align}
This construction can be generalized to the case of $r$ NS5 and $s$ NS5$'$, whose geometry reads
\begin{align}
x y = z^s w^r \; .
\end{align}
Adjoint fields arise between parallel 5-branes of the same type, but a relative rotation corresponds to giving them a mass, and the effective field theory is obtained below the mass scale. Thus, we can start with a system of $3n$ NS5s, corresponding to $\mathbb{C}^3/\mathbb{Z}'_{3n}$, and rotate the NS5 in an alternating scheme. If $3n$ is even, we can give  mass to all  the adjoints fields and integrate them out, whereas in case of $3n$ odd two 5-branes remain always parallel. This is exactly the process we have described in the previous section, and we end up with $L^{\frac{3n}{2},\frac{3n}{2},\frac{3n}{2}}$ or $L^{\frac{3n-1}{2},\frac{3n+1}{2},\frac{3n-1}{2}}$, see for example $L^{3,3,3}$ in Fig.~\ref{fig:D4L333}.

When a pair of ${\Omega}$6$^{\pm}$ planes is present in the system, it induces a $\mathbb{Z}_2$ indentification among 5-branes and stacks of D4. For $n$ odd, one of the orientifold planes lies in correspondence of a 5-brane which separates a stack of D4 labelled by $i$ and another one labelled by $j$. In this case, fields $X_{ij}$, $X_{ji}$ are projected onto symmetric or antisymmetric representations, both in the same one in case of an NS5, in opposite ones for an NS5$^\prime$\footnote{This is because the NS5$^\prime$ divide the ${\Omega}$6 WV into two regions. Crossing the 5-branes, the RR-charge changes sign.}. The other ${\Omega}$6$^{\pm}$ lies on a stack of D4, projecting the gauge group onto $SO/Sp$ and the adjoint field onto a symmetric or antisymmetric representation. This happens also for $n$ even, where both orientifold planes acts in this way. Note that the mode of fractional branes along parallel 5-branes is projected out, since the D4s are stuck at the orientifold singularity: they can not be moved along $x^4$, $x^5$ since the orientifold plane does not wrap these directions.

In these elliptic models, Seiberg duality is described as the reordering of the 5-branes~\cite{Elitzur:1997fh, Uranga:1998vf}. Consider, for example, the situation in which node $(a-1)$ has an adjoint and we want to dualize node $(a)$. As described in Sec.~\ref{sec:Seiberg} and Figs.~\ref{fig:SeibergDualAdj}-\ref{fig:SeibergDualAdjDimer} the adjoint moves from node $(a-1)$ to $(a+1)$, while the rest of the quiver remains untouched. Let us look at the process in the ellipctic model, where the same configuration corresponds to a stack of D4 labelled by $(a-1)$ delimited by two parallel 5-branes, say two NS5. The stack $(a)$ is delimited then by an NS5 and an NS5$'$, and finally $(a+1)$ by an NS5$'$ and an NS5, in this order. Seiberg duality on $(a)$ means that one exchanges the 5-branes at its ends, so no adjoint field is generated. On the other hand, the 5-branes delimiting $(a-1)$ are now orthogonal, whereas for $(a+1)$ the 5-branes are parallel, explaining the presence of the adjoint field in the magnetic theory. 

Moreover, this construction explains why Seiberg duality picks the conformal point of $(\mathrm{SPP}/\mathbb{Z}'_2)^{\Omega}$ at $p=4\tau_0$ for solution A $(\tau_0=-\tau_{00})$ and at $p=2\tau_0$ for solution B $(\tau_0=+\tau_{00})$. In fact, moving the NS5 towards the NS5$'$ and eventually crossing it, results in the dual configuration where the number of D4 does not change~\cite{Elitzur:1997fh, Uranga:1998vf}, since in the system no D6 are present\footnote{If we add D6-branes along directions 0123789 in between the 5-branes, new D4-branes are created while exchanging the position of the 5-branes, in order to preserve supersymmetry~\cite{Elitzur:1997fh}.}. For the field theory associated to $(\mathrm{SPP}/\mathbb{Z}'_2)^{\Omega}$, this means that dualizing either node 1 or 2 yields $N_1=\widetilde{N}_1$ or $N_2=\widetilde{N}_2$, which is true only for $p=4\tau_0$ in solution A (see Eq.~\eqref{eq:RankZ2A}) and $p=2\tau_0$ in solution B (see Eq.~\eqref{eq:RanksZ2B}). The same mechanism of reordering of 5-branes occurs in dualizing a gauge node in $(\mathrm{SPP}/\mathbb{Z}'_n)^{\Omega}$, which results in $\widetilde{N_i}=N_i$, as already noted in Sec.~\ref{sec:DualSPPZn}. For $n>2$ the electric theory has the unique solution at $r_{00}=0$, value obtained also for solution B of $(\mathrm{SPP}/\mathbb{Z}'_2)^{\Omega}$ at $p=2\tau_0$. Note that this solution B has the same choice for $\tau$ signs as for $n>2$.

In theories without an orientifold plane and each gauge node connected to the adjacents by a chiral and an anti-chiral field, the condition that, after Seiberg duality, the rank of the gauge group remains the same is quite natural. In fact, the electric theory is conformal when all ranks of the gauge factors are equal, hence, in the conformal window $\widetilde{N}_i=N_i$ for the magnetic theory. On the other hand, the presence of an orientifold plane shifts the ranks of gauge factors, for anomaly cancellation, and the condition for the ranks to remain equal is no longer obvious.

The message from $(\mathrm{SPP}/\mathbb{Z}'_n)^{\Omega}$ with $n\geq 2$ is that for theories where each gauge factor has both chiral and anti-chiral matter and in presence of orientifold planes, the conformal window is given by the condition that the magnetic dual theory yields $\widetilde{N}_i = N_i$, as for theories without orientifold. Moreover, it seems that in this class of unoriented theories at the conformal point $r_{00}=0$.

Checking these statements for SPP$^{\Omega}$ is quite complicated, both on the elliptic model and on the field theory side. In the former, we can either move an NS5$'$ brane and its image towards the orientifold plane or deal with an NS5 at the position of the orientifold plane. From the field theory perspective, we may construct a magnetic dual theory by means of the deconfinement trick~\cite{Berkooz:1995km, Pouliot:1995me, Sakai:1997xs, Luty:1996cg, Garcia-Etxebarria:2015hua}, in which the tensorial matter is deconfined into fundamental fields of an auxiliary gauge group, leaving the original unitary group with only (anti)fundamental fields. Applying this technique in SPP$^{\Omega}$ with antisymmetric matter, the extra gauge group is symplectic. Indeed, for $r_{00}=0$ the rank of the Seiberg dual of $SU(N_1)$ remains unchanged, as in the previous cases, while this does not happen with symmetric matter and an orthogonal extra group. However, after the duality the auxiliary group can not be confined back. While most of the family $(\mathrm{SPP}/\mathbb{Z}'_n)^{\Omega}$ have the conformal point at $r_{00}=0$, all we can say for SPP$^{\Omega}$ is that we are not able to exclude fixed points with $r_{00}\neq 0$. From the result we got for $(\textrm{SPP}/\mathbb{Z}'_n)^{\Omega}$, in which we have analytical tools, we expect that the same should hold also for $\textrm{SPP}^{\Omega}$, namely, that the conformal point exists only for a specific value of $p$ for which $r_{00}=0$. This fact will be investigated further in the future.

\section{Discussion}\label{Sec:Discussion}

In this paper we have found an infinite class of pairs of non-chiral unoriented theories whose conformal fixed points have the same central charge $a^{\Omega}$, 't Hooft anomalies and superconformal index. These theories are $\left(\textrm{SPP}/\mathbb{Z}'_n \right)^{\Omega}$ and the orientifold $\Omega$ of $ L^{\frac{3n-1}{2},\frac{3n+1}{2},\frac{3n-1}{2}}$ for $n$ odd or $L^{\frac{3n}{2},\frac{3n}{2},\frac{3n}{2}}$ for $n$ even, whose parent theories can be constructed from the non-chiral orbifold $\mathbb{C}^3/\mathbb{Z}'_{3n}$ by mass deformation of pairs of adjoint fields~\cite{Bianchi:2013gka}. 

For $n$ odd, we find that both theories in each pair belong to the third scenario, whereby the orientifold projections, realised with fixed points on the dimer, break the conformal invariance of the parent theories even at large $N$, and they flow to a new conformal fixed point in the infrared~\cite{Antinucci:2020yki}. For $n=1$ the two theories are actually the same, namely $\left( \textrm{SPP}\right)^{\Omega}$. Imposing that all $\beta$-functions vanish, it seems that for any value of the number of fractional branes $p$ the theory has a fixed point. However, for some values of $p$, the central charge $a$ must be corrected taking into account operators that decouple before reaching the conformal point. The effect is that the ratio $a^{\Omega}/a$ tends to increase. Considering that the $\Omega$ projection is a $\mathbb{Z}_2$ involution of the parent theory, any point at which $a^{\Omega}/a>1/2$ seems unphysical. Moreover, for $p\geq5$, $\textrm{Tr }R\neq 0$ already at leading order, spoiling the holographic duality, at least in its simple form. Being the parent theory a holographic theory, beyond this value the comparison with the $a$ charge of the parent theory is unreliable. A possibility is that for $p\geq 5$ the field theory has a conformal point, with no gravity dual. Furthermore, following the prescription in~\cite{Benvenuti:2017lle}, for $p=1$ all the terms in the superpotential must be eliminated. Eventually, for $n=1$ all we can say from the field theory side is that in the range $2\leq p \leq 4$ $a$-maximization yields a maximum in the third scenario. 

For $n$ even, while for $\textrm{SPP}/\mathbb{Z}'_{n}$ the orientifold involution is performed with fixed points and belongs to the third scenario, for  $L^{\frac{3n}{2},\frac{3n}{2},\frac{3n}{2}}$ it is performed by fixed lines and belongs to the first scenario, realizing the same mechanism discussed for the first time in~\cite{Antinucci:2020yki}. This is a stronger evidence of the fact that the conformal points that we find are physically relevant. For $n=2$, apparently conformal invariance does not fix the relative rank $p$ of the gauge groups, but Seiberg duality selects the value of $p=2\tau_0$, which yields $r_{00}=0$.  Interestingly, for $n$ even the orientifold projection of the theory $\mathbb{C}^3/\mathbb{Z}'_{3n}$ belongs to the third scenario and only for $p=2\tau_0$ it shares the same central charge $a^{\Omega}$, 't Hooft anomalies and superconformal index of the unoriented SPP$/\mathbb{Z}'_{n}$ and, in turn, $L^{\frac{3n}{2},\frac{3n}{2},\frac{3n}{2}}$.

On the gravity side, all these theories can be described in Type IIA by means of elliptic models~\cite{Uranga:1998vf} where the geometry of the singularity depends on the configuration of NS5-branes, while stacks of D4-branes suspended between them provide the field theory. Rotating the 5-brane, one gives a geometric meaning to mass deformation and the flow to SPP$/\mathbb{Z}'_n$ and $L^{k,n-k,k}$ mentioned above. In this context, Seiberg duality gives a clearer picture of the conformal point. Reordering the 5-branes does not change the number of D4, since no D6-branes are present. Back on the field theory side, this means that the ranks in the magnetic theory remain unchanged, and this happens only for the unique value of $p$ that gives $r_{00}=0$. Nonetheless, we have not been able to fully understand why one should pick that unique value in the case of $n=1$, an issue related to the presence of tensorial matter. We leave this as an open problem.

The fact that each pair of theories share the same central charge $a^{\Omega}$, 't Hooft anomalies and superconformal index, along with the fact that they share the same global symmetries, implies that they are connected by an exactly marginal deformation. Since the endpoints of this exactly marginal deformation are orientifold projections of toric theories, one can conjecture that the two theories are actually dual, as suggested in~\cite{Antinucci:2020yki} for the case of the orientifold of PdP$_{3c}$ and PdP$_{3b}$. The geometric interpretation of this infrared duality is presently lacking, as already mentioned in the introduction, but having found an infinite class of theories in which this duality is realised gives hope that a geometric picture could emerge. In particular, note that as opposed to the case investigated in~\cite{Antinucci:2020yki}, the metrics of the parent theories of all these models are known.

\section*{Acknowledgments}

We thank I. García Etxebarria for a careful reading of the draft and suggestions. A.A. thanks S. Benvenuti and S. Bajeot for fruitful discussions. M.B. acknowledges fruitful discussions with U.~Bruzzo, D.~Bufalini, P.~Fr\'e, D.~Martelli, J.~F.~Morales, and R.~Savelli. S.M. thanks M. Sacchi for discussions and A. Hanany for discussions and suggestions at various stages. The work of M.B. is partially supported by Grant ID1202 “Strong Interactions: from Lattice QCD to Strings, Branes and Holography” within the ‘Beyond Borders 2019’ scheme of the University of Roma “Tor Vergata”.

\bibliographystyle{elsarticle-num}
\bibliography{biblio}

\begin{thebibliography}{10}
\expandafter\ifx\csname url\endcsname\relax
  \def\url#1{\texttt{#1}}\fi
\expandafter\ifx\csname urlprefix\endcsname\relax\def\urlprefix{URL }\fi
\expandafter\ifx\csname href\endcsname\relax
  \def\href#1#2{#2} \def\path#1{#1}\fi

\bibitem{Beasley_2000}
C.~Beasley, B.~R. Greene, C.~Lazaroiu, M.~Plesser, D3-branes on partial
  resolutions of abelian quotient singularities of calabi–yau threefolds,
  Nuclear Physics B 566~(3) (2000).
\newblock \href {http://arxiv.org/abs/hep-th/9907186}
  {\path{arXiv:hep-th/9907186}}, \href
  {https://doi.org/10.1016/s0550-3213(99)00646-x}
  {\path{doi:10.1016/s0550-3213(99)00646-x}}.

\bibitem{Maldacena_1999}
J.~Maldacena, The large n limit of superconformal field theories and
  supergravity, International Journal of Theoretical Physics 38~(4) (1999).
\newblock \href {http://arxiv.org/abs/hep-th/9711200}
  {\path{arXiv:hep-th/9711200}}, \href
  {https://doi.org/10.1023/a:1026654312961}
  {\path{doi:10.1023/a:1026654312961}}.

\bibitem{Gubser_1998}
S.~Gubser, I.~Klebanov, A.~Polyakov, Gauge theory correlators from non-critical
  string theory, Physics Letters B 428~(1-2) (1998).
\newblock \href {http://arxiv.org/abs/hep-th/9802109}
  {\path{arXiv:hep-th/9802109}}, \href
  {https://doi.org/10.1016/s0370-2693(98)00377-3}
  {\path{doi:10.1016/s0370-2693(98)00377-3}}.

\bibitem{Witten:1998qj}
E.~Witten, {Anti-de Sitter space and holography}, Adv. Theor. Math. Phys. 2
  (1998) 253--291.
\newblock \href {http://arxiv.org/abs/hep-th/9802150}
  {\path{arXiv:hep-th/9802150}}, \href
  {https://doi.org/10.4310/ATMP.1998.v2.n2.a2}
  {\path{doi:10.4310/ATMP.1998.v2.n2.a2}}.

\bibitem{Morrison:1998cs}
D.~R. Morrison, M.~R. Plesser, {Nonspherical horizons. 1.}, Adv. Theor. Math.
  Phys. 3 (1999) 1--81.
\newblock \href {http://arxiv.org/abs/hep-th/9810201}
  {\path{arXiv:hep-th/9810201}}, \href
  {https://doi.org/10.4310/ATMP.1999.v3.n1.a1}
  {\path{doi:10.4310/ATMP.1999.v3.n1.a1}}.

\bibitem{Klebanov:1998hh}
I.~R. Klebanov, E.~Witten, {Superconformal field theory on three-branes at a
  Calabi-Yau singularity}, Nucl. Phys. B536 (1998) 199--218.
\newblock \href {http://arxiv.org/abs/hep-th/9807080}
  {\path{arXiv:hep-th/9807080}}, \href
  {https://doi.org/10.1016/S0550-3213(98)00654-3}
  {\path{doi:10.1016/S0550-3213(98)00654-3}}.

\bibitem{Intriligator:2003jj}
K.~A. Intriligator, B.~Wecht, {The Exact superconformal R symmetry maximizes
  a}, Nucl. Phys. B667 (2003) 183--200.
\newblock \href {http://arxiv.org/abs/hep-th/0304128}
  {\path{arXiv:hep-th/0304128}}, \href
  {https://doi.org/10.1016/S0550-3213(03)00459-0}
  {\path{doi:10.1016/S0550-3213(03)00459-0}}.

\bibitem{Bertolini:2004xf}
M.~Bertolini, F.~Bigazzi, A.~Cotrone, {New checks and subtleties for AdS/CFT
  and a-maximization}, JHEP 12 (2004) 024.
\newblock \href {http://arxiv.org/abs/hep-th/0411249}
  {\path{arXiv:hep-th/0411249}}, \href
  {https://doi.org/10.1088/1126-6708/2004/12/024}
  {\path{doi:10.1088/1126-6708/2004/12/024}}.

\bibitem{Gubser_1998amax}
S.~S. Gubser, Einstein manifolds and conformal field theories, Physical Review
  D 59~(2) (1998).
\newblock \href {http://arxiv.org/abs/hep-th/9807164}
  {\path{arXiv:hep-th/9807164}}, \href
  {https://doi.org/10.1103/PhysRevD.59.025006}
  {\path{doi:10.1103/PhysRevD.59.025006}}.

\bibitem{Sagnotti:1987tw}
A.~Sagnotti, {Open Strings and their Symmetry Groups}, in: {NATO Advanced
  Summer Institute on Nonperturbative Quantum Field Theory (Cargese Summer
  Institute)}, 1987, pp. 0521--528.
\newblock \href {http://arxiv.org/abs/hep-th/0208020}
  {\path{arXiv:hep-th/0208020}}.

\bibitem{Pradisi:1988xd}
G.~Pradisi, A.~Sagnotti, {Open String Orbifolds}, Phys. Lett. B 216 (1989)
  59--67.
\newblock \href {https://doi.org/10.1016/0370-2693(89)91369-5}
  {\path{doi:10.1016/0370-2693(89)91369-5}}.

\bibitem{Bianchi:1990yu}
M.~Bianchi, A.~Sagnotti, {On the systematics of open string theories}, Phys.
  Lett. B247 (1990) 517--524.
\newblock \href {https://doi.org/10.1016/0370-2693(90)91894-H}
  {\path{doi:10.1016/0370-2693(90)91894-H}}.

\bibitem{Bianchi:1990tb}
M.~Bianchi, A.~Sagnotti, {Twist symmetry and open string Wilson lines}, Nucl.
  Phys. B 361 (1991) 519--538.
\newblock \href {https://doi.org/10.1016/0550-3213(91)90271-X}
  {\path{doi:10.1016/0550-3213(91)90271-X}}.

\bibitem{Polchinski:1995mt}
J.~Polchinski, {Dirichlet Branes and Ramond-Ramond charges}, Phys. Rev. Lett.
  75 (1995) 4724--4727.
\newblock \href {http://arxiv.org/abs/hep-th/9510017}
  {\path{arXiv:hep-th/9510017}}, \href
  {https://doi.org/10.1103/PhysRevLett.75.4724}
  {\path{doi:10.1103/PhysRevLett.75.4724}}.

\bibitem{Angelantonj:2002ct}
C.~Angelantonj, A.~Sagnotti, {Open strings}, Phys. Rept. 371 (2002) 1--150,
  [Erratum: Phys. Rept.376,no.6,407(2003)].
\newblock \href {http://arxiv.org/abs/hep-th/0204089}
  {\path{arXiv:hep-th/0204089}}, \href
  {https://doi.org/10.1016/S0370-1573(02)00273-9,
  10.1016/S0370-1573(03)00006-1} {\path{doi:10.1016/S0370-1573(02)00273-9,
  10.1016/S0370-1573(03)00006-1}}.

\bibitem{Argurio:2017upa}
R.~Argurio, M.~Bertolini, {Orientifolds and duality cascades: confinement
  before the wall}, JHEP 02 (2018) 149.
\newblock \href {http://arxiv.org/abs/1711.08983} {\path{arXiv:1711.08983}},
  \href {https://doi.org/10.1007/JHEP02(2018)149}
  {\path{doi:10.1007/JHEP02(2018)149}}.

\bibitem{Bianchi:2013gka}
M.~Bianchi, G.~Inverso, J.~F. Morales, D.~Ricci~Pacifici, {Unoriented Quivers
  with Flavour}, JHEP 01 (2014) 128.
\newblock \href {http://arxiv.org/abs/1307.0466} {\path{arXiv:1307.0466}},
  \href {https://doi.org/10.1007/JHEP01(2014)128}
  {\path{doi:10.1007/JHEP01(2014)128}}.

\bibitem{Antinucci:2020yki}
A.~Antinucci, S.~Mancani, F.~Riccioni, {Infrared duality in unoriented Pseudo
  del Pezzo}, Phys. Lett. B 811 (2020) 135902.
\newblock \href {http://arxiv.org/abs/2007.14749} {\path{arXiv:2007.14749}},
  \href {https://doi.org/10.1016/j.physletb.2020.135902}
  {\path{doi:10.1016/j.physletb.2020.135902}}.

\bibitem{Feng_2003}
B.~Feng, S.~Franco, A.~Hanany, Y.~He, Unhiggsing the del pezzo, Journal of High
  Energy Physics 2003~(08) (2003).
\newblock \href {http://arxiv.org/abs/hep-th/0209228}
  {\path{arXiv:hep-th/0209228}}, \href
  {https://doi.org/10.1088/1126-6708/2003/08/058}
  {\path{doi:10.1088/1126-6708/2003/08/058}}.

\bibitem{Feng_2004}
B.~Feng, Y.~He, F.~Lam, On correspondences between toric singularities and
  -webs, Nuclear Physics B 701~(1-2) (2004).
\newblock \href {http://arxiv.org/abs/hep-th/0403133}
  {\path{arXiv:hep-th/0403133}}, \href
  {https://doi.org/10.1016/j.nuclphysb.2004.08.048}
  {\path{doi:10.1016/j.nuclphysb.2004.08.048}}.

\bibitem{Hanany:2012hi}
A.~Hanany, R.~Seong, Brane tilings and reflexive polygons, Fortsch. Phys. 60
  (2012) 695--803.
\newblock \href {http://arxiv.org/abs/1201.2614} {\path{arXiv:1201.2614}},
  \href {https://doi.org/10.1002/prop.201200008}
  {\path{doi:10.1002/prop.201200008}}.

\bibitem{Kutasov:2003iy}
D.~Kutasov, A.~Parnachev, D.~A. Sahakyan, {Central charges and U(1)(R)
  symmetries in N=1 superYang-Mills}, JHEP 11 (2003) 013.
\newblock \href {http://arxiv.org/abs/hep-th/0308071}
  {\path{arXiv:hep-th/0308071}}, \href
  {https://doi.org/10.1088/1126-6708/2003/11/013}
  {\path{doi:10.1088/1126-6708/2003/11/013}}.

\bibitem{Kutasov:1995np}
D.~Kutasov, A.~Schwimmer, {On duality in supersymmetric Yang-Mills theory},
  Phys. Lett. B 354 (1995) 315--321.
\newblock \href {http://arxiv.org/abs/hep-th/9505004}
  {\path{arXiv:hep-th/9505004}}, \href
  {https://doi.org/10.1016/0370-2693(95)00676-C}
  {\path{doi:10.1016/0370-2693(95)00676-C}}.

\bibitem{Intriligator:1995ff}
K.~A. Intriligator, {New RG fixed points and duality in supersymmetric SP(N(c))
  and SO(N(c)) gauge theories}, Nucl. Phys. B 448 (1995) 187--198.
\newblock \href {http://arxiv.org/abs/hep-th/9505051}
  {\path{arXiv:hep-th/9505051}}, \href
  {https://doi.org/10.1016/0550-3213(95)00296-5}
  {\path{doi:10.1016/0550-3213(95)00296-5}}.

\bibitem{Leigh:1995qp}
R.~G. Leigh, M.~J. Strassler, {Duality of Sp(2N(c)) and S0(N(c)) supersymmetric
  gauge theories with adjoint matter}, Phys. Lett. B 356 (1995) 492--499.
\newblock \href {http://arxiv.org/abs/hep-th/9505088}
  {\path{arXiv:hep-th/9505088}}, \href
  {https://doi.org/10.1016/0370-2693(95)00871-H}
  {\path{doi:10.1016/0370-2693(95)00871-H}}.

\bibitem{Intriligator:1995ax}
K.~A. Intriligator, R.~G. Leigh, M.~J. Strassler, {New examples of duality in
  chiral and nonchiral supersymmetric gauge theories}, Nucl. Phys. B 456 (1995)
  567--621.
\newblock \href {http://arxiv.org/abs/hep-th/9506148}
  {\path{arXiv:hep-th/9506148}}, \href
  {https://doi.org/10.1016/0550-3213(95)00473-1}
  {\path{doi:10.1016/0550-3213(95)00473-1}}.

\bibitem{Bianchi:2014qma}
M.~Bianchi, S.~Cremonesi, A.~Hanany, J.~F. Morales, D.~Ricci~Pacifici,
  R.~Seong, {Mass-deformed Brane Tilings}, JHEP 10 (2014) 027.
\newblock \href {http://arxiv.org/abs/1408.1957} {\path{arXiv:1408.1957}},
  \href {https://doi.org/10.1007/JHEP10(2014)027}
  {\path{doi:10.1007/JHEP10(2014)027}}.

\bibitem{Bianchi:2020fuk}
M.~Bianchi, D.~Bufalini, S.~Mancani, F.~Riccioni, {Mass deformations of
  unoriented quiver theories}, JHEP 07 (2020).
\newblock \href {http://arxiv.org/abs/2003.09620} {\path{arXiv:2003.09620}},
  \href {https://doi.org/10.1007/JHEP07(2020)015}
  {\path{doi:10.1007/JHEP07(2020)015}}.

\bibitem{Cvetic:2005ft}
M.~Cvetic, H.~Lu, D.~N. Page, C.~N. Pope, {New Einstein-Sasaki spaces in five
  and higher dimensions}, Phys. Rev. Lett. 95 (2005) 071101.
\newblock \href {http://arxiv.org/abs/hep-th/0504225}
  {\path{arXiv:hep-th/0504225}}, \href
  {https://doi.org/10.1103/PhysRevLett.95.071101}
  {\path{doi:10.1103/PhysRevLett.95.071101}}.

\bibitem{Martelli:2005wy}
D.~Martelli, J.~Sparks, {Toric Sasaki-Einstein metrics on S**2 x S**3}, Phys.
  Lett. B 621 (2005) 208--212.
\newblock \href {http://arxiv.org/abs/hep-th/0505027}
  {\path{arXiv:hep-th/0505027}}, \href
  {https://doi.org/10.1016/j.physletb.2005.06.059}
  {\path{doi:10.1016/j.physletb.2005.06.059}}.

\bibitem{Franco:2005sm}
S.~Franco, A.~Hanany, D.~Martelli, J.~Sparks, D.~Vegh, B.~Wecht, {Gauge
  theories from toric geometry and brane tilings}, JHEP 01 (2006) 128.
\newblock \href {http://arxiv.org/abs/hep-th/0505211}
  {\path{arXiv:hep-th/0505211}}, \href
  {https://doi.org/10.1088/1126-6708/2006/01/128}
  {\path{doi:10.1088/1126-6708/2006/01/128}}.

\bibitem{Bianchi:2007wy}
M.~Bianchi, F.~Fucito, J.~F. Morales, {D-brane instantons on the T**6 / Z(3)
  orientifold}, JHEP 07 (2007) 038.
\newblock \href {http://arxiv.org/abs/0704.0784} {\path{arXiv:0704.0784}},
  \href {https://doi.org/10.1088/1126-6708/2007/07/038}
  {\path{doi:10.1088/1126-6708/2007/07/038}}.

\bibitem{Franco:2007ii}
S.~Franco, A.~Hanany, D.~Krefl, J.~Park, A.~M. Uranga, D.~Vegh, {Dimers and
  orientifolds}, JHEP 09 (2007) 075.
\newblock \href {http://arxiv.org/abs/0707.0298} {\path{arXiv:0707.0298}},
  \href {https://doi.org/10.1088/1126-6708/2007/09/075}
  {\path{doi:10.1088/1126-6708/2007/09/075}}.

\bibitem{Argurio:2020dko}
R.~Argurio, M.~Bertolini, S.~Franco, E.~Garc\'\i{}a-Valdecasas, S.~Meynet,
  A.~Pasternak, V.~Tatitscheff, {Dimers, Orientifolds and Anomalies}, JHEP 02
  (2021) 153.
\newblock \href {http://arxiv.org/abs/2009.11291} {\path{arXiv:2009.11291}},
  \href {https://doi.org/10.1007/JHEP02(2021)153}
  {\path{doi:10.1007/JHEP02(2021)153}}.

\bibitem{Garcia-Valdecasas:2021znu}
E.~Garc\'\i{}a-Valdecasas, S.~Meynet, A.~Pasternak, V.~Tatitscheff, {Dimers in
  a Bottle}, JHEP 04 (2021) 274.
\newblock \href {http://arxiv.org/abs/2101.02670} {\path{arXiv:2101.02670}},
  \href {https://doi.org/10.1007/JHEP04(2021)274}
  {\path{doi:10.1007/JHEP04(2021)274}}.

\bibitem{Benvenuti:2017lle}
S.~Benvenuti, S.~Giacomelli, {Supersymmetric gauge theories with decoupled
  operators and chiral ring stability}, Phys. Rev. Lett. 119~(25) (2017)
  251601.
\newblock \href {http://arxiv.org/abs/1706.02225} {\path{arXiv:1706.02225}},
  \href {https://doi.org/10.1103/PhysRevLett.119.251601}
  {\path{doi:10.1103/PhysRevLett.119.251601}}.

\bibitem{Uranga:1998vf}
A.~M. Uranga, {Brane configurations for branes at conifolds}, JHEP 01 (1999)
  022.
\newblock \href {http://arxiv.org/abs/hep-th/9811004}
  {\path{arXiv:hep-th/9811004}}, \href
  {https://doi.org/10.1088/1126-6708/1999/01/022}
  {\path{doi:10.1088/1126-6708/1999/01/022}}.

\bibitem{Elitzur:1997fh}
S.~Elitzur, A.~Giveon, D.~Kutasov, {Branes and N=1 duality in string theory},
  Phys. Lett. B 400 (1997) 269--274.
\newblock \href {http://arxiv.org/abs/hep-th/9702014}
  {\path{arXiv:hep-th/9702014}}, \href
  {https://doi.org/10.1016/S0370-2693(97)00375-4}
  {\path{doi:10.1016/S0370-2693(97)00375-4}}.

\bibitem{Berkooz:1995km}
M.~Berkooz, {The Dual of supersymmetric SU(2k) with an antisymmetric tensor and
  composite dualities}, Nucl. Phys. B 452 (1995) 513--525.
\newblock \href {http://arxiv.org/abs/hep-th/9505067}
  {\path{arXiv:hep-th/9505067}}, \href
  {https://doi.org/10.1016/0550-3213(95)00400-M}
  {\path{doi:10.1016/0550-3213(95)00400-M}}.

\bibitem{Pouliot:1995me}
P.~Pouliot, {Duality in SUSY SU(N) with an antisymmetric tensor}, Phys. Lett. B
  367 (1996) 151--156.
\newblock \href {http://arxiv.org/abs/hep-th/9510148}
  {\path{arXiv:hep-th/9510148}}, \href
  {https://doi.org/10.1016/0370-2693(95)01427-6}
  {\path{doi:10.1016/0370-2693(95)01427-6}}.

\bibitem{Sakai:1997xs}
T.~Sakai, {Duality in supersymmetric SU(N) gauge theory with a symmetric
  tensor}, Mod. Phys. Lett. A 12 (1997) 1025--1034.
\newblock \href {http://arxiv.org/abs/hep-th/9701155}
  {\path{arXiv:hep-th/9701155}}, \href
  {https://doi.org/10.1142/S0217732397001047}
  {\path{doi:10.1142/S0217732397001047}}.

\bibitem{Luty:1996cg}
M.~A. Luty, M.~Schmaltz, J.~Terning, {A Sequence of duals for Sp(2N)
  supersymmetric gauge theories with adjoint matter}, Phys. Rev. D 54 (1996)
  7815--7824.
\newblock \href {http://arxiv.org/abs/hep-th/9603034}
  {\path{arXiv:hep-th/9603034}}, \href
  {https://doi.org/10.1103/PhysRevD.54.7815}
  {\path{doi:10.1103/PhysRevD.54.7815}}.

\bibitem{Garcia-Etxebarria:2015hua}
I.~García-Etxebarria, B.~Heidenreich, {Strongly coupled phases of
  $\mathcal{N}=1$ S-duality}, JHEP 09 (2015) 032.
\newblock \href {http://arxiv.org/abs/1506.03090} {\path{arXiv:1506.03090}},
  \href {https://doi.org/10.1007/JHEP09(2015)032}
  {\path{doi:10.1007/JHEP09(2015)032}}.

\end{thebibliography}

\end{document}